\documentclass[10pt,a4paper]{article}
\pdfoutput=1

\usepackage[super, numbers, comma, sort&compress]{natbib}
\usepackage{setspace}

\usepackage{times}
\usepackage[textwidth=17.8cm, textheight=24.5cm]{geometry}
\usepackage{siunitx}
\usepackage{hyperref}
\hypersetup{urlcolor=blue, citecolor=blue, linkcolor=blue, colorlinks=true}

\usepackage[table]{xcolor}

\usepackage{times}
\usepackage{lscape}
\usepackage{rotating} 
   
\usepackage{graphicx} 
\usepackage[shortcuts]{extdash} 
\usepackage{tabularx, booktabs}
\usepackage{caption} 
\usepackage{titling} 
\usepackage{wasysym}

\usepackage{multicol}
\usepackage{dblfloatfix}    
\usepackage{float}	

\usepackage{lineno}

\begin{document} 

\noindent\LARGE Metals by micro-scale additive manufacturing: comparison of microstructure and mechanical properties\\
\smallskip

\noindent\large Alain Reiser$^1$, 
Lukas Koch$^1$,
Kathleen A. Dunn$^2$,
Toshiki Matsuura$^3$, Futoshi Iwata$^3$,
Ofer Fogel$^4$, Zvi Kotler$^4$,
Nanjia Zhou$^{5,6}$, 
Kristin Charipar$^7$, Alberto Piqu\'e$^7$, 
Patrik Rohner$^8$, Dimos Poulikakos$^8$, 
Sanghyeon Lee$^9$, Seung K. Seol$^{9, 10}$, 
Ivo Utke$^{11}$, 
Cathelijn van Nisselroy$^{12}$, Tomaso Zambelli$^{12}$,
Jeffrey M. Wheeler$^1$, Ralph Spolenak$^{1\ast}$
\bigskip

\footnotesize
\noindent$^{1}$Laboratory for Nanometallurgy, Department of Materials, ETH Z\"urich, Vladimir-Prelog-Weg 1-5/10, 8093 Z\"urich, Switzerland.\\
{$^{2}$College of Nanoscale Science \& Engineering, SUNY Polytechnic Institute, 257 Fuller Road, Albany, NY, 12203 USA.}\\
{$^{3}$Graduate School of Integrated Science and Technology, Shizuoka University, Johoku, Naka-ku, Hamamatsu, 432-8561, Japan.}\\
{$^{4}$Additive Manufacturing Laboratory, Orbotech Ltd., P.O. Box 215, Yavne 81101, Israel.}\\
{$^{5}$ Laboratory of 3D Micro/Nano Fabrication and Characterization of Zhejiang Province, School of Engineering, Westlake University, 18 Shilongshan Road, Hangzhou 310024, Zhejiang Province, China.}\\
{$^{6}$ Institute of Advanced Technology, Westlake Institute for Advanced Study, 18 Shilongshan Road, Hangzhou 310024, Zhejiang Province, China.}\\
{$^{7}$Naval Research Laboratory, Materials Science and Technology Division, 4555 Overlook Ave. SW, Washington, DC 20375 USA.}\\
{$^{8}$Laboratory of Thermodynamics in Emerging Technologies, Department of Mechanical and Process Engineering, ETH Z\"urich, Sonneggstr. 3, 8092 Z\"urich, Switzerland.}\\
{$^{9}$Nano Hybrid Technology Research Center, Korea Electrotechnology Research Institute (KERI), Changwon-Si, Gyeongsangnam-do, 51543, Republic of Korea.}\\
{$^{10}$Electrical Functionality Materials Engineering, University of Science and Technology (UST), Changwon-Si, Gyeongsangnam-do, 51543, Republic of Korea.}\\
{$^{11}$Laboratory of Mechanics for Materials and Nanostructures, Empa,  Feuerwerkerstrasse 39, 3602 Thun, Switzerland.}\\
{$^{12}$Laboratory of Biosensors and Bioelectronics, Department of Information Technology and Electrical Engineering, ETH Z\"urich, Gloriastrasse 35, 8092 Z\"urich, Switzerland.}\\
\smallskip

\noindent{$^\ast$To whom correspondence should be addressed; E-mail: ralph.spolenak@mat.ethz.ch}

\normalsize
\singlespacing



\bf
 Many emerging applications in microscale engineering rely on the fabrication of three-dimensional architectures in inorganic materials. Small-scale additive manufacturing (AM) aspires to provide flexible and facile access to these geometries. Yet, the synthesis of device-grade inorganic materials is still a key challenge towards the implementation of AM in microfabrication. Here, we present a comprehensive overview of the microstructural and mechanical properties of metals fabricated by most state-of-the-art AM methods that offer a spatial resolution $\leq$\SI{10}{\micro\meter}. Standardized sets of samples were studied by cross-sectional electron microscopy, nanoindentation and microcompression. We show that current microscale AM techniques synthesize metals with a wide range of microstructures and elastic and plastic properties, including materials of dense and crystalline microstructure with excellent mechanical properties that compare well to those of thin-film nanocrystalline materials. The large variation in materials performance can be related to the individual microstructure, which in turn is coupled to the various physico-chemical principles exploited by the different printing methods. The study provides practical guidelines for users of small-scale additive methods and establishes a baseline for the future optimization of the properties of printed metallic objects~\==~a significant step towards the potential establishment of AM techniques in microfabrication.
 \normalfont
\bigskip

\begin{multicols}{2}
\section{Introduction}
Complex three-dimensional structures with feature sizes in the micro- and nanometer range currently enable new areas of exciting growth in materials science and engineering. 3D geometries are often superior to their established, planar counterparts and enable a wide range of emerging applications in fields ranging from microelectromechanical systems (MEMS) and electronics to biomedicine and metamaterials\cite{Onses2015a,Hohmann2015,Rogers2016,Zhang2017a,Hirt2017,Montemayor2015,Soukoulis2011,Lee2012c}.
As progress in 3D engineering follows from the ability to realize relevant structures from suitable materials\cite{Zhang2017a}, the development of innovative small-scale 3D fabrication methods is essential. At the required scales, additive manufacturing (AM) promises the greatest flexibility and closest control of synthesized geometries in comparison to alternative fabrication approaches, e.g., self assembly and stress-controlled deformation\cite{Zhang2017a}. Hence, a large number of microscale AM techniques with a minimum feature size $<$\SI{10}{\micro\meter} have emerged in the past decade\cite{Lewis2006,Hohmann2015,Hirt2017,Onses2015a}, 
paving the way to additively manufactured 3D metamaterials\cite{Gansel2009,Ergin2010,Schaedler2011,Jang2013NM,Xia2019}, printed electrical circuit elements\cite{Ahn2009a,Hu2010,Wang2010,Schneider2016,Zhou2017}, or small-scale sensors\cite{Gavagnin2014,Luo2017,Arnold2018,Sattelkow2019}.
\par
Yet, to enable above-mentioned progress, AM techniques must provide the materials that today's microfabrication processes require: combinations of device-grade, inorganic materials that provide structural, electronic, and optical functionality. This need is illustrated by the widespread use of two-photon lithography (TPL) as a mere templating tool: after structuring an organic photoresist, the photosensitive material is exchanged with inorganic materials necessary for proper functioning of the final structure, e.g., materials with a high dielectric constant for optical metamaterials\cite{Gansel2009,Ergin2010}, or materials with a high stiffness or strength for mechanical metamaterials\cite{Schaedler2011,Jang2013NM,Xia2019}. These materials are invariably deposited by established microfabrication processes such as atomic layer deposition\cite{Jang2013NM} or electrodeposition\cite{Gansel2009,Schaedler2011}. While this symbiosis is an ideal approach for many applications, direct patterning of high-performance materials is preferable for a number of reasons, including a reduction in the number of processing steps, the realization of geometries that are not dictated by requirements for subsequent template inversion (proper filling of complex templates can be demanding\cite{Schurch2018}) and~\==~ maybe most interesting for advanced designs of devices and materials~\==~the possibility of spatially varying properties (not accessible with homogeneous coating or filling). 
\par
Unfortunately, direct AM of inorganic materials often results in materials that cannot yet satisfy the standards of modern microfabrication\cite{Hirt2017}. This shortcoming, in combination with challenging scalability and limited throughput, is a major handicap for incorporating AM in advanced micro- and nanofabrication routines. Materials engineering is thus necessary to improve the quality of printed inorganic materials. 
A first and crucial step towards this goal is the knowledge of the deposited materials' microstructure and its effect on the materials' properties. {Here, we identify a deficit in existing literature: while some recent studies have covered materials optimization of particular AM techniques\cite{Winter2016,Daryadel2018,Arnold2018,Sattelkow2019}, characterization of general microstructure-property relationships of printed inorganic materials is in its infancy~\==~a shortfall the present paper tackles, focusing on the mechanical performance of metals.}
\par
The indispensability of metals for the microfabrication of high-performance 3D devices has motivated the development of multiple direct metal AM processes in the past decade\cite{Hirt2017}. As these techniques explore different physico-chemical principles, the microstructure of the deposited metals varies greatly with respect to purity, defect density and porosity. As a result, a large variance in materials properties is observed. Studies on electrical properties report typical conductivities of \num{e-5}~\==~0.55 $\times$ bulk conductivity\cite{Ahn2009a,Skylar-Scott2016,Winter2016,Botman2009a,Utke2008,Hu2010,Tanaka2006,Maruo2008}. Similarly, elastic and plastic properties of additively fabricated metals vary over a range of $\approx$ 0.05~\==~1 $\times$ bulk values (Table \ref{tab:techniquesoverview}). While these studies enable isolated insights into the materials synthesized by different laboratories and techniques, their entirety does not allow for a properly founded overview of today's materials quality and related issues~\==~the exemplary nature of previous work, the large variation in sample geometry and the different characterization techniques used prevent a fair comparison between techniques and their materials.
\par
Here, we present a comprehensive overview of the microstructure of metals printed by most contemporary small-scale AM processes, and relate the microstructure of these metals to their mechanical performance. Mechanical properties of 3D metal structures are a key for many applications at small scales, both in traditional MEMS\cite{VanSpengen2003} and for projected implementations of printed metals, {e.g.}, unsuspended interconnects for flexible electronics\cite{Ahn2009a}, out-of-plane, high-aspect-ratio switches\cite{Yi2016,Farahani2014}, sensors\cite{Luo2017,Arnold2018} and probes\cite{Gavagnin2014,Sattelkow2019}, or printed actuators and manipulators\cite{Farahani2014,Fogel2018}. Studying a standardized set of samples by cross-sectional electron microscopy, nanoindentation and microcompression, we show that a range of characteristic microstructures causes a significant variation in elastic and plastic properties within one order of magnitude. On one hand, the standardized approach used in this study allows a fair comparison of the capabilities of the various small-scale AM methods (within the boundaries discussed). On the other, the presented overview is the groundwork for future optimization of materials performance.


\section{Overview of included AM methods and samples}
\begin{figure*}[htbp] 
   \centering
   \includegraphics[width=178mm]{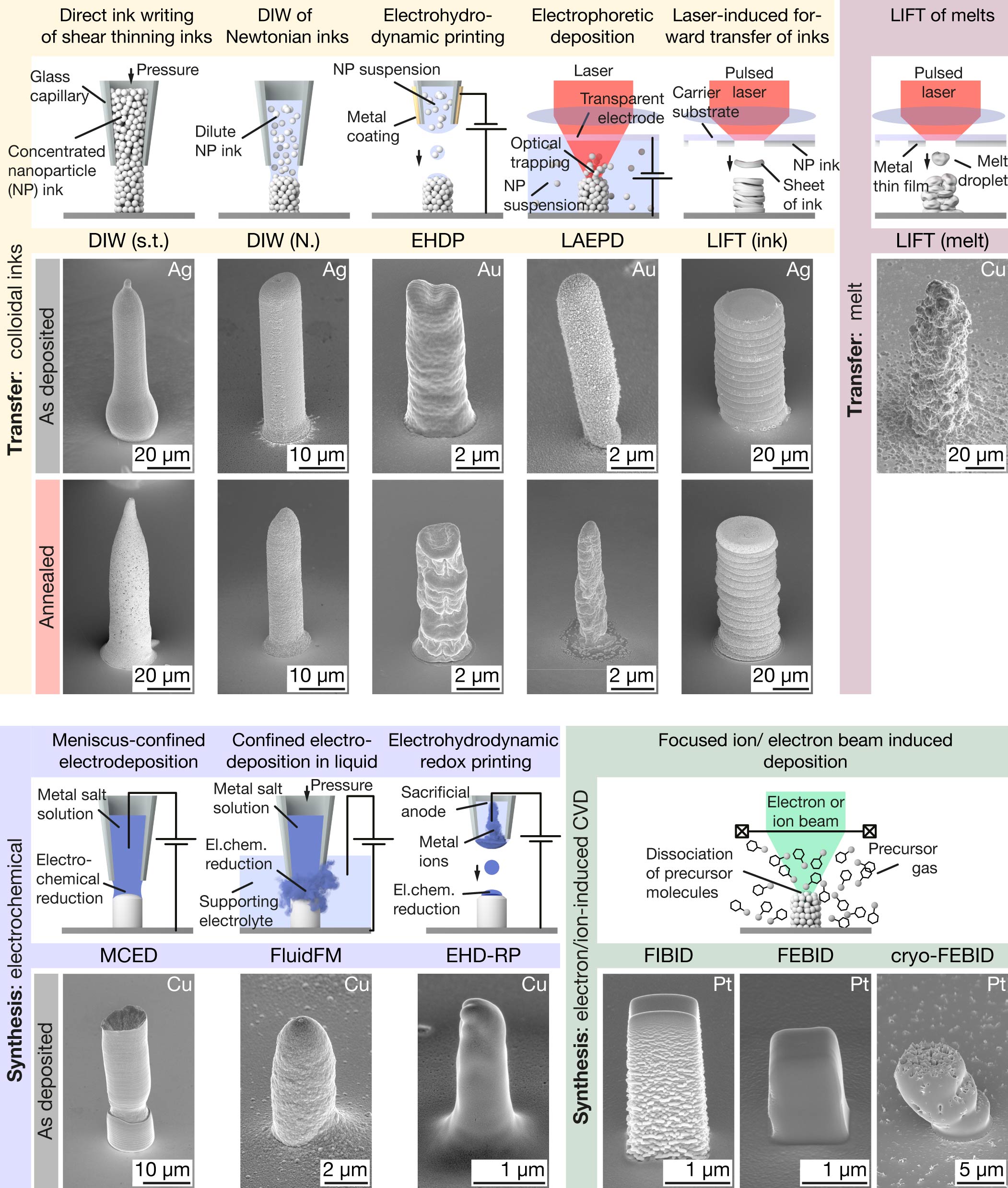} 
   \caption{\textbf{Small-scale metal AM methods included in this study and pillars printed by these techniques.} Small-sale AM methods are grouped into \textit{transfer} and \textit{synthesis} methods, based on their principle of metal deposition. Subgroups include: Transfer of colloids, transfer of melts, electrochemical synthesis and synthesis via electron/ion-induced CVD. The SE micrographs show representative pillars printed with each of the techniques tested in this study. As-printed and annealed pillars are shown if thermal annealing was performed (not the same samples). Samples for DIW and LIFT (ink) are printed from Ag inks, and samples from EHDP and LAEPD from suspensions of Au nanoparticles. The LIFT (melt) pillar is Cu (for Au, see Supplementary Figure~\ref{fig:SI_Morphology_LIFTmelt}), as are the structures for MCED, the FluidFM and EHD-RP. FIBID and (cryo-)FEBID pillars were deposited from Methylcyclopentadienyl platinum (IV) trimethyl (MeCpPt(Me)$_3$). Tilt angle of all micrographs: \SI{55}{\degree}.}
\label{fig:geometry_pillars}
\end{figure*}

The microscale metal AM techniques included in this study are listed in Table \ref{tab:techniquesoverview}, and schematics of the respective working principles are shown in Figure~\ref{fig:geometry_pillars}. The processes are classified according to their underlying principle of material deposition to emphasize the fundamental similarities that determine the properties of the resulting materials. We group the techniques in two main categories: \textit{transfer} and \textit{synthesis} techniques. Transfer techniques require the previous synthesis of metallic materials before the actual AM process~\==~the subsequent deposition simply transfers the pre-synthesized material to the location of interest. In contrast, the synthesis methods rely on the growth of the metal at the location of interest during the AM process. Both categories are further divided into the sub-groups of methods that do or do not use colloidal inks (transfer techniques), and methods that use wet electrochemistry or local electron/ion-initiated surface reactions with physisorbed molecules supplied from the gas phase (synthesis methods). Relevant techniques not covered in this paper are two-photon-induced reduction of metal ions\cite{Tanaka2006}, pyrolysis of TPL-structured, metal-containing resins\cite{Vyatskikh2018}, and implosion fabrication\cite{Oran2018}.
\par
\begin{sidewaystable*} 
\caption{\textbf{Studied small-scale metal AM techniques.} Techniques and printed metals tested in this study, including literature values and measured data for the mechanical performance of the printed materials: $E$: Young's modulus, $H$: hardness, $\sigma_\text{y}$: yield stress, $\sigma_\text{0.07}$: flow stress at \SI{7}{\percent} strain. For inks, as-deposited (ad.) and thermally annealed (ann.) metals are discriminated. References in the first column refer to historically important publications and notable review articles. Consult Supplementary Table \ref{tab:EH} for average values of the mechanical properties of all tested samples including standard deviations. $^\ast$Note: Pt nanoparticles embedded in a carbonaceous matrix. 
}\label{tab:techniquesoverview}
\centering
\resizebox{\textheight}{!}{
\begin{tabular}{l  l | c | c | c | c | c | c | c}
\toprule
\textbf{Technique}							&											& \textbf{Printed metal}			& \multicolumn{6}{l}{\textbf{Mechanical properties}} 		\\
										&											&(This study)				& \multicolumn{3}{l}{Literature}		 						& \multicolumn{3}{l}{\textbf{\cellcolor{gray}This study}}		\\ 
										&											& 					& $E$ [GPa]			& $H$ [GPa]	& $\sigma_\text{y}$ [GPa] & \cellcolor{lightgray}$E$ [GPa]					& \cellcolor{lightgray}$H$ [GPa]				& \cellcolor{lightgray}$\sigma_\text{0.07}$ [GPa]	\\\hline
\rule{0pt}{1.2em}\textbf{Transfer techniques}		&											&					&					&			&					&									&								& 	\\
\rule{0pt}{1.2em}Colloidal inks		&\textbf{DIW (s.t.)}~\==~Direct ink writing of shear-thinning inks\cite{Ahn2009a,Zhou2017}	& Ag		&					&			&					& 12.6~\==~33.3 (ad.), 31.1~\==~42.1 (ann.)	& 0.141 (ad.), 0.435~\==~0.461 (ann.)	& 0.265 (ad.), 0.168~\==~0.289 (ann.)	\\	 
				&\textbf{DIW (N.)}~\==~Direct ink writing of Newtonian inks\cite{Lee2017}	 			& Ag					&					&			&					& 4.62~--~35.7 (ad.), 48.8~--~65.7 (ann.)		& 0.062 (ad.), 0.952 (ann.)				& 0.315 (ad.), 0.422 (ann.) \\
				&\textbf{EHDP}~\==~Electrohydrodynamic printing of nanoparticle inks\cite{Park2007,Galliker2012,Onses2015a}	& Au	& \num{2E-4} (ad.)\cite{Galliker2014a}, 6 (ann.)\cite{Schneider2013}	& &	& 0.52~--~1.09 (ad.), 31.3~--~39.8 (ann.)	& 0.048 (ad.)	& 0.017 (ad.), 0.246~--~0.411 (ann.)\\
				&\textbf{LAEPD}~\==~Laser-assisted electrophoretic deposition of nanoparticles\cite{Takai2014}	& Au				& 1.5 (ad.)\cite{Takai2014}&			&					& 7.29~--~9.09 (ad.), 47.1~--~54.2 (ann.)		& 0.369 (ad.), 1.42 (ann.)				& 0.146 (ad.), 0.366 (ann.)\\
				&\textbf{LIFT (ink)}~\==~Laser-induced forward transfer of nanoparticle inks\cite{Pique2008,Wang2010,Mathews2013,Pique2016a}	& Ag					& 54.8\cite{Birnbaum2010c}	& 1.37\cite{Birnbaum2010c}	&	& 1.16~--~1.94 (ad.), 12.8~--~36.4 (ann.)		& 0.108 (ad.), 0.400~--~0.661 (ann.)		& 0.013 (ad.), 0.088~--~0.118 (ann.) \\
\rule{0pt}{2em}Melts	&\textbf{LIFT (melt)}~\==~Laser-induced forward transfer of thin films melts\cite{Zenou2015a}	& Au, Cu		& 12 (Cu), 9 (Au)\cite{Fogel2018}	&	&					& 49.8~--73.2 (Cu), 24.3~--~28.3 (Au)		& 1.66 (Cu), 0.293 (Au)				& 0.415 (Cu), 0.186 (Au)\\
\hline
\rule{0pt}{1.2em}\textbf{Synthesis techniques}	&							&					&					&					&						&		&	\\
\rule{0pt}{1.2em}Electrochemical	&\textbf{MCED}~\==~Meniscus-confined electrodeposition\cite{Hu2010,Seol2015a,Behroozfar2017,Daryadel2018,Lin2019}	& Cu		& 128\cite{Behroozfar2017} & 2\cite{Behroozfar2017}	& 0.63~\==~0.962\cite{Daryadel2018, Daryadel2018a} & 114.2~--~121.8	& 2.71	& 0.774	\\
 	&\textbf{FluidFM}~\==~Confined Electrodeposition in Liquid\cite{Hirt2016,Momotenko2016,Ercolano2019}				& Cu					&					&		& 0.7~\==~0.9\cite{Hirt2016}	&134.0~--~138.4						& 2.28							& 0.962\\
	&\textbf{EHD-RP}~\==~Electrohydrodynamic redox printing\cite{Reiser2019}							& Cu					& 81\cite{Reiser2019}	&	& 1~\==~1.5\cite{Reiser2019}		& 80.4~--~81.7							& 2.22							& 1.10~--~1.38\\
\rule{0pt}{2em}Electron/ion-induced CVD &\textbf{FIBID}~\==~Focused ion beam-induced deposition\cite{Utke2008}		& Pt$^\ast$				&					& 			&					& 95.3~--~140.0						& 9.42							& 2.64	\\
	&\textbf{FEBID}~\==~Focused electron beam-induced deposition\cite{Utke2008,Fowlkes2016,Arnold2018}	& Pt$^\ast$		& 10~\==~100\cite{Utke2006,Friedli2009,Lewis2017,Arnold2018}	& 3.7~\==~7.6\cite{Wich2008}	& 1~\==~\SI{2} (tensile) \cite{Utke2006}	& 59.3~--~75.5	& 6.01	& 2.65	\\
	&\textbf{cryo-FEBID}~\==~FEBID at cryogenic temperatures\cite{Bresin2013}						& Pt$^\ast$					&					&			&					& 3.85~--~13.8							& 0.843							& 0.100	\\

\bottomrule
\end{tabular}
}
\end{sidewaystable*} 

For a detailed description of the working principle of the different AM techniques and their performance characteristics, please refer to our review\cite{Hirt2017}. In brief, printing metals from colloidal inks is typically a two-step process that requires post-print thermal annealing to render metallic materials, burning off organic components and sintering the particles. The principles of ink transfer to the substrate vary between the techniques (DIW, EHDP, LAEPD and LIFT (ink), see Table \ref{tab:techniquesoverview} for an explanation of the acronyms), but resulting materials characteristics are comparable: an agglomeration of surfactant-coated micro- or nanoparticles in the as-deposited state, 
and a crystalline, sintered microstructure in the annealed state. LIFT of metallic melts is an ink-free transfer method that does not demand any post-print treatment, as it transfers molten droplets of metals that solidify as pure metals on the substrate. Electrochemical methods (MCED, FluidFM, EHD-RP) use different approaches to localize growth, but all rely on the electrochemical reduction of metal ions and generally offer as-deposited dense and crystalline metals. Focused particle beam methods (FIBID, FEBID, cryo-FEBID) make use of electron-induced dissociation of physisorbed gaseous precursors (typically organometallic compounds) to synthesize metal-carbon composites with metallic characteristics~\==~as the process can be likened to a localized non-thermal CVD process, we use the term "electron/ion-induced CVD" to summarize the working principle.
\par
Samples for this study consisted of a standardized set of pads and pillars that was printed with each technique to assess both the microstructural as well as the mechanical properties of the deposited metals. Due to current process limitations, the elemental nature of the studied metals varies between techniques: Au and Ag for ink-based methods, Cu for electrochemical methods, and Pt for FEBID and FIBID (Table \ref{tab:techniquesoverview}). Further, the dimensions of the samples were adjusted to the spatial resolution and volumetric print speed of the individual technique. Pads ranged from 10 to \SI{200}{\micro\meter} in width and \SI{400}{\nano\meter} to $>$\SI{10}{\micro\meter} in thickness. Pillar diameters were typically $\geq$\SI{1}{\micro\meter}, with maximum diameters of $\approx$\SI{45}{\micro\meter}. A set of as-deposited and annealed samples was prepared for nanoparticle inks, because thermal annealing is generally required for functional performance of these materials. In contrast, only as-deposited but no post-processed (annealed, cured, purified) FIBID and FEBID structures were included, because the vast majority of reported applications is limited to as-deposited materials\cite{Hirt2017} and existing purification and annealing procedures\cite{Botman2009a} are not yet applied by a broad range of practitioners.

\section{Results}
\subsection{Morphology of printed samples}
\begin{figure*}[htbp] 
   \centering
   \includegraphics[width=178mm]{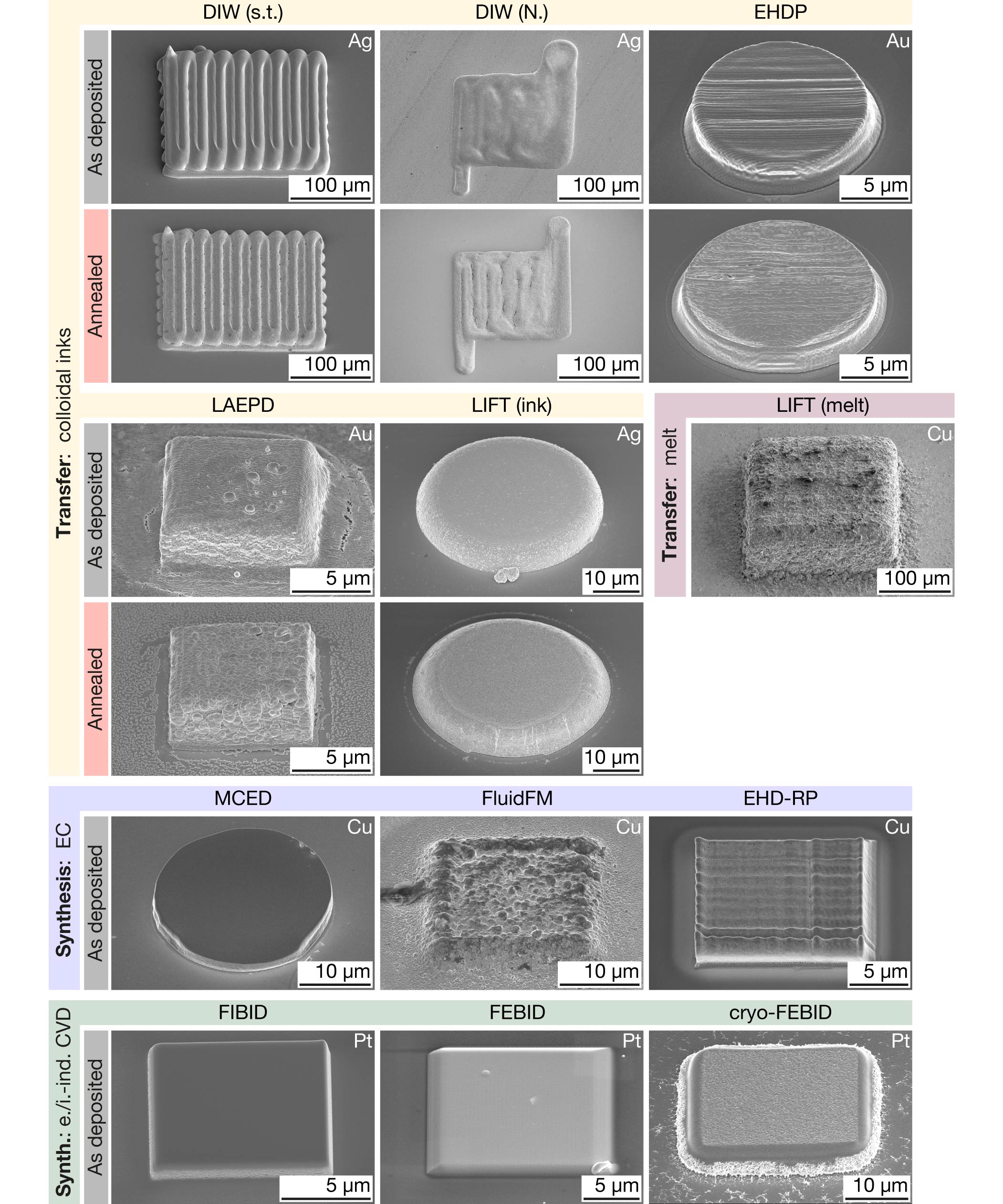} 
   \caption{\textbf{Printed pads.} SE micrographs of representative pads printed with each of the techniques. If thermal annealing was performed, samples of as-printed and annealed pads are shown. Metals: DIW, LIFT (ink): Ag; EHDP, LAEPD: Au. LIFT (melt), MCED, FluidFM, EHD-RP: Cu. FIBID, (cryo-)FEBID: MeCpPt(Me)$_3$. For Au deposited by LIFT (melt), see Supplementary Figure \ref{fig:SI_Morphology_LIFTmelt}. Tilt angle: \SI{45}{\degree}.}
\label{fig:geometry_pads}
\end{figure*}

Representative samples printed with each of the techniques are displayed in Figures~\ref{fig:geometry_pillars} and \ref{fig:geometry_pads}. If multiple annealing states were prepared, the sample annealed the longest at the highest temperature is shown. Cu and Au structures were tested for LIFT (melt), but only Cu is shown here (for Au, refer to Supplementary Figure~\ref{fig:SI_Morphology_LIFTmelt}). Note that the dimensions of the printed structures were dictated by the requirements of mechanical testing: pads were required to be at least a few hundred nanometers in thickness and $\geq$\SI{10}{\micro\meter} in width (without any upper boundaries for thickness and size) and were supposed to have a smallest possible surface roughness. Pillars had to be at least \SI{1}{\micro\meter} in diameter and 2~\==~\SI{3}{\micro\meter} in height (with a maximum diameter of \SI{50}{\micro\meter}). No flat top was required. Any conclusions regarding resolution and smallest printable geometries should be made with these boundary conditions in mind~\==~none of the methods strived to synthesize the smallest geometries for this study.
\par

\par
In general, the morphology and volume of the printed pads and pillars reflect the deposition principle of each individual technique. FEBID and FIBID geometries show highest resolution and lowest surface roughness 
but their build volume is limited to a few cubic micrometers. In contrast, structures printed by LIFT (melt) feature the highest microscale surface roughness 
but their volume is only rivaled by DIW samples. DIW pillars drawn from a single filament are smooth, but the pads' surface modulation with a modulation length of tens of micrometers clearly reveals the in-plane hatch pattern. Techniques with submicrometer feature size (EHDP, EHD-RP, FIBID, FEBID and cryo-FEBID) enable printing of micrometer-sized pillars with in-plane hatching and hence control the pillar's shape (for example square in the case of EHD-RP, FIBID and FEBID, Supplementary Figure \ref{fig:SI_Morphology_squarepillars}). Commonly, pads were fabricated with in-plane hatching. The few pads deposited without in-plane hatching show lowest surface roughness (LIFT (ink), MCED), but the stacking of layers results in a pronounced roughness of LIFT (ink) printed pillars. In cryo-FEBID pillars, substrate drift during the cycling between precursor deposition and patterning caused a shift between individual layers. 
\par
\begin{figure*}[htbp] 
   \centering
   \includegraphics[width=178mm]{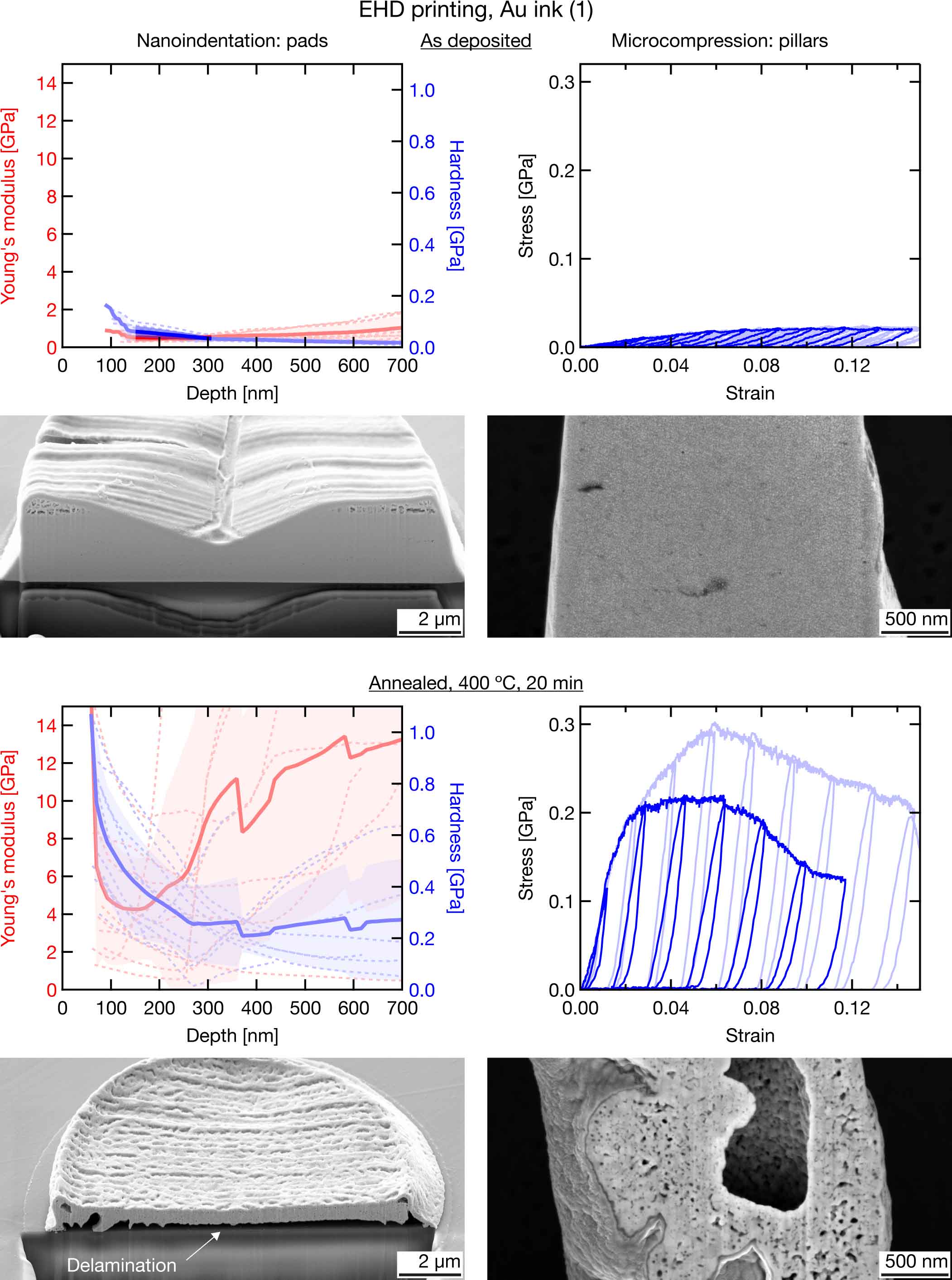} 
   \caption{Electrohydrodynamic printing, Au ink (EHDP). The as-deposited pad was indented before taking the image, hence the pyramidal indent. Nanoindentation data of annealed pads was not analyzed, as the large-scale delamination of the pads upon annealing prevented the collection of representative data.}
\label{fig:SI_EHD1}
\end{figure*}
\begin{figure*}[htbp] 
   \centering
   \includegraphics[width=178mm]{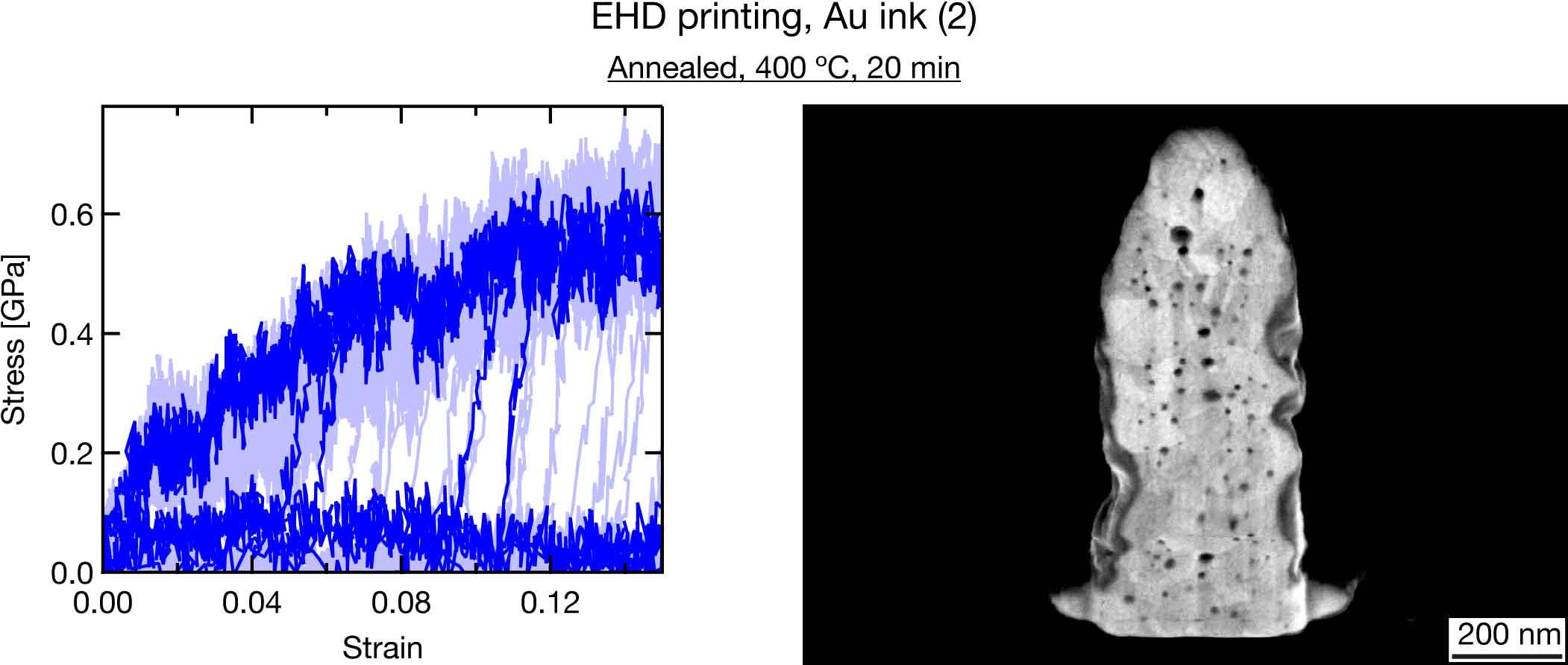} 
   \caption{Electrohydrodynamic printing, Au ink (EHDP). Microcompression data and representative cross-section of annealed pillars 300~\==~\SI{350}{\nano\meter} in diameter.}
\label{fig:SI_EHD2}
\end{figure*}

Thermal annealing of DIW and LIFT (ink) geometries exhibits good shape retention. In contrast, sintering of EHDP and LAEPD structures caused shrinkage and warping. In the case of EHDP, the shrinkage even resulted in unsuspended pads that prohibited reliable indentation testing (Supplementary Figure~\ref{fig:SI_EHD1}). The warping of EHDP pillars was less pronounced for small diameters (printed without in-plane hatching and lower flow rates, Supplementary Figure~\ref{fig:SI_EHD2}).

\subsection{Microstructure}
\begin{figure*}[htbp] 
   \centering
   \includegraphics[width=178mm]{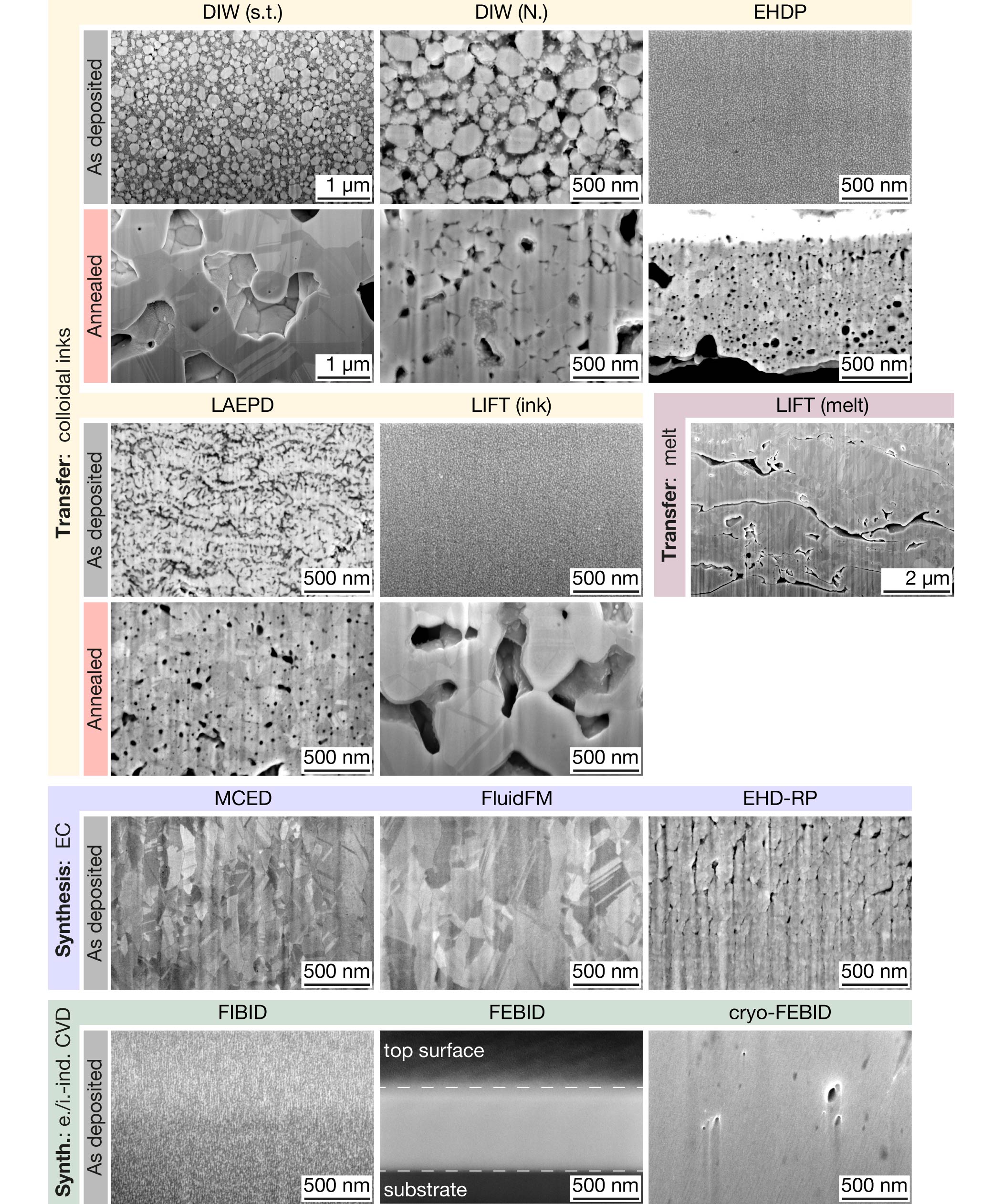} 
   \caption{\textbf{Microstructure of printed pads.} Representative cross-section micrographs of printed pads. Four distinctive microstructures are observed: (1) Agglomerations of metal colloids and organic constituents (as-deposited inks), (2) polycrystalline and porous metals (transfer techniques, annealed), (3) polycrystalline and dense metals (electrochemical techniques), and (4) dense metal-carbon composites (FIBID and FEBID). Note: FIB curtaining effects are observed in the following micrographs and are imaging artefacts rather than real features: DIW (N.), LAEPD, LIFT (ink), LIFT (melt), EHD-RP and cryo-FEBID. For micrographs of lower magnification, see Supplementary Figure~\ref{fig:SI_MicrostructurePads}. All images are tilt-corrected.}
\label{fig:microstructurePads}
\end{figure*}

\begin{figure*}[htbp] 
   \centering
   \includegraphics[width=178mm]{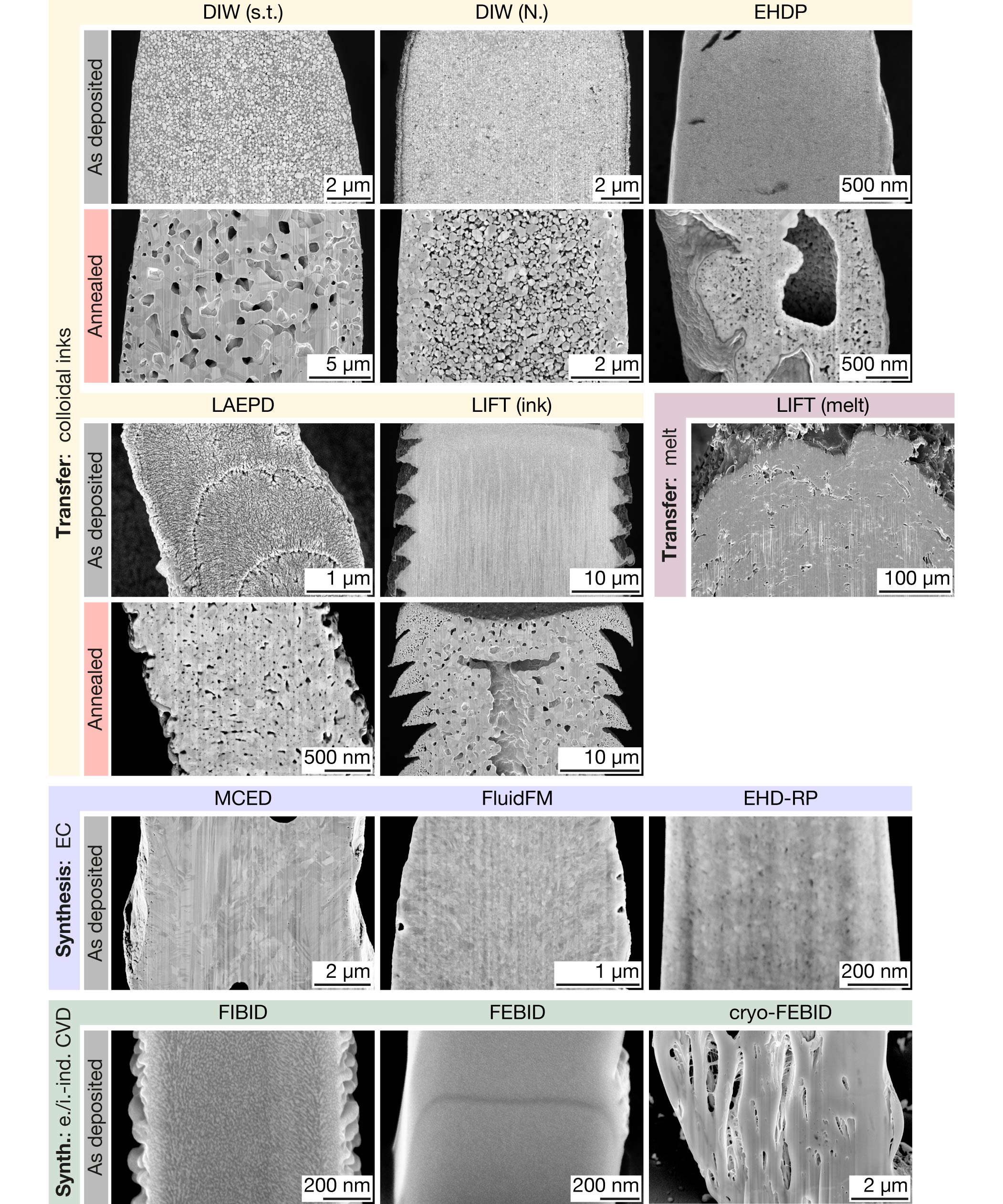} 
   \caption{\textbf{Microstructure of printed pillars.} Representative cross-section micrographs of printed pillars. The microstructure of printed pillars is generally comparable to the microstructure of the corresponding pads. However, radial microstructure gradients are obvious in most annealed pillars printed from colloidal inks (DIW, EHDP, LIFT (ink)), whereas pillars printed by synthesis methods and LIFT (melt) are more homogeneous. {The line feature in the FEBID pillar was caused by an interruption of the automated exposure pattern.} All images are tilt-corrected.}
\label{fig:microstructurePillars}
\end{figure*}

The microstructure of printed metals was assessed by cross-sectional electron microscopy of both pads (Fig.~\ref{fig:microstructurePads}) and pillars (Fig.~\ref{fig:microstructurePillars}). If multiple annealing states were prepared, the sample annealed the longest at the highest temperature is shown. Only Cu is shown for LIFT (melt). For a complete collection of the microstructure of all samples and micrographs of pads with lower magnification, please refer to Supplementary Section~\ref{sec:SI_mech} and Supplementary Figure~\ref{fig:SI_MicrostructurePads}. Supplementary Figure~\ref{fig:SI_EDX} provides a qualitative overview of the chemical composition of printed pads analyzed by EDX spectroscopy.
\par

Four types of microstructure are identified if the micrographs are reduced to their common denominators: (1) agglomerations of metal colloids and organic constituents, (2) polycrystalline and porous metals, (3) polycrystalline and dense metals, and (4) dense carbon-metal composites. All as-printed colloidal inks belong to category (1). The non-metallic contrast (dark) in the micrographs in Figures~\ref{fig:microstructurePads} and \ref{fig:microstructurePillars} is interpreted as organic components, based on the chemical analysis that suggests significant amounts of carbon in as-printed inks (Supplementary Figure~\ref{fig:SI_EDX}) and the fact that all inks in this study are composed of micro- and nanoparticles coated with organic surfactants and mixed with organic solvents. All transfer methods eventually synthesize materials of category (2) (crystalline and porous), although the degree and homogeneity of the observed porosity is subject to large variations. The microstructure synthesized upon thermal consolidation of colloidal inks is pronouncedly porous. The pore size ranges from tens of nanometers to approximately one micron, and the porosity is usually homogeneous without a preferential orientation. Metals printed by LIFT of Au and Cu melts are also porous, although the density of Cu is higher than that of Au (Supplementary Figure~\ref{fig:SI_LIFT}). In contrast to annealed inks, the pore distribution is pronouncedly inhomogeneous, the pore size in the microscale range, and the pore shape non-spherical (pores are often elongated inter-droplet gaps). Additionally, the grain size can vary by one order of magnitude within a printed structure. Category (3) (polycrystalline and dense) is populated by Cu synthesized by additive electrochemical methods, with a typically microstructure that is nano- to microcrystalline and dense (that is, no obvious porosity is detected at the studied length scale). Yet, EHD-RP pillars feature nanoscale voids, and pads evolve mostly vertical gaps at a height of $\approx$\SI{500}{\nano\meter} (Fig.~\ref{fig:microstructurePads}) or already after a few layers (not shown). Below that height, the grown Cu is dense. 
Room-temperature FEBID and FIBID deposits belong to category (4) (carbon-metal composites): they are dense, but, as established in literature, consist of metal nanoparticles embedded in an amorphous carbon matrix\cite{Utke2008}. A typical metal content achieved with commercial setups at room temperature is approximately 10~\==~15~at.\%\cite{Botman2009a} for MeCpPt(Me)$_\text{3}$, the precursor used in this study. FEBID deposits grown at cryogenic temperatures have been reported to have a similar composition\cite{Bresin2011}. In contrast to room-temperature FEBID materials, cryo-FEBID pillars feature pronounced, vertically elongated pores (Fig.~\ref{fig:microstructurePillars}).
\par
The homogeneity of the deposited microstructure, that is the spatial distribution of grain and pore size and pore fraction, varies between techniques and also printed structures. Annealed inks often feature gradients in porosity and variations of grain sizes within a deposit. This phenomenon is observed mostly in pillars (Figure~\ref{fig:microstructurePillars}), and less frequently in pads (Supplementary Figure~\ref{fig:SI_Microstructure_padspillars}). The micrographs in Figure~\ref{fig:microstructurePillars} specifically suggest that the outer region of pillars is often denser and features smaller pore sizes (DIW, EHDP, LIFT (ink)). In extreme cases, annealed pillars developed hollow centers (EHDP, LIFT (ink)). In contrast, synthesis techniques produce a homogeneous microstructure in both pads and pillars (apart from the large pores observed in the pillars by MCED, cryo-FEBID, and the change in microstructure as a function of height in the pads by EHD-RP).

\subsection{Mechanical properties}
\begin{figure*}[htbp] 
   \centering
   \includegraphics[width=178mm]{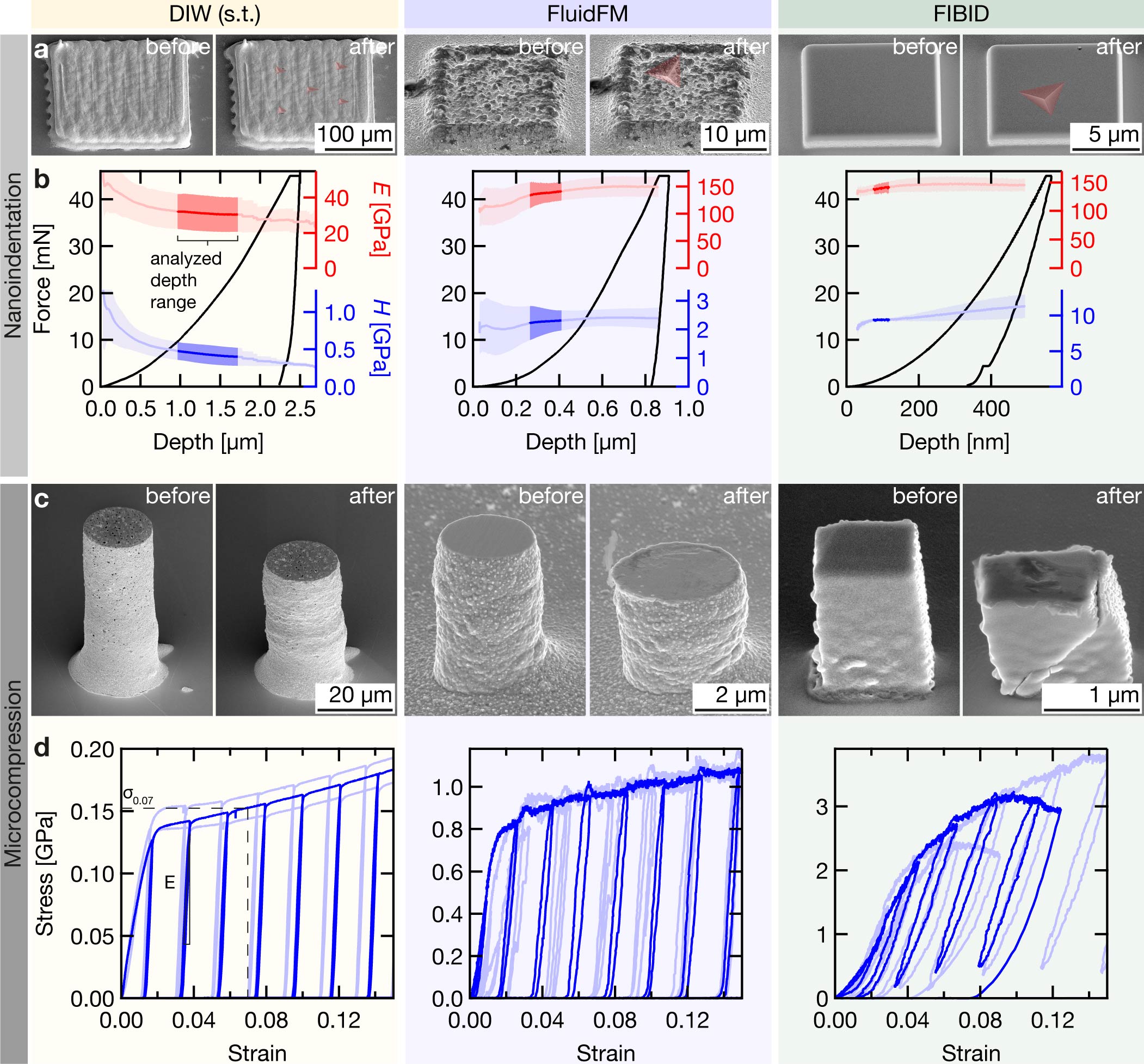} 
   \caption{\textbf{Mechanical testing.} Nanoindentation and microcompression data of samples printed by three exemplary techniques: DIW (s.t.) (representing transfer techniques), FluidFM (electrochemical techniques), and FIBID (electron/ion-induced CVD). \textbf{a}, Representative pads before (left) and after (right) nanoindentation. The indents are highlighted in red. Tilt angle: \SI{45}{\degree}. \textbf{b}, Single indentation curves (black) and average Young's modulus $E$ (red) and hardness $H$ (blue) as a function of indentation depth. The solid line is the mean value derived from all measured indents, the shaded area the standard deviation. Elastic and plastic properties are reported from the highlighted depth range. \textbf{c}, Pillars before (left) and after (right) microcompression. Tilt angle: \SI{55}{\degree}. \textbf{d}, Three representative stress-strain curves measured for each technique. One curve is highlighted for clarity.}
\label{fig:data}
\end{figure*}

The mechanical properties of the printed materials were analyzed by nanoindentation of pads and microcompression of micropillars, deriving the Young's modulus $E$ (once from nanoindentation, once from microcompression), the hardness $H$ (nanoindentation) and the flow stress at \SI{7}{\percent} strain $\sigma_\text{0.07}$ (microcompression). Figure~\ref{fig:data} presents indentation and microcompression data of samples printed by three techniques: DIW (s.t.) (representing transfer techniques), FluidFM (electrochemical synthesis) and FIBID (electron/ion-induced CVD). The complete dataset including all samples of all techniques is provided in Supplementary Section~\ref{sec:SI_mech}. The averaged values of $E$, $H$ and $\sigma_\text{0.07}$ are plotted in Figure~\ref{fig:EH}a, b and listed in Table~\ref{tab:techniquesoverview} and, in more detail, in Supplementary Table~\ref{tab:EH}. Fig.~\ref{fig:EH}c, d normalizes the measured values to literature values of thin films fabricated by traditional deposition techniques {(PVD and electrodeposition, denoted with the subscript \emph{PVD}, Supplementary Table~\ref{tab:SI_litval})} to enable a comparison between different printed metals and a general assessment of the overall materials quality. The comparison to thin-film instead of bulk metals is motivated by the fact that these are the materials and properties established in microfabrication.
\par
Both the elastic and plastic properties vary by two orders of magnitude between the individual techniques if all as-printed samples are considered, and by one order of magnitude if as-printed inks are excluded. The normalized modulus $E/E_\text{PVD}$ of annealed inks and samples printed by LIFT is always $<1$, while their hardness $H/H_\text{PVD}$ can be $\geq$1. Electrochemical techniques enable $E/E_\text{PVD}=1$ (except EHD-RP with $E/E_\text{PVD}<1$) and \mbox{$H/H_\text{PVD}\geq1$}. Room-temperature FEBID and FIBID structures exhibit a lower elastic modulus than Pt, but a hardness that is higher than that of most metals routinely used in microfabrication ($H$: 6~\==~\SI{9}{\giga\pascal}, typical hardness of Pt thin films $\approx$1.5~\==\SI{5}{\giga\pascal}\cite{Mencik2006,Mall2009}). Thus, the normalized hardness $H/H_\text{PVD}$ of FEBID and FIBID structures is $>$1, although $E/E_\text{PVD}<1$ (A comparison to hardest metals, for example W with a hardness of 6~\==~\SI{18}{\giga\pascal}\cite{Zhang2014,Abadias2006} would yield $H_\text{PVD}\leq1$). An exception are cryo-FEBID deposits, which are pronouncedly more compliant and softer than their room-temperature counterparts.

\begin{figure*}[htbp] 
   \centering
   \includegraphics[width=178mm]{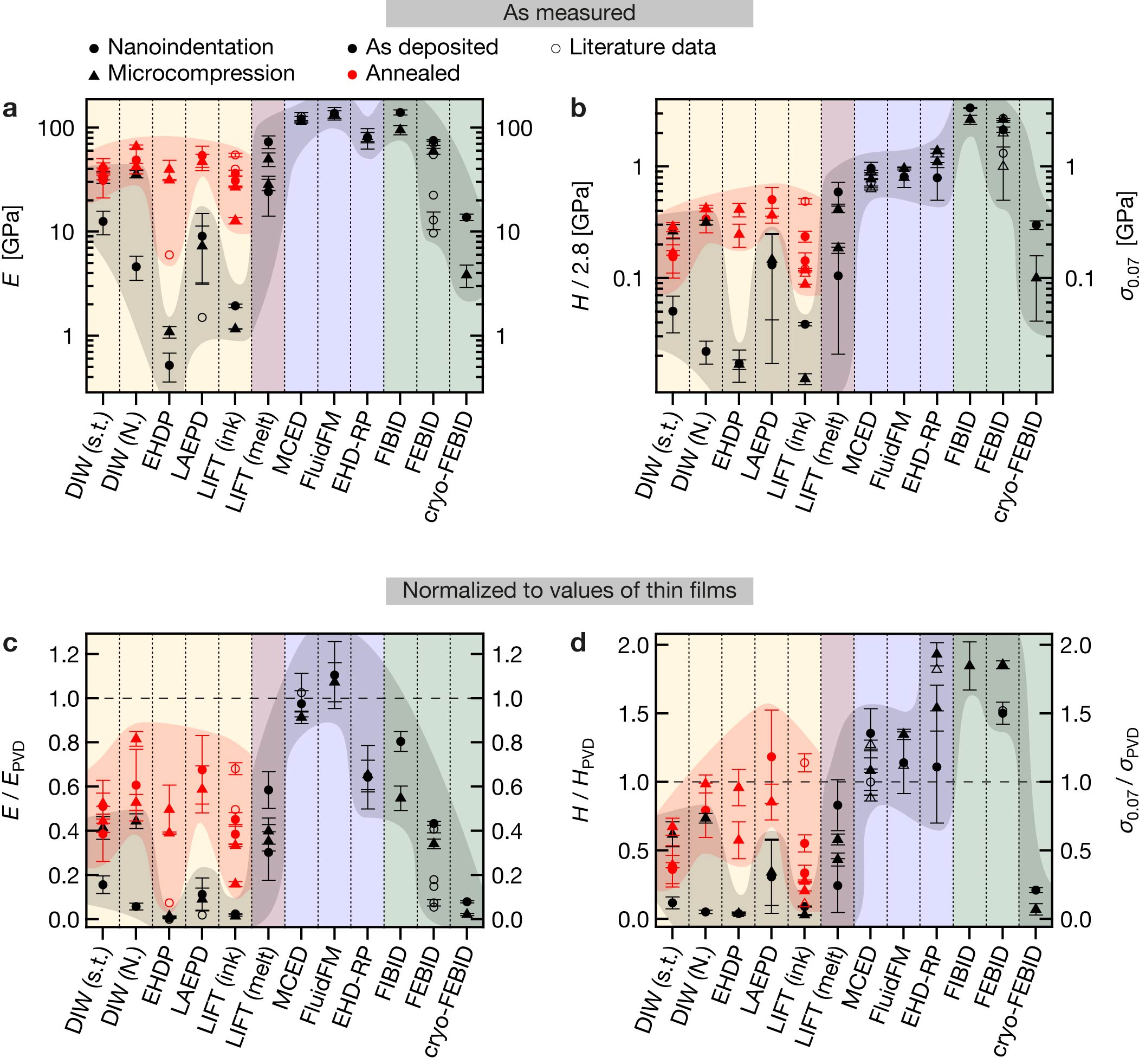} 
   \caption{\textbf{Young's modulus $E$, hardness $H$ and flow stress $\sigma_\text{0.07}$} of printed metals. \textbf{a}, $E$ measured by nanoindentation (circles) and microcompression (triangles) for as-deposited (black) and annealed (red) samples of all techniques. Literature data: open symbols. \textbf{b}, $H$ (circles) and $\sigma_\text{0.07}$ (triangles). Hardness data is divided by a constraint factor of 2.8 for a direct comparison with the flow stress. Literature data: open symbols. \textbf{c, d}, Measured data normalized by literature data for thin films deposited by traditional microfabrication methods (Supplementary Table~\ref{tab:SI_litval}). References for literature data: EHDP\cite{Schneider2013}, LAEPD\cite{Takai2014}, LIFT (ink)\cite{Birnbaum2010a,Birnbaum2010c,Birnbaum2011a}, MCED\cite{Behroozfar2017,Daryadel2018}, FluidFM\cite{Hirt2016}, EHD-RP\cite{Reiser2019}, FEBID\cite{Utke2006,Wich2008,Friedli2009,Lewis2017,Arnold2018}.}
\label{fig:EH}
\end{figure*}

\section{Discussion}
\subsection{Limitations of the study and validity and general applicability of the results}
Generally, materials properties are a function of a materials' chemistry and microstructure. This implies that no absolute value for any property can be assigned to the individual small-scale AM methods, because each technique can potentially synthesize {materials with a range of microstructures and compositions} by making respective adjustments to the feedstock materials and the process parameters. Thus, the generalization of the microstructural and mechanical data reported here is constrained by the following considerations: within the limits discussed in Supplementary section \ref{sec:SI_limitations}, the reported values are representative and reproducible for each technique. Yet, these limits~\==~imposed by the narrow range of metals, the absence of any materials' optimization, and the non-ideal test geometries~\==~underline the necessity to treat the here-reported data as ranges rather than absolute numbers: the measured values are mere snapshots of the materials synthesized today, and should be perceived as a visualization of commonalities and differences between individual techniques rather than as absolute and static maxima for any given method.

\subsection{Morphology }
The samples of this study were not fabricated for benchmarking the resolution or surface roughness of the individual small-scale AM techniques. For a general overview of the geometrical capabilities of the respective techniques, please refer to our recent review article\cite{Hirt2017}. Nevertheless, two points are noteworthy and further discussed in Supplementary Section \ref{sec:SI_Morphology}: first, thermal annealing of inks can cause distortion of printed objects due to shrinkage upon coalescence of individual particles, but such a distortion is not always observed. Second, the different printing strategies utilized by the different techniques~\==~without or with in-plane hatching depending on whether the technique does or does not offer an adaptable feature size~\==~not only affect the print speed but also the sample morphology.

\newpage
\subsection{Microstructure and resulting mechanical properties}
Four types of microstructure were identified:
\begin{enumerate}
\item Agglomeration of metal colloids and organic constituents (as-printed inks)
\item Polycrystalline and porous metals (annealed inks and LIFT of melts)
\item Polycrystalline and dense metals (electrochemical synthesis)
\item Dense carbon-metal composites (electron/ion-induced CVD)
\end{enumerate}
These microstructures result in four distinctive classes of mechanical performance:
\begin{enumerate}
\item $E/E_\text{PVD}\ll1$, $H/H_\text{PVD}\ll1$ (as-printed inks)
\item $E/E_\text{PVD}<1$, $H/H_\text{PVD}\leq1$ (annealed inks and LIFT of melts) 
\item $E/E_\text{PVD}\leq1$, $H/H_\text{PVD}\geq1$ (electrochemical synthesis)
\item $E/E_\text{PVD}<1$, $H/H_\text{PVD}>1$ (electron/ion-induced CVD)
\end{enumerate}

As a note: the Young's modulus of thin films $E_\text{PVD}$ is consistent in literature and usually comparable to the bulk Young's modulus $E$, because the elastic properties are mostly independent of a metals' microstructure (apart from porosity). Hence, the normalized data for $E$ is representative. In contrast, literature values for $H_\text{PVD}$ easily vary by a factor of two due to variations in microstructure of the deposited films ($H_\text{Ag}$: 0.7~\==~1.5\cite{Shugurov2004a,Panin2005}, $H_\text{Au}$: 1~\==~\SI{2}{\giga\pascal}\cite{Okuda1999,Volinsky2004}, $H_\text{Cu}$: 1.6~\==~\SI{3.5}{\giga\pascal}\cite{Beegan2003,Lu2005,Beegan2007,Chang2007}, $H_\text{Pt}$: 1.5~\==~8.6\cite{Mall2009,Nili2014,Mencik2006}). Consequently, the normalized $H$ data should be treated with care, although we tried to select a value that is most representative of the available literature (Ag: \SI{1.2}{\giga\pascal}, Au: \SI{1.2}{\giga\pascal}, Cu: \SI{2}{\giga\pascal}, Pt: \SI{4}{\giga\pascal}) to normalize our data.

\subsubsection{Transfer techniques: inks}
As-deposited inks are composed of individual micro- or nanoparticles. The volumetric metal content of the inks is a function of particle size distribution and amount of nonvolatile organic ink constituents. Due to organic capping layers and residual solvent, the filling factors of these colloidal systems are necessarily below the theoretical maximal values for hard spheres. Especially in the case of nanoparticles, the surfactant layers can occupy a considerable volume: for illustration, Richner {et al.}\cite{Richner2016a} have calculated theoretic volume filling factors as small as 0.21~\==~0.26 for nanoparticle inks (random loose packing, \SI{5}{\nano\meter} particle dia.). Nevertheless, inks can be compared to colloidal crystals in first approximation, despite their lower filling factors. The mechanical behavior of all as-deposited inks ($E$:~0.5~\==~\SI{10}{\giga\pascal}, $H$:~30~\==~\SI{600}{\mega\pascal}) is similar to that of nanoparticle colloidal crystals ($E$:~0.1~\==~\SI{6}{\giga\pascal}\cite{Podsiadlo2010,Yan2013}, $H$:~10~\==~\SI{450}{\mega\pascal}\cite{Podsiadlo2010}). The considerable variation is presumably owed to the large difference in particle packing density, the differences in particle shape and size distribution which will both affect the flow properties, the chemical instability of as-printed inks (and thus changing properties over time, also when kept in inert atmosphere) and the differences in ink preparation (some inks were actively dried after printing, others not). Two exceptions from this general picture of colloidal inks were observed: first, the strength of DIW inks determined by microcompression was significantly higher than the respective hardness derived from nanoindentation of the same materials (briefly discussed in Supplementary Section~\ref{sec:SI_mech_DIW}). {Second, we hypothesize that electrophoretically printed Au nanoparticles are fused upon laser-assisted deposition (Figure~\ref{fig:microstructurePads}), resulting in the high strength observed for as-deposited LAEPD samples (Supplementary Section~\ref{sec:SI_microstructureLAEPD}).}

\par
The microstructure of annealed inks is dominated by inter-granular porosity resulting from incomplete densification upon sintering, although the absolute pore fraction varies between techniques (Figure~\ref{fig:microstructurePads}) and annealing states (Supplementary Section~\ref{sec:SI_mech}), and can range from $\approx$10~\==~\SI{35}{\percent} (estimated areal fraction by threshold analysis of cross-section micrographs). Additional to the mere presence of porosity, its inhomogeneous distribution is often observed in printed pillars (Figure~\ref{fig:microstructurePillars}). These gradients in pore fraction, pore size and grain size are probably a result of local variations in sintering conditions (heat conductance and heat capacitance) in combination with mechanical constraints exerted by both the substrate and previously sintered material (as discussed below). Apart from the microstructure, the shrinkage upon annealing can affect the printed geometries by warping or partial delamination.
\par
In general, the observed behavior of annealed inks is in qualitative and quantitative agreement with previous mechanical studies of Ag inks reporting moduli and hardnesses of 10~\==~\SI{100}{\giga\pascal} and 0.03~\==~\SI{1.4}{\giga\pascal}, respectively\cite{Greer2007,Lee2010,Birnbaum2010c} and decreasing strength with longer annealing due to coarsening of the porosity (ligament size effect)\cite{Dou2010}. The porosity and its characteristic length scale in combination with the nanoscale grain size results in the peculiar observation that the modulus of annealed inks is invariably inferior to that of dense thin films, but their normalized strength can be close to unity or larger: while the elastic properties of a porous material are mainly determined by the fraction and shape of its pores\cite{Hashin1983,Roberts2000}, always resulting in a decreased stiffness with increased porosity, the yield strength is additionally modulated by the grain size as well as the ligament size (in highly porous solids such as inks after short annealing\cite{Hodge2007,Dou2010}). Consequently, a porous material can be strong if the grain size or ligament size is sufficiently small so that Hall-Petch strengthening or ligament size effects compensate the loss in strength due to the lower density. The strength of these materials typically decreases upon prolonged annealing due to coarsening of grains and ligaments (compare microcompression results in Supplementary Figures~\ref{fig:SI_Nanjia1} and \ref{fig:SI_Nanjia2}, and nanoindentation hardness in Supplementary Figures~\ref{fig:SI_Pique1} and \ref{fig:SI_Pique2}).
\par
The pronounced porosity and the large spread in elastic and plastic properties reported in this study as well as in the literature highlight a major challenge for ink-based techniques: shrinkage upon annealing is inherent to the concept of colloidal inks. Accompanying effects such as the evolution of porosity, warping of printed geometries or stress-induced failure predominantly characterize the properties of the printed materials and structures. Thus, the management of the densification and coarsening processes upon annealing is inevitable for most applications. The same conclusions apply to metals synthesized by pyrolysis of metal-containing resists structured by TPL\cite{Vyatskikh2018}, one of the metal AM methods not included in this study. Here, a linear shrinkage of \SI{80}{\percent} upon pyrolysis highlights the challenge for dealing with volume loss in extremis.
\par
The observed porosity of the annealed inks contrasts with the low residual porosity usually reported for macroscale structures fabricated by DIW of metal and metal oxide inks (with \SI{50}{\percent} isotropic linear shrinkage)\cite{Jakus2015}. Nearly \SI{100}{\percent} density is achieved in a range of sintered metal colloids, including Fe and Ni\cite{Taylor2016}, W\cite{Rothlisberger2016,Calvo2018}, or CoCrFeNi high-entropy alloys\cite{Kenel2019}. To explain the marked difference, a comparison of the boundary conditions of micro- and macroscale sintering is insightful. At large scales, the absence of mechanical constraints is a most important factor in preventing residual porosity: without the mechanical constraint of a substrate, the sintering of macroscale DIW geometries facilitates unconstrained, homogeneous shrinkage and thus densification. In contrast, at small scales, printed microscale objects are not free to shrink and densify~\==~they are (and usually need to be) invariably attached to a substrate. Accordingly, shrinkage is constrained by their support but also the material that has been sintered before and is already firmly attached to the substrate (such as the outer regions of annealed ink pillars or the surface of EHDP pads). Hence, complete densification cannot be achieved, as the evolving stresses counteract the coalescence of pores or result in failure or delamination. In this light, other concepts that typically enable high densification at larger scales are probably also less effective at small scales, including higher sintering temperatures (homologous temperatures used for sintering of large-scale structures are typically in the range of $\approx$0.35~\==~0.9\cite{Taylor2016,Calvo2018}, and were $\approx$0.2~\==~0.4 for the present study) and the use of sintering aids\cite{Rothlisberger2016}.
\par
Consequently, a compromise between maximum densification and minimal shrinkage is probably necessary for ink-printed geometries at small scales. For highest strength combined with best shape retention, a brief annealing that guarantees neck formation between individual nanoparticles but avoids coarsening of the microstructure is probably the most promising approach, as it avoids large shrinkage but offers high strength via ligament size effect strengthening. High densification via thermal sintering will, in most cases, be accompanied by pronounced grain growth and hence a weakening of the metal. Thus, alternative strengthening mechanisms such as precipitation hardening should be considered for high strength if nanoporous structures are not an option. Unfortunately, it is likely that sintering protocols need to be adjusted to every geometry and ink-substrate combination: the observed gradients in microstructure and differences between the microstructure of pillars and pads subjected to the same thermal treatment (Supplementary Figure~\ref{fig:SI_Microstructure_padspillars}) suggest that the local densification is influenced by local boundary conditions that vary for every geometry and materials combination. Nevertheless, although this optimization is not a trivial task, a large body of established sintering knowledge from powder metallurgy could, in principle, guide strategies for improved annealing of printed inks. 

\subsubsection{Transfer techniques: LIFT of melts}
LIFT of metal melt droplets renders as-printed functional materials that do not require any post-print processing. Nevertheless, as with inks, residual porosity is a characteristic feature of the synthesized materials. This observation is in agreement with literature reports of porosities of 5~\==~\SI{15}{\percent} in Cu\cite{Zenou2015b,Winter2016,Feinaeugle2018} and up to \SI{50}{\percent} in Au\cite{Feinaeugle2018}. Grain and pore sizes measured in this study are large compared to that of inks (early annealing stages), but grain sizes $<$\SI{100}{\nano\meter} have been reported for LIFT-printed Cu\cite{Fogel2019}. The porosity in combination with a medium grain size result in normalized elastic and plastic properties that are lower than those of dense thin films, although the measured elastic moduli of 50~\==~\SI{75}{\giga\pascal} (Cu) and 25~\==~\SI{30}{\giga\pascal} (Au) are higher than previously reported values (Cu: \SI{12}{\giga\pascal}, Au: \SI{9}{\giga\pascal}\cite{Fogel2018}). 
\par
The large pore size, the inhomogeneous pore distribution and the comparably high surface roughness are a challenge for the use of these materials in mechanical applications at small scales. To some extent, the porosity and roughness can be minimized by alternative printing algorithms and a higher laser fluence, increasing the impact energy and temperature of the droplets\cite{Winter2016}. As a complication, the observed microstructural inhomogeneities (variations in density and grain size) suggest that the boundary conditions for solidification are subjected to strong local variations. Consequently, microstructure engineering is a task with many variables. Nevertheless, optimization for higher strength by increasing the defect density of the melt-solidified metal could be achieved, especially Cu with its comparably low porosity. Because direct control of the local cooling rates and hence the evolving grain size is inaccessible, alternative approaches to microstructure engineering should be considered. As the LIFT process facilitates alloying\cite{Zenou2015b}, solid solution hardening or precipitation of an immiscible phase for hardening or grain boundary pinning could be a viable path to increase the defect density and thus the strength, regardless of local cooling rates.





\subsubsection{Synthesis techniques: electrochemical}
Cu synthesized by MCED and the FluidFM, the two techniques that use traditional aqueous electrochemistry, is dense and crystalline, with grain sizes of $<$100~\==~\SI{500}{\nano\meter}. Comparable microstructural data for electrochemical AM techniques has been published earlier\cite{Suryavanshi2006,Suryavanshi2007,Hirt2016,Behroozfar2017,Daryadel2018}. As a result of the high density, the measured elastic and plastic properties ($E$:~114~\==~\SI{138}{\giga\pascal}, $H$:~2.3~\==~\SI{2.7}{\giga\pascal}) can easily compete with properties of Cu films prepared by sputtering or electroplating ($E$:~\SI{130}{\giga\pascal}, $H$:~\SI{2}{\giga\pascal} for a grain size of \SI{100}{\nano\meter}\cite{Chang2007}). The high materials quality of electrochemically 3D printed Cu is confirmed by previous studies of mechanical\cite{Hirt2016,Behroozfar2017,Daryadel2018,Daryadel2018a} and electrical\cite{Suryavanshi2007,Hu2010} properties.
\par
In contrast, samples printed by EHD-RP are not completely dense: pillars are homogeneous but feature nanoscale pores, and all pads developed vertical gaps (often resembling columnar growth), either after a few layers or only after an initially dense growth (Supplementary Figure~\ref{fig:SI_MicrostructurePads}). Grain sizes in both geometries are $<$\SI{50}{\nano\meter} and thus pronouncedly smaller than those by the other electrochemical techniques. The combination of residual porosity and nanoscale grain size results in a high strength ($H$:~\SI{2.2}{\giga\pascal}, $\sigma_\text{0.07}$:~1.1~\==~\SI{1.38}{\giga\pascal}) but a lowered elastic modulus ($E$:~$\approx$\SI{80}{\giga\pascal}). These findings are in line with previously reported mechanical data\cite{Reiser2019}. Despite the use of an organic solvent, the measured carbon content is as low as that of MCED and FluidFM (Supplementary Figure~\ref{fig:SI_EDX}). Thus, it is assumed that the porosity or potentially oxide phases rather than residual organic solvents are responsible for the low stiffness. Thermal annealing would possibly promote densification but has not yet been studied. While previous results showed a dependence of the synthesized grain size on the deposition parameters\cite{Reiser2019}, the present study adds the observation that the microstructure is further influenced by the printing strategy itself: the evolution of vertical gaps in pads is likely the result of growth instabilities in combination with preferred growth at protrusions due to field-focusing effects. The reason for the exclusive presence of these instabilities in the pads is unclear. Yet, the initially dense growth of the pads indicate that the synthesis of dense Cu is possible given the printing parameters are optimized. 
\par
In general, electrochemical small-scale AM offers extended options for microstructure engineering. First steps in this direction have already been made, demonstrating a range of grain sizes\cite{Daryadel2018}, twin densities\cite{Behroozfar2017,Daryadel2018a} and local alloying and tuning of local porosity within printed geometries\cite{Reiser2019}. In its extent, the range of microstructures that could be accessed with electrochemical AM is unique amongst all small-scale AM methods. This advantage could and should be exploited for maximizing materials properties and adjusting the synthesized microstructure to the specific needs of different applications (for example electrical conductivity versus mechanical strength).
\par
In principle, the high density and purity and resulting high stiffness and strength of electrodeposited metals makes these materials the obvious choice for mechanical and also electrical applications. Unfortunately, electrochemical methods are currently limited in broad applicability: first, the electrochemical reduction necessitates an electrically conductive substrate. Second, the techniques are typically restricted to small build volumes due to a low deposition speed compared to transfer methods (however, the low volumetric throughput has been addressed recently: an increase in deposition speed by one order of magnitude by EHD-RP\cite{Reiser2019}, parallelization demonstrated by MCED\cite{Lin2019} and a variable voxel size shown with the FluidFM\cite{Ercolano2019} are strategies that have increased the volumetric throughput and enable deposition of 3D geometries more than hundred microns in length, width and height with submicron resolution\cite{Ercolano2019}).
 Third, the electrochemical strategies are currently limited to metals only, while ink-, laser-transfer- and ion/electron-beam-based techniques allow for the deposition of all materials classes\cite{Hirt2017}. Consequently, electrochemical methods are~\==~for the time being~\==~restricted to applications that demand highest materials properties in metals but can adjust to the mentioned limitations.

\subsubsection{Synthesis techniques: electron/ion-induced CVD}
It is generally accepted that FEBID and FIBID deposits obtained from MeCpPt(Me)$_\text{3}$ as used in this study feature a composite structure, {i.e.} metal nanoparticles embedded in an amorphous hydrogenated carbon (a-C:H) matrix\cite{Utke2008}. Typical metal contents achieved with MeCpPt(Me)$_\text{3}$ in commercial setups at room temperature are approximately 10~\==~15~at.\% for FEBID\cite{Botman2009a,Utke2008} and around 45~at.\% for FIBID\cite{Utke2008}, although other volatile metal compounds that achieve much higher metal content have been demonstrated\cite{Huth2018,Utke2008}. As a result of the low metal content, the mechanical behavior is dominated by the carbonaceous matrix. The properties of a-C:H are highly dependent on its sp$^2$/sp$^3$ ratio, with the stiffness and strength increasing with increasing sp$^3$ content\cite{Jiang1989a,Weiler1996}. {In FEBID and FIBID, this ratio as well as the density of the matrix are variable with the primary electron or ion energy and dose and further depend on the balance of precursor flux versus electron/ion flux\cite{Li2007c,Friedli2009,Arnold2018}.} As the structure of the matrix is thus sensitive to the deposition parameters, our measurements ($E$:~60~\==~\SI{140}{\giga\pascal}, $H$:~6~\==~\SI{9.4}{\giga\pascal}, $\sigma_{0.07}$:~\SI{2.6}{\giga\pascal}) and reported mechanical properties of FIBID and FEBID deposits ($E$: 10~\==~\SI{100}{\giga\pascal}\cite{Utke2006,Friedli2009,Lewis2017,Arnold2018}, $H$: 3.6~\==~\SI{7.6}{\giga\pascal}\cite{Ding2005,Wich2008}, tensile strength $\sigma_{tensile}$: 1~\==~\SI{2}{\giga\pascal}\cite{Utke2006}) cover a large range. {Interestingly, the measured elastic modulus of the FEBID deposits is higher than previously reported moduli of as-deposited Pt-C structures\cite{Friedli2009,Lewis2017,Arnold2018} and similar to that of e-beam-annealed FEBID deposits\cite{Arnold2018}, suggesting that the here-used deposition conditions imparted an {in-situ} e-beam curing, which is suggested to lower the hydrogen content of the matrix and improve the reticulation of the carbon network\cite{Arnold2018}.} Nevertheless, the mechanical properties are considerably lower than data of a-C:H synthesized by conventional plasma deposition techniques ($E$: 40~\==~\SI{760}{\giga\pascal}\cite{Jiang1989a,Weiler1996,Cho2005}, $H$: 5~\==~\SI{30}{\giga\pascal}\cite{Jiang1989a,Weiler1996}, $\sigma_{tensile}$: \SI{7}{\giga\pascal}\cite{Cho2005}), {presumably due to the contained softer Pt phase and the fact the FEBID and FIBID carbon matrix is still hydrogenated and features a high sp$^2$/sp$^3$ ratio (typically, sp$^2$/sp$^3$=1~\==~9\cite{Bret2005a,Li2007c}).
}
\par
The low volume fraction of metal is not a disadvantage for mechanical performance~\==~the stiffness compares well to that of metals synthesized with the other small-scale AM techniques, and the strength outperforms the electrochemically deposited metals by a factor 2~\==~3. Further, variation of the  sp$^\text{2}$/sp$^\text{3}$ ratio enables optimization of the stiffness and strength for a given application\cite{Arnold2018,Sattelkow2019}. A drawback of amorphous carbon is its brittleness compared to metals. However, as long as fracture toughness is not an issue, there is no need to increase the metal loading from a mechanical perspective. Of course, an increase of the metal content is necessary for an electrical resistance that can compete with that of metals: the room-temperature electrical resistivity of as-deposited Pt-carbon material is around \SI{1e-6}{\ohm\meter} for FIBID (46 at.\% Pt) and 1~\==~\SI{10e-2}{\ohm\meter} for FEBID (10-15 at.\% Pt), corresponding to approximately 10$\times$ (FIBID) and \num{e5}$\times$ (FEBID) the resistivity of bulk metals \cite{Botman2009a}.
\par
The pronouncedly lower mechanical properties measured for {cryo-FEBI deposits} are likely owed to the porosity of the structures (especially of the pillars). This porosity results from evaporation of non-irradiated precursor-gas upon heating the deposits from cryogenic to room temperature\cite{Bresin2011}. As the amount of non-irradiated precursor is a function of total fluence\cite{Bresin2011}, optimization of the exposure parameters could probably result in an increased density and thus improved mechanical properties.






\section{Conclusion and Outlook}
We have surveyed the microstructure and resulting mechanical properties of metals fabricated by almost all contemporary small-scale AM methods. This study provides a comprehensive overview of metals accessible to these techniques today, and thus provides groundwork for optimization of the materials required for tomorrow. Both commonalities and differences between the synthesized metals were found: metals deposited by transfer techniques (from colloidal inks or metal melts) are typically nano- to microporous. In contrast, materials produced by synthesis techniques are usually dense, and in the case of electrochemical methods, often nanocrystalline. Thus, modern microscale AM techniques synthesize metallic materials with a wide range of microstructures and can supply materials for various potential applications: dense and nanocrystalline metals for mechanical applications, dense and microcrystalline metals for high electrical or thermal conductivity, or highly porous materials for the use in catalysis or optical metamaterials.
\par
Two main consequences are drawn for the future development of the techniques and their applications. First, not all 3D printed metals are equal~\==~the characteristic microstructure associated with each different technique varies considerably. Consequently, users that request specific materials properties need to select an appropriate AM method. For applications requiring materials of high stiffness or strength, electrochemical techniques or FEBID and FIBID are recommended. These techniques deliver materials with dense microstructure and excellent mechanical properties, equal to those of metals used in state-of-the-art microfabrication. For applications that require low defect density, electrochemical methods are the first choice (potentially in combination with thermal annealing for grain coarsening). If porous microstructures can be tolerated or are even wanted, and high throughput is required, printing of inks or LIFT is typically the simplest and most efficient choice.
\par
Second, the materials challenges faced by ink-based methods and LIFT need to be tackled~\==~relying on electrochemical and electron/ion-beam techniques only is not an option. Although electrochemical and electron/ion-beam approaches offer high-performance metals, they suffer from other drawbacks such as comparably low volumetric growth rates and a limited range of materials. {As high throughput (ideally in combination with high materials quality) is essential for many industrial applications, many microscale AM techniques strive towards higher deposition rates, parallel printing or deposition with variable voxel sizes. While parallelization as well as on-the-fly adjustments of the voxel size have been demonstrated with both synthesis\cite{Post2011,Lin2019,Ercolano2019} and transfer techniques\cite{Lee2008c,Pan2017,Auyeung2015,Wang2010}, transfer-techniques typically offer deposition rates normalized by resolution that can be orders of magnitude higher than that of synthesis methods\cite{Hirt2017} (partial exceptions are EHD-RP and cryo-FEBID).} Hence, transfer techniques generally have an inherent advantage for high throughput. Additionally, transfer methods offer a much wider range of accessible materials that also include non-metals, which are not yet available with electrochemical techniques. For these two reasons, transfer-methods are and will remain important for many applications of small-scale printing, despite their shortcomings in materials quality. Consequently, there is a strong need to improve the materials synthesized by these methods in order to guarantee levels of performance required in typical applications. Hence, we advocate for more detailed studies of printed materials' microstructure and, most importantly, optimization of thermal post-print processing. To this end, the large body of established sintering strategies employed in powder metallurgy could be tapped to guide attempts towards an improved densification of printed inks. On the other hand, electrochemical and electron/ion-beam-based methods need to expand the available materials, their compatibility with non-conductive substrates, and their volumetric growth rates.
\par
In summary, some small-scale AM methods can already provide device-grade metals, but the narrow range of accessible chemistries and the comparably small build volumes of these methods still limit the application of additive techniques in modern microfabrication. However, if the porosity of ink-derived metals and the limited applicability of electrochemical and electron-beam-based methods are addressed, microscale AM of inorganic materials has the potential to offer a large range of device-grade materials and to become a versatile and powerful alternative to lithography-based 3D manufacturing at small scales.
\bigskip

\bigskip

\bigskip

\section{Experimental}
\subsection{Sample preparation}

\subsubsection{Direct ink writing of shear-thinning Ag inks}
The concept and setup for DIW of Ag structures, as well as the synthesis of the shear-thinning Ag inks, have been described earlier\cite{Ahn2009a,Zhou2017}. In brief, the coagulated ink consisted of 75~\==~85~wt.\% polyacrylic acid (PAA)-coated Ag particles in DI water. Pads and pillars were printed on a Regenovo Bio-Architect WS printer at speeds of 30~\==~\SI{80}{\micro\meter\per\second} and extrusion pressures of 17~\==~\SI{24}{\bar}. Pads were printed with a layer-by-layer approach, while pillars were deposited by continuously retracting the nozzle from the substrate. All samples were printed directly onto soda lime glass substrates. After printing, samples were heated on a hotplate in air. As-deposited samples were dried at \SI{100}{\celsius} (30~min), and annealed samples were sintered at a final temperature of \SI{300}{\celsius} after a controlled temperature ramp (\SI{100}{\celsius} (\SI{10}{\minute}), \SI{150}{\celsius} (\SI{10}{\minute}), \SI{200}{\celsius} (\SI{10}{\minute}), \SI{250}{\celsius} (\SI{10}{\minute}), and finally, \SI{300}{\celsius} for \SI{30}{\minute} or \SI{2}{\hour}). All samples were sputter-coated with a conductive layer of Pt-Pd (\SI{8}{\nano\meter}, CCU-010 safematic) prior to SEM imaging and mechanical testing to avoid charging in the electron microscope.

\subsubsection{Direct ink writing of Newtonian Ag inks}
The basic process and setup for printing, as well as the synthesis of the Ag inks, was previously described by Lee {et al.}\cite{Lee2017}. The synthesized ink (PAA-coated Ag particles, 25~wt.\% in DI water) shows Newtonian fluid characteristics at shear rates of 0~\==~\SI{e3}{\per\second}, with a viscosity of \SI{6.8e-3}{\pascal\second}. Glass nozzles with an opening of \SI{10}{\micro\meter} were fabricated with a P-97 pipette puller (Sutter Instruments). The ink was filled from the back of the nozzle and drawn to the nozzle tip by capillary forces without applied pressure. Pads were printed with a layer-by-layer strategy (hatch distance \SI{10}{\micro\meter}, layer height \SI{10}{\micro\meter}) at a speed of \SI{10}{\micro\meter\per\second}. Pillars were printed by retracting the nozzle at \SI{10}{\micro\meter\per\second}. All samples were printed on (111) Si wafers coated with a \SI{1}{\micro\meter} thick Pt film. Annealing was performed in an ambient-atmosphere furnace (ov-11, JEIO-Tech), either at \SI{150}{\celsius} for 1 hour or \SI{450}{\celsius} for 12 hours.

\subsubsection{Electrohydrodynamic printing of Au inks}
The printing procedure was previously described by Galliker {et al.}\cite{Galliker2012} Similarly, the ink has been reported elsewhere.\cite{Richner2016a}. In brief, Au nanoparticles with Decanethiol ligands, $\approx$\SI{5}{\nano\meter} in diameter, were synthesized with the Stucky method\cite{Zheng2006} and dispersed in Tetradecane. The maximum ink concentration was $\approx$\SI{10}{\milli\gram\per\milli\liter}. All samples were printed on diced SiO$_\text{2}$ wafers (UniversityWafer, USA). The printing parameters are similar to the process presented by Schneider {et al.}\cite{Schneider2016}. Yet, the printing process was optimized for high mass flow for printing the large structures presented herein. Therefore, the \SI{250}{\hertz} AC actuation voltage was increased to 260~--~360 V$_\text{p}$ and larger nozzles with an aperture diameter of 1.8~--~\SI{2}{\micro\meter} were used. These parameters resulted in an ejection frequency of $\approx$\SI{1}{\kilo\hertz} of droplets $\approx$\SI{200}{\nano\meter} in diameter. In general, the samples were printed in a layer-by-layer fashion, where each layer was deposited in a serpentine-like printing path. With a translation speed of \SI{5}{\micro\meter\per\second} and a hatch distance (line pitch) of \SI{300}{\nano\meter}, approximately 30 layers are required to deposit a pad of \SI{5}{\micro\meter} in height. In contrast, pillars (especially small pillars) were printed with lower mass flow, using lower actuation voltages and smaller nozzles. As-printed samples were annealed for \SI{10}{\minute} at \SI{400}{\celsius} in a constant gas flow of O$_\text{2}$ at ambient pressure in a rapid thermal processing furnace (As-One 150, Annealsys, France)\cite{Schneider2016}.

\subsubsection{Laser-assisted electrophoretic deposition of Au nanoparticles}
The principle, setup and typical printing parameters for laser-assisted electrophoretic printing have been described in detail in literature\cite{Iwata2009,Takai2014}. For deposition of Au structures, an aqueuous solution of Au nanoparticles (0.25 wt.\%, \SI{3}{\nano\meter} dia., Tanaka Kikinzoku Kogyo K. K.) was dispensed between an ITO-coated cover glass and an ITO-coated glass substrate (\SI{200}{\nano\meter} ITO on soda lime glass, Geomatec inc.). An electric field of \SI{12}{\kilo\volt\per\meter} was applied between these electrodes for electrophoresis (electrode distance: \SI{160}{\micro\meter}). A CW Nd:YVO$_4$ laser (\SI{532}{\nano\meter}, {\SI{5}{\milli\watt},} Spectra physics, Millenia Pro) was used for optical trapping of the Au particles (Objective lens: x60, NA=1.2, Olympus, UPLSAPO 60XW). For printing of pillars, the stage was lowered at \SI{0.67}{\micro\meter\per\second}. Pads were printed by depositing an array of overlapping pillars. Printed structures were annealed for \SI{1}{\hour} at \SI{300}{\celsius} in atmosphere in a furnace (AFM-10, ASWAN). 

\subsubsection{Laser-induced forward transfer of Ag nanoparticle inks}
The details of the method have been described in previous publications\cite{Pique2008,Wang2010,Charipar2018}. {Donor substrates were prepared via doctor-blading a commercial Ag ink (80~wt.\%, particle dia.: 7~\==~\SI{12}{\nano\meter}, viscosity $\approx$\SI{90}{\pascal\second}, NPS Nanopaste, Harima Chemicals Group) onto lithographically defined wells (\SI{4}{\milli\meter}~$\times$~\SI{2}{\centi\meter}) in glass substrates.} {Laser printing was performed using a Nd:YVO$_\text{4}$ (JDSU, Q301-HD) pulsed laser ($\lambda$ = \SI{355}{\nano\meter}, \SI{30}{\nano\second} FWHM) with a laser fluence of $\approx$30~\==~\SI{100}{\milli\joule\per\square\centi\meter}.} Pads were printed by laser-transferring individual voxels of Ag ink in the shape of circular disks, and pillars were deposited by stacking multiple individual voxel disks\cite{Charipar2018}. All samples were printed onto diced (100), p-type Si wafers (University Wafer, Inc.). Thermal annealing was performed in a furnace (ambient atmosphere) for one hour at temperatures in the range of 150~\==~\SI{230}{\celsius}. 

\subsubsection{Laser-induced forward transfer of melts of Au and Cu thin films}
The printing setup for LIFT and the general principle have been reported in literature\cite{Zenou2015a,Zenou2016}. A \SI{3}{\watt} laser with a pulse duration of \SI{0.8}{\nano\second} and a wavelength of \SI{532}{\nano\meter} (Picospark, Teem Photonics) is used for the LIFT process. The laser was deflected by a scanning mirror, and its spot size was \SI{28\pm0.5}{\micro\meter} (4-sigma) at the donor interface. Soda lime glass slides coated with 500 nm of either Cu or Au were used as donors. Substrates were soda lime glass coated with \SI{10}{\nano\meter} Ti / \SI{100}{\nano\meter} Au. Printing was performed in ambient atmosphere with a donor-substrate gap of \SI{300}{\micro\meter}. All samples were printed with droplet overlaps of \SI{3}{\micro\meter}\cite{Zenou2016} and a total of five layers. The pulse energy used was 5.5~\==~\SI{7.5}{\micro\joule} for Cu and 4~\==~\SI{6.5}{\micro\joule} for Au.

\subsubsection{Meniscus-confined electrodeposition of Cu}
The method and setup was described by Seol {et al.}\cite{Seol2015a}. In short, Cu was deposited from an electrolyte of \SI{1.05}{M} CuSO$_\text{4}$$\cdot$5H$_\text{2}$O (99~\%, Samchun Chem.) in an aqueous solution of \SI{0.8}{M} H$_\text{2}$SO$_\text{4}$ (95~\%, Daejung Chem.) using a two-electrode setup with a Pt anode. All samples were grown by pulsed electrodeposition (pulse profile: \SI{-1.7}{\volt} (\SI{1}{\second}), \SI{1}{\volt} (\SI{0.5}{\second})) at growth rates of \SI{0.02}{\micro\meter\per\second} and \SI{0.4}{\micro\meter\per\second} for pads and pillars respectively. Both, pads and pillars, were printed by retracting the nozzles (diameter: \SI{25}{\micro\meter} for pads, \SI{10}{\micro\meter} for pillars) from the substrate in steps of \SI{10}{\micro\meter}. All samples were deposited on a  \SI{1}{\micro\meter}-thick film of Pt on cut (111) Si wafers. No thermal annealing was performed.

\subsubsection{FluidFM electrodeposition of Cu}
The FluidFM principle and general printing parameters have been reported elsewhere\cite{Hirt2016,Ercolano2019}. {All structures were printed with FluidFM Nanopipette probes (\SI{300}{\nano\meter} opening, Cytosurge AG) mounted on either a FluidFM BOT (Cytosurge AG) in the case of pads, or on a FluidFM ADD-ON (Cytosurge AG) for classical AFM systems (Nanowizard I, JPK) in the case of pillars.} For printing, the probe was filled with a Cu electrolyte solution (1 M CuSO$_\text{4}$ (Sigma Aldrich) in aq. 0.8 M H$_\text{2}$SO$_\text{4}$ (Sigma Aldrich), filtered using a Millex-VV Syringe filter (\SI{0.1}{\micro\meter}, PVDF, Sigma Aldrich)) and immersed in a supporting electrolyte bath (aq. H$_\text{2}$SO$_\text{4}$, pH 3). The electrochemical cell was equipped with a {graphite (BOT) or a Pt (ADD-ON)} counter electrode and a Ag/AgCl reference electrode. Samples were deposited onto Ti(\SI{3}{\nano\meter})/Au(\SI{25}{\nano\meter}) films coated either on soda lime glass (Menzel Gl\"aser, Thermoscientific) in the case of pillars, or Si substrates ((100), Microchemicals) in the case of pads. Pads were printed with a layer-by-layer strategy at a potential of \num{-0.45}~\==~\SI{-0.6}{\volt} vs. Ag/AgCl, an applied pressure of 4~\==~\SI{8}{\milli\bar} and a layer height of 0.4~\==~\SI{0.8}{\micro\meter}. {Pillars were printed in a layer-by-layer fashion, at a potential of \SI{-0.67}{\volt} vs. Ag/AgCl, an applied pressure of \SI{7}{\milli\bar} and a voxel height of \SI{0.25}{\micro\meter}. No thermal annealing was performed.}

\subsubsection{Electrohydrodynamic redox printing of Cu}
{The basic concept and setup for EHD-RP have been published earlier\cite{Reiser2019}.} In brief, a sacrificial Cu anode (0.1 mm dia. wire, 99.9985~\%, Alfa Aesar) was immersed in a printing nozzle filled with acetonitrile (Optima, Fisher Chemical). The nozzles were pulled (P-2000 micropipette puller system, Sutter Instrument) from quartz capillaries (QF100-70-15, Sutter Instrument) to an aperture diameter of 135~\==~\SI{145}{\nano\meter}. Cu wires were etched in pure nitric acid (Sigma Aldrich) for 15~s) prior to use. Printing was conducted in Ar-atmosphere ($<$40~ppm O$_\text{2}$, as measured by a Module ISM-3 oxygen sensor (PBI Dansensor)). Samples were printed onto Si substrates ((100), SiMat) with a Ti(\SI{3}{\nano\meter})/Au(\SI{20}{\nano\meter}) coating (deposited in our laboratory sputter facility by DC magnetron sputtering (PVD Products Inc.)). 15$\times$\SI{15}{\micro\meter} pads were printed layer by layer, with a serpentine-like print path, a hatch distance (line pitch) of \SI{100}{\nano\meter} and a rotation of the hatch direction by \SI{90}{\degree} in every subsequent layer. The voltage applied to the sacrificial anode was 110~\==~\SI{150}{\volt}. The nozzle-substrate separation was kept constant at \SI{7.5}{\micro\meter} by retracting the nozzle after every printed layer by a distance equal to an estimated layer height. Different combinations of in-plane speed, layer height and number of layers were used. Typical values were: in-plane speed: 10~\==~\SI{40}{\micro\meter\per\second}; layer height: 25~\==\SI{100}{\nano\meter}; layers: 10~\==~80. Nominally 1$\times$\SI{1}{\micro\meter} wide, square pillars were printed with the same protocol as the pads. Printing parameters were: hatch distance: \SI{100}{\nano\meter}; voltage: \SI{100}{\volt}; nozzle-substrate separation: \SI{7.5}{\micro\meter}; in-plane speed: \SI{5}{\micro\meter\per\second}; layer height: \SI{200}{\nano\meter}; layers: 14. Two pillars 160 and \SI{170}{\nano\meter} in diameter were printed with the same electric field but with no relative motion between substrate and nozzle.

\subsubsection{Focused ion and electron beam induced deposition of Pt}
A Tescan Lyra FIB-SEM (background pressure \SI{9e-6}{\milli\bar}, operating pressure \SI{1e-5}{\milli\bar}) with electron field emission gun, Orsay Physics Ga-ion source, and Orsay Physics gas injection system (GIS) was used for FIBID and FEBID of Pt deposits. The GIS nozzle exit (inner diameter: \SI{350}{\micro\meter}) was placed at a distance of approximately \SI{200}{\micro\meter} to the area of deposition. We used Methylcyclopentadienyl platinum (IV) trimethyl (MeCpPt(Me)$_\text{3}$) as a precursor gas, heated to \SI{80}{\celsius}. The local precursor pressure above the substrate was calculated by Empa's freeware GIS simulator to be 0.4 and \SI{0.8}{\pascal} (\SI{2e3}{} and \SI{4e3}{ML\per\second}) for FIB and FEB deposition, respectively. All samples were deposited onto diced (100) Si wafers (Semiwafer). The deposition parameters for FIBID and FEBID pillars and pads are given as (beam current, beam energy, beam size, scan pattern, dwell time, pixel distance, refresh time, total time, and deposition rate). FIBID pillars: (\SI{5}{\pico\ampere}, \SI{10}{\kilo\volt}, \SI{32}{\nano\meter}, fly back, \SI{80}{\nano\second}, \SI{25.6}{\nano\meter}, \SI{122}{\micro\second}, \SI{10}{\minute}~\SI{28}{\second}, \SI{0.7}{\cubic\micro\meter\per\nano\ampere\per\second}). FIBID pads: (\SI{285}{\pico\ampere}, \SI{10}{\kilo\volt}, \SI{32}{\nano\meter}, fly back, \SI{80}{\nano\second}, \SI{25.6}{\nano\meter}, \SI{12}{\milli\second}, \SI{19}{\minute}~\SI{34}{\second}, \SI{0.3}{\cubic\micro\meter\per\nano\ampere\per\second}). FEBID pillars: (\SI{1}{\nano\ampere}, \SI{5}{\kilo\volt}, \SI{29}{\nano\meter}, rotating leading edge (RLE), \SI{80}{\nano\second}, \SI{29}{\nano\meter}, \SI{92.5}{\micro\second}, \SI{16}{\minute}~\SI{38}{\second}, \SI{0.0026}{\cubic\micro\meter\per\nano\ampere\per\second}). FEBID pads: (\SI{3.9}{\nano\ampere}, \SI{5}{\kilo\volt}, \SI{180}{\nano\meter}, RLE, \SI{1}{\micro\second}, \SI{180}{\nano\meter}, \SI{6.9}{\milli\second}, \SI{70}{\minute}, \SI{0.0064}{\cubic\micro\meter\per\nano\ampere\per\second}).

\subsubsection{Cryo-FEBID of Pt}
The set-up has been described in detail by Bresin {et al.}\cite{Bresin2011,Bresin2013}, including the important processes governing morphology, size, microstructure and composition. All experiments were conducted in a FEI Nova 600 Nanolab dual beam system (base pressure \SI{6.7e-5}{\pascal}). For deposition, the MeCpPt(Me)$_\text{3}$ precursor gas is condensed on the substrate at \SI{-190}{\celsius} (custom-built cryogenic stage cooled by liquid nitrogen in a heat exchanger) and subsequently patterned with the electron beam. Square pads were printed by injecting the precursor gas for \SI{30}{\second} (crucible temperature: \SI{27}{\celsius}) and scanning the electron beam (\SI{0.62}{\nano\ampere},\SI{20}{\kilo\volt}) in a normal imaging raster with a pixel dwell time of \SI{100}{\nano\second}. Pillars were deposited by three cycles of condensation and e-beam patterning. In contrast to the pads, the precursor was injected for \SI{60}{\second}, and beam parameters were \SI{0.6}{\nano\ampere}, \SI{18}{\kilo\volt}. Further, the individual layers were exposed with a stationary rather than a scanning beam, exposing a nominal area of \SI{10}{\nano\meter} for \SI{30}{\second}. All structures were deposited onto p-type (100) silicon with 100 nm of thermal oxide (plasma cleaned prior to deposition (\SI{20}{\percent} O$_\text{2}$ in Ar, Fischione Model 1020)).

\subsection{Analysis}
The morphology of the printed samples was studied with a FEG-SEM (Magellan 400, FEI, USA). Cross-sections of samples for microstructure analysis were cut and imaged with a dual-beam FIB-SEM (NVision40, Zeiss, Germany) using the InLens detection mode and an acceleration voltage of \SI{5}{\kilo\volt} and final polishing currents of 40~\==~\SI{80}{\pico\ampere}. The chemical composition was qualitatively analyzed by SEM EDX spectroscopy (Quanta 200F, FEI, equipped with an Octane Super EDX system, EDAX, software: Genesis, EDAX). All spectra were recorded with identical acquisition conditions: acceleration voltage: \SI{10}{\kilo\volt}, amplification time: \SI{7.68}{\micro\second}, live time: \SI{60}{\second}, dead time: 25~\==~\SI{30}{\percent}, scan size: 5$\times$\SI{5}{\square\micro\meter} for small pads, 15$\times$\SI{15}{\square\micro\meter} for large pads, no specimen tilt. In general, all samples were stored either under vacuum in a desiccator or at atmospheric pressures in a low-humidity Ar cabinet.
 
        \par
The mechanical properties of the printed materials were measured by nanoindentation of pads and microcompression of micropillars. Nanoindentation was performed with three different testing setups: a Ultra Nanoindentation Tester (Anton Paar Tritec SA, Switzerland) or a iNano Nanoindenter (Nanomechanics, Inc., USA) were used to test samples of low surface roughness. Both systems were fitted with a diamond Berkovich indenter (Synton-MDP, Switzerland). An Alemnis indenter (Alemnis GmbH, Thun, Switzerland)\cite{Wheeler2013} fitted with either a Berkovich or a cube-corner indenter (Synton-MDP, Switzerland) was used for indentation of LIFT  (melt)-printed pads, as these samples showed higher surface roughness and thus required higher indentation depths. Typical linear loading rates were 5000~--~\SI{10000}{\micro\newton\per\minute} and \SI{10}{\milli\newton\per\minute} for the Anton Paar Tritec and the Alemnis systems, respectively. The Nanomechanics system was used in constant strain rate mode, with a target strain rate of 0.1.
Hardness and elastic modulus were recorded as a function of depth, either by performing progressive load-unload cycles (Anton Paar Tritec and Nanomechanics system) or a continuous measurement by superimposing a sinus oscillation on the load profile (\SI{110}{\hertz}, Nanomechanics system). Reported values are as-measured hardness and modulus values averaged over a depth range adjusted to the thickness, width and roughness of the tested pads. No model to account for substrate effects was applied. The following Poisson's ratios were used for the conversion of indentation modulus to Young's modulus: Ag: 0.37, Au: 0.42, Cu: 0.36\cite{KOSTER2014}. For all Pt FIBID and FEBID samples, we calculated with the Poisson's ratio of the matrix material (glassy carbon, 0.2) instead of the value for Pt. A constraint factor of 2.8\cite{Maier2011} was used for conversion of hardness to strength.
	\par
Microcompression testing was performed using an in-situ SEM Indenter (Alemnis GmbH, Thun, Switzerland)\cite{Wheeler2013} mounted in a Vega3 SEM (Tescan, Brno, Czech Republic). This facilitated both sample positioning and direct observation of the deformation process. The indenter was fitted with diamond flat punches of 1.5~\==~\SI{50}{\micro\meter} in diameter, depending on the width of the tested pillars. Pillars were compressed in progressive loading-unloading cycles to a total strain of 10~\==~\SI{25}{\percent}. Compression was performed under displacement control at a rate proportional to the pillar's height to produce a strain rate on the order of \SI{1e-3}{\per\second}. For determination of the yield stress, load-displacement curves were converted to stress-strain curves based on the average diameter of the deformed portion of the pillar. Due to the non-uniform pillar diameter and the blurred onset of yield, no classic yield criterion was applied. Instead, we report the flow stress at \SI{7}{\percent} strain, which corresponds directly to the representative strain under a Berkovich indenter\cite{Leitner2016}. For the calculation of the Young's modulus, stress-strain curves based on the average load-bearing cross-section of the whole pillar were used. The Young's modulus was usually measured from the first unloading segment after the plastic yield point. No volume conservation was applied. Sneddon's model\cite{Sneddon1965} was used to correct for pillar sink-in, using the following elastic moduli of the different substrates: soda lime glass: \SI{72}{\giga\pascal}; fused silica: \SI{73}{\giga\pascal}; Si $<$100$>$: \SI{130}{\giga\pascal}; Si $<$111$>$: \SI{186}{\giga\pascal}. Metallic coating layers between pillar and substrate were ignored in the model, as the elastic field was expected to extend far beyond those layers. Prior to testing, the top portions of most pillars were cut using a dual-beam FIB-SEM (NVision40, Zeiss, Germany) to produce flat pillar tops and to trim the pillars to an aspect ratio 2~\==~~3. This helped to avoid buckling and to achieve a clearer plastic yield point which is less influenced by the rounded top of the pillars. Final Ga$^\text{+}$-ion milling currents of {\SI{80}{\pico\ampere}~\==~\SI{3}{\nano\ampere} were used, depending on the diameter of the pillars.
	\par
The measured mechanical data were normalized to literature values of polycrystalline thin films deposited by PVD or electrochemical deposition to decouple microstructural from chemical influences on the mechanical properties (To allow a comparison between the different chemical elements used by different AM methods) and to enable a facile comparison of the quality of printed metals to that of thin-film materials synthesized by established deposition techniques. The following values were used (Table \ref{tab:SI_litval}): $E_\text{Ag}$:~\SI{80.5}{\giga\pascal}, $H_\text{Ag}$:~\SI{1.2}{\giga\pascal}, $E_\text{Au}$:~\SI{80.2}{\giga\pascal}, $H_\text{Au}$:~\SI{1.2}{\giga\pascal}, $E_\text{Cu}$:~\SI{125}{\giga\pascal}, $H_\text{Cu}$:~\SI{2}{\giga\pascal}, $E_\text{Pt}$:~\SI{174}{\giga\pascal}, $H_\text{Pt}$:~\SI{4}{\giga\pascal}. Literature data for $E$ of Co and W FEBID deposits were normalized to $E_\text{Co}$:~\SI{204}{\giga\pascal} and $E_\text{W}$:~\SI{369}{\giga\pascal}, while the hardness values were not normalized. The measured flow stress $\sigma_\text{0.07}$ was normalized by literature values for $H/2.8$. All FEBID and FIBID values were normalized to literature data of Pt to highlight the difference in mechanical behavior compared to metals, although we acknowledge that FEBID and FIBID deposits are mostly composed of amorphous carbon.

\section{Acknowledgments}
A.R., J.W. and R.S. thank S. Ganzeboom (ETH Z\"urich) for experimental support and gratefully thank D. Momotenko (Laboratory of Biosensors and Bioelectronics, ETH Z\"urich) for access to his nozzle-puller. Electron-microscopy analysis was performed at ScopeM, the microscopy platform of ETH Z\"urich. A.R., J.W. and R.S. acknowledge the financial support by Grant no. ETH 47 14-2. 
The work by C.v.N. and T.Z. was supported by Swiss Agency for Technology and Innovation Innosuisse (Project Nr: PNFM-NM 18511.1). C.v.N. would like to thank L. Hirt (formerly ETH Z\"urich) for experimental support. 
P.R. post-treated his samples at the Binnig and Rohrer Nanotechnology Center (BRNC) at IBM Zurich and analyzed them at the ETH Center for Mirco- and Nanoscience (FIRST). P.R. acknowledges funding by the SFA Advanced Manufacturing program under the Powder Focusing project.
K.C. and A.P. acknowledge that this work was funded by the Office of Naval Research (ONR) through the Naval Research Laboratory Basic Research Program.
The contribution of S.L. and S.K.S. was supported in part by Korea Electrotechnology Research Institute (KERI) Primary research program (No. 19-12-N0101-27) through the National Research Council of Science \& Technology (NST) funded by Ministry of Science and ICT.
I.U. would like to thank G. B\"urki (Empa) and P. L. Gal (Tescan / Orsay Physics) for experimental support.

\section{Author contributions}
A.R. contrived and coordinated the study. R.S. supervised the project. A.R. provided SEM and FIB and nanoindentation analysis. J.M.W. and A.R. performed microcompression analysis. The following authors were responsible for sample fabrication with the different techniques: L.K. (EHD-RP), K.A.D. (cryo-FEBID), T.M. and F.I. (LAEDP), O.F. and Z.K. (LIFT (melt)), N.Z. and J.L. (DIW (s.t.)), K.C. and A.P. (LIFT (ink)), P.R. and D.P. (EHDP), S.L. and S.K.S. (DIW (N.) and MCED), I.U. (FEBID and FIBID), C.v.N. and T.Z. (FluidFM). A.R., J.M.W. and R.S. interpreted the results, while all authors discussed the results. A.R. wrote the manuscript and visualized the data. All authors reviewed and commented the manuscript.

\section {Conflicts of interest}
D.P. is the cofounder of SCRONA. All other authors declare no conflict of interest.

\bibliographystyle{nature_alain.bst}
\bibliography{PhDThesisAlain.bib}

\begin{thebibliography}{100}

\bibitem{Onses2015a}
Onses, M.~S., Sutanto, E., Ferreira, P.~M., Alleyne, A.~G. \& Rogers, J.~A.
\newblock {Mechanisms, Capabilities, and Applications of High-Resolution
  Electrohydrodynamic Jet Printing}.
\newblock \emph{Small} \textbf{11}, 4237--4266 (2015).

\bibitem{Hohmann2015}
Hohmann, J.~K., Renner, M., Waller, E.~H. \& von Freymann, G.
\newblock {Three-Dimensional $\mu$-Printing: An Enabling Technology}.
\newblock \emph{Adv. Opt. Mater.} \textbf{3}, 1488--1507 (2015).

\bibitem{Rogers2016}
Rogers, J., Huang, Y., Schmidt, O.~G. \& Gracias, D.~H.
\newblock {Origami MEMS and NEMS}.
\newblock \emph{MRS Bull.} \textbf{41}, 123--129 (2016).

\bibitem{Zhang2017a}
Zhang, Y. \emph{et~al.}
\newblock {Printing, folding and assembly methods for forming 3D mesostructures
  in advanced materials}.
\newblock \emph{Nat. Rev. Mater.} \textbf{2}, 17019 (2017).

\bibitem{Hirt2017}
Hirt, L., Reiser, A., Spolenak, R. \& Zambelli, T.
\newblock {Additive Manufacturing of Metal Structures at the Micrometer Scale}.
\newblock \emph{Adv. Mater.} \textbf{29}, 1604211 (2017).

\bibitem{Montemayor2015}
Montemayor, L., Chernow, V. \& Greer, J.~R.
\newblock {Materials by design: Using architecture in material design to reach
  new property spaces}.
\newblock \emph{MRS Bull.} \textbf{40}, 1122--1129 (2015).

\bibitem{Soukoulis2011}
Soukoulis, C.~M. \& Wegener, M.
\newblock {Past achievements and future challenges in the development of
  three-dimensional photonic metamaterials}.
\newblock \emph{Nat. Photonics} \textbf{5}, 523--530 (2011).

\bibitem{Lee2012c}
Lee, J.-H., Singer, J.~P. \& Thomas, E.~L.
\newblock {Micro-/Nanostructured Mechanical Metamaterials}.
\newblock \emph{Adv. Mater.} \textbf{24}, 4782--4810 (2012).

\bibitem{Lewis2006}
Lewis, J.~A.
\newblock {Direct Ink Writing of 3D Functional Materials}.
\newblock \emph{Adv. Funct. Mater.} \textbf{16}, 2193--2204 (2006).

\bibitem{Gansel2009}
Gansel, J.~K. \emph{et~al.}
\newblock {Gold Helix Photonic Metamaterial as Broadband Circular Polarizer}.
\newblock \emph{Science (80-. ).} \textbf{325}, 1513--1515 (2009).

\bibitem{Ergin2010}
Ergin, T., Stenger, N., Brenner, P., Pendry, J.~B. \& Wegener, M.
\newblock {Three-Dimensional Invisibility Cloak at Optical Wavelengths}.
\newblock \emph{Science (80-. ).} \textbf{328}, 337--339 (2010).

\bibitem{Schaedler2011}
Schaedler, T.~A. \emph{et~al.}
\newblock {Ultralight Metallic Microlattices}.
\newblock \emph{Science (80-. ).} \textbf{334}, 962--965 (2011).

\bibitem{Jang2013NM}
Jang, D., Meza, L.~R., Greer, F. \& Greer, J.~R.
\newblock {Fabrication and deformation of three-dimensional hollow ceramic
  nanostructures}.
\newblock \emph{Nat. Mater.} \textbf{12}, 893--898 (2013).

\bibitem{Xia2019}
Xia, X. \emph{et~al.}
\newblock {Electrochemically reconfigurable architected materials}.
\newblock \emph{Nature} \textbf{573}, 205--213 (2019).

\bibitem{Ahn2009a}
Ahn, B.~Y. \emph{et~al.}
\newblock {Omnidirectional Printing of Flexible, Stretchable, and Spanning
  Silver Microelectrodes}.
\newblock \emph{Science (80-. ).} \textbf{323}, 1590--1593 (2009).

\bibitem{Hu2010}
Hu, J. \& Yu, M.-F.
\newblock {Meniscus-Confined Three-Dimensional Electrodeposition for Direct
  Writing of Wire Bonds}.
\newblock \emph{Science (80-. ).} \textbf{329}, 313--316 (2010).

\bibitem{Wang2010}
Wang, J., Auyeung, R. C.~Y., Kim, H., Charipar, N.~A. \& Piqu{\'{e}}, A.
\newblock {Three-Dimensional Printing of Interconnects by Laser Direct-Write of
  Silver Nanopastes}.
\newblock \emph{Adv. Mater.} \textbf{22}, 4462--4466 (2010).

\bibitem{Schneider2016}
Schneider, J. \emph{et~al.}
\newblock {Electrohydrodynamic NanoDrip Printing of High Aspect Ratio Metal
  Grid Transparent Electrodes}.
\newblock \emph{Adv. Funct. Mater.} \textbf{26}, 833--840 (2016).

\bibitem{Zhou2017}
Zhou, N., Liu, C., Lewis, J.~A. \& Ham, D.
\newblock {Gigahertz Electromagnetic Structures via Direct Ink Writing for
  Radio-Frequency Oscillator and Transmitter Applications}.
\newblock \emph{Adv. Mater.} \textbf{29}, 1605198 (2017).

\bibitem{Gavagnin2014}
Gavagnin, M. \emph{et~al.}
\newblock {Free-Standing Magnetic Nanopillars for 3D Nanomagnet Logic}.
\newblock \emph{ACS Appl. Mater. Interfaces} \textbf{6}, 20254--20260 (2014).

\bibitem{Luo2017}
Luo, J. \emph{et~al.}
\newblock {Printing Functional 3D Microdevices by Laser-Induced Forward
  Transfer}.
\newblock \emph{Small} \textbf{13}, 1602553 (2017).

\bibitem{Arnold2018}
Arnold, G. \emph{et~al.}
\newblock {Tunable 3D Nanoresonators for Gas‐Sensing Applications}.
\newblock \emph{Adv. Funct. Mater.} \textbf{28}, 1707387 (2018).

\bibitem{Sattelkow2019}
Sattelkow, J. \emph{et~al.}
\newblock {Three-Dimensional Nanothermistors for Thermal Probing}.
\newblock \emph{ACS Appl. Mater. Interfaces} \textbf{11}, 22655--22667 (2019).

\bibitem{Schurch2018}
Sch{\"{u}}rch, P., Peth{\"{o}}, L., Schwiedrzik, J., Michler, J. \& Philippe,
  L.
\newblock {Additive Manufacturing through Galvanoforming of 3D Nickel
  Microarchitectures: Simulation-Assisted Synthesis}.
\newblock \emph{Adv. Mater. Technol.} \textbf{1800274}, 1800274 (2018).

\bibitem{Winter2016}
Winter, S., Zenou, M. \& Kotler, Z.
\newblock {Conductivity of laser printed copper structures limited by
  nano-crystal grain size and amorphous metal droplet shell}.
\newblock \emph{J. Phys. D. Appl. Phys.} \textbf{49}, 165310 (2016).

\bibitem{Daryadel2018}
Daryadel, S., Behroozfar, A. \& Minary-Jolandan, M.
\newblock {Toward Control of Microstructure in Microscale Additive
  Manufacturing of Copper Using Localized Electrodeposition}.
\newblock \emph{Adv. Eng. Mater.} \textbf{21}, 1800946 (2019).

\bibitem{Skylar-Scott2016}
Skylar-Scott, M.~A., Gunasekaran, S. \& Lewis, J.~A.
\newblock {Laser-assisted direct ink writing of planar and 3D metal
  architectures}.
\newblock \emph{Proc. Natl. Acad. Sci.} \textbf{113}, 6137--6142 (2016).

\bibitem{Botman2009a}
Botman, A., Mulders, J. J.~L. \& Hagen, C.~W.
\newblock {Creating pure nanostructures from electron-beam-induced deposition
  using purification techniques: a technology perspective}.
\newblock \emph{Nanotechnology} \textbf{20}, 372001 (2009).

\bibitem{Utke2008}
Utke, I., Hoffmann, P. \& Melngailis, J.
\newblock {Gas-assisted focused electron beam and ion beam processing and
  fabrication}.
\newblock \emph{J. Vac. Sci. Technol. B Microelectron. Nanom. Struct.}
  \textbf{26}, 1197 (2008).

\bibitem{Tanaka2006}
Tanaka, T., Ishikawa, A. \& Kawata, S.
\newblock {Two-photon-induced reduction of metal ions for fabricating
  three-dimensional electrically conductive metallic microstructure}.
\newblock \emph{Appl. Phys. Lett.} \textbf{88}, 081107 (2006).

\bibitem{Maruo2008}
Maruo, S. \& Saeki, T.
\newblock {Femtosecond laser direct writing of metallic microstructures by
  photoreduction of silver nitrate in a polymer matrix.}
\newblock \emph{Opt. Express} \textbf{16}, 1174--1179 (2008).

\bibitem{VanSpengen2003}
{Merlijn van Spengen}, W.
\newblock {MEMS reliability from a failure mechanisms perspective}.
\newblock \emph{Microelectron. Reliab.} \textbf{43}, 1049--1060 (2003).

\bibitem{Yi2016}
Yi, Z. \emph{et~al.}
\newblock {Vertical, capacitive microelectromechanical switches produced via
  direct writing of copper wires}.
\newblock \emph{Microsystems Nanoeng.} \textbf{2}, 16010 (2016).

\bibitem{Farahani2014}
Farahani, R.~D., Chizari, K. \& Therriault, D.
\newblock {Three-dimensional printing of freeform helical microstructures: a
  review}.
\newblock \emph{Nanoscale} \textbf{6}, 10470 (2014).

\bibitem{Fogel2018}
Fogel, O. \emph{et~al.}
\newblock {3D printing of functional metallic microstructures and its
  implementation in electrothermal actuators}.
\newblock \emph{Addit. Manuf.} \textbf{21}, 307--311 (2018).

\bibitem{Vyatskikh2018}
Vyatskikh, A. \emph{et~al.}
\newblock {Additive manufacturing of 3D nano-architected metals}.
\newblock \emph{Nat. Commun.} \textbf{9}, 593 (2018).

\bibitem{Oran2018}
Oran, D. \emph{et~al.}
\newblock {3D nanofabrication by volumetric deposition and controlled shrinkage
  of patterned scaffolds}.
\newblock \emph{Science (80-. ).} \textbf{362}, 1281--1285 (2018).

\bibitem{Lee2017}
Lee, S. \emph{et~al.}
\newblock {Three-dimensional Printing of Silver Microarchitectures Using
  Newtonian Nanoparticle Inks}.
\newblock \emph{ACS Appl. Mater. Interfaces} \textbf{9}, 18918--18924 (2017).

\bibitem{Park2007}
Park, J.-U. \emph{et~al.}
\newblock {High-resolution electrohydrodynamic jet printing}.
\newblock \emph{Nat. Mater.} \textbf{6}, 782--789 (2007).

\bibitem{Galliker2012}
Galliker, P. \emph{et~al.}
\newblock {Direct printing of nanostructures by electrostatic autofocussing of
  ink nanodroplets}.
\newblock \emph{Nat. Commun.} \textbf{3}, 890 (2012).

\bibitem{Galliker2014a}
Galliker, P., Schneider, J. \& Poulikakos, D.
\newblock {Dielectrophoretic bending of directly printed free-standing
  ultra-soft nanowires}.
\newblock \emph{Appl. Phys. Lett.} \textbf{104}, 073105 (2014).

\bibitem{Schneider2013}
Schneider, J. \emph{et~al.}
\newblock {A novel 3D integrated platform for the high-resolution study of cell
  migration plasticity}.
\newblock \emph{Macromol. Biosci.} \textbf{13}, 973--983 (2013).

\bibitem{Takai2014}
Takai, T., Nakao, H. \& Iwata, F.
\newblock {Three-dimensional microfabrication using local electrophoresis
  deposition and a laser trapping technique}.
\newblock \emph{Opt. Express} \textbf{22}, 28109 (2014).

\bibitem{Pique2008}
Piqu{\'{e}}, A., Auyeung, R.~C., Kim, H., Metkus, K. \& Mathews, S.~A.
\newblock {Digital Microfabrication by Laser Decal Transfer}.
\newblock \emph{J. Laser Micro/Nanoengineering} \textbf{3}, 163--169 (2008).

\bibitem{Mathews2013}
Mathews, S.~A., Auyeung, R. C.~Y., Kim, H., Charipar, N.~A. \& Piqu{\'{e}}, A.
\newblock {High-speed video study of laser-induced forward transfer of silver
  nano-suspensions}.
\newblock \emph{J. Appl. Phys.} \textbf{114}, 064910 (2013).

\bibitem{Pique2016a}
Piqu{\'{e}}, A., Kim, H., Auyeung, R.~C., Beniam, I. \& Breckenfeld, E.
\newblock {Laser-induced forward transfer (LIFT) of congruent voxels}.
\newblock \emph{Appl. Surf. Sci.} \textbf{374}, 42--48 (2016).

\bibitem{Birnbaum2010c}
Birnbaum, A.~J., Wahl, K.~J., Auyeung, R. C.~Y. \& Piqu{\'{e}}, A.
\newblock {Nanoporosity-induced effects on Ag-based metallic nano-inks for
  non-lithographic fabrication}.
\newblock \emph{J. Micromechanics Microengineering} \textbf{20}, 077002 (2010).

\bibitem{Zenou2015a}
Zenou, M., Sa'ar, A. \& Kotler, Z.
\newblock {Laser jetting of femto-liter metal droplets for high resolution 3D
  printed structures}.
\newblock \emph{Sci. Rep.} \textbf{5}, 17265 (2015).

\bibitem{Seol2015a}
Seol, S.~K. \emph{et~al.}
\newblock {Electrodeposition-based 3D Printing of Metallic Microarchitectures
  with Controlled Internal Structures}.
\newblock \emph{Small} \textbf{11}, 3896--3902 (2015).

\bibitem{Behroozfar2017}
Behroozfar, A. \emph{et~al.}
\newblock {Microscale 3D Printing of Nanotwinned Copper}.
\newblock \emph{Adv. Mater.} \textbf{30}, 1705107 (2018).

\bibitem{Lin2019}
Lin, Y.-p., Zhang, Y. \& Yu, M.-f.
\newblock {Parallel Process 3D Metal Microprinting}.
\newblock \emph{Adv. Mater. Technol.} \textbf{4}, 1800393 (2019).

\bibitem{Daryadel2018a}
Daryadel, S. \emph{et~al.}
\newblock {Localized Pulsed Electrodeposition Process for Three-Dimensional
  Printing of Nanotwinned Metallic Nanostructures}.
\newblock \emph{Nano Lett.} \textbf{18}, 208--214 (2018).

\bibitem{Hirt2016}
Hirt, L. \emph{et~al.}
\newblock {Template-Free 3D Microprinting of Metals Using a Force-Controlled
  Nanopipette for Layer-by-Layer Electrodeposition}.
\newblock \emph{Adv. Mater.} \textbf{28}, 2311--2315 (2016).

\bibitem{Momotenko2016}
Momotenko, D., Page, A., Adobes-Vidal, M. \& Unwin, P.~R.
\newblock {Write–Read 3D Patterning with a Dual-Channel Nanopipette}.
\newblock \emph{ACS Nano} \textbf{10}, 8871--8878 (2016).

\bibitem{Ercolano2019}
Ercolano, G. \emph{et~al.}
\newblock {Multiscale Additive Manufacturing of Metal Microstructures}.
\newblock \emph{Adv. Eng. Mater.} \textbf{1900961}, 1900961 (2019).

\bibitem{Reiser2019}
Reiser, A. \emph{et~al.}
\newblock {Multi-metal electrohydrodynamic redox 3D printing at the submicron
  scale}.
\newblock \emph{Nat. Commun.} \textbf{10}, 1853 (2019).

\bibitem{Fowlkes2016}
Fowlkes, J.~D. \emph{et~al.}
\newblock {Simulation-Guided 3D Nanomanufacturing via Focused Electron Beam
  Induced Deposition}.
\newblock \emph{ACS Nano} \textbf{10}, 6163--6172 (2016).

\bibitem{Utke2006}
Utke, I. \emph{et~al.}
\newblock {Tensile strengths of metal-containing joints fabricated by focused
  electron beam induced deposition}.
\newblock \emph{Adv. Eng. Mater.} \textbf{8}, 155--157 (2006).

\bibitem{Friedli2009}
Friedli, V., Utke, I., M{\o}lhave, K. \& Michler, J.
\newblock {Dose and energy dependence of mechanical properties of focused
  electron-beam-induced pillar deposits from Cu(C 5 HF 6 O 2 ) 2}.
\newblock \emph{Nanotechnology} \textbf{20}, 385304 (2009).

\bibitem{Lewis2017}
Lewis, B.~B. \emph{et~al.}
\newblock {Growth and nanomechanical characterization of nanoscale 3D
  architectures grown via focused electron beam induced deposition}.
\newblock \emph{Nanoscale} \textbf{9}, 16349--16356 (2017).

\bibitem{Wich2008}
Wich, T.
\newblock {Nanostructuring and Nanobonding by EBiD}.
\newblock In S.~Fatikow, editor, \emph{Autom. Nanohandling by Microrobots},
  pages 295--340,  (Springer London, London2008).

\bibitem{Bresin2013}
Bresin, M., Toth, M. \& Dunn, K.~A.
\newblock {Direct-write 3D nanolithography at cryogenic temperatures}.
\newblock \emph{Nanotechnology} \textbf{24}, 035301 (2013).

\bibitem{Bresin2011}
Bresin, M., Thiel, B., Toth, M. \& Dunn, K.
\newblock {Focused electron beam-induced deposition at cryogenic temperatures}.
\newblock \emph{J. Mater. Res.} \textbf{26}, 357--364 (2011).

\bibitem{Mencik2006}
Mencik, J. \& Swain, M.~V.
\newblock {Mechanical Properties of Platinum Films on Silicon and Glass
  Determined by Ultra-Microindentation}.
\newblock \emph{MRS Proc.} \textbf{356}, 729 (1994).

\bibitem{Mall2009}
Mall, S., Lee, H., Leedy, K.~D. \& Coutu, R.~A.
\newblock {Interrelationship Between Hardness and Resistivity of Metal Alloy
  Films as Contact Materials in MEMS Switches}.
\newblock In \emph{Part B Magn. Storage Tribol. Manuf. Tribol. Nanotribology;
  Eng. Surfaces; Biotribology; Emerg. Technol. Spec. Symp. Contact Mech. Spec.
  Symp. Nanotribology}, pages 1377--1378,  (ASME2006).

\bibitem{Zhang2014}
Zhang, Z., Chen, D., Han, W. \& Kimura, A.
\newblock {Irradiation hardening in pure tungsten before and after
  recrystallization}.
\newblock \emph{Fusion Eng. Des.} \textbf{98-99}, 2103--2107 (2015).

\bibitem{Abadias2006}
Abadias, G., Dub, S. \& Shmegera, R.
\newblock {Nanoindentation hardness and structure of ion beam sputtered TiN, W
  and TiN/W multilayer hard coatings}.
\newblock \emph{Surf. Coatings Technol.} \textbf{200}, 6538--6543 (2006).

\bibitem{Birnbaum2010a}
Birnbaum, A.~J. \emph{et~al.}
\newblock {Laser printed micron-scale free standing laminate composites:
  Process and properties}.
\newblock \emph{J. Appl. Phys.} \textbf{108}, 1--7 (2010).

\bibitem{Birnbaum2011a}
Birnbaum, A.~J., Zalalutdinov, M.~K., Wahl, K.~J. \& Pique, A.
\newblock {Fabrication and Response of Laser-Printed Cavity-Sealing Membranes}.
\newblock \emph{J. Microelectromechanical Syst.} \textbf{20}, 436--440 (2011).

\bibitem{Shugurov2004a}
Shugurov, A., Panin, A., Hui-Gon, C. \& Oskomov, K.
\newblock {Size effects on the mechanical properties of thin metallic films
  studied by nanoindentation}.
\newblock \emph{Sci. Technol. 2004. KORUS 2004. Proceedings. 8th Russ. Int.
  Symp.} \textbf{3}, 168--172 vol. 3 (2004).

\bibitem{Panin2005}
Panin, A., Shugurov, A. \& Oskomov, K.
\newblock {Mechanical properties of thin Ag films on a silicon substrate
  studied using the nanoindentation technique}.
\newblock \emph{Phys. Solid State} \textbf{47}, 2055--2059 (2005).

\bibitem{Okuda1999}
Okuda, S., Kobiyama, M. \& Inami, T.
\newblock {Mechanical Properties and Thermal Stability of Nanocrystalline Gold
  Prepared by Gas Deposition Method}.
\newblock \emph{Mater. Trans. JIM} \textbf{40}, 412--415 (1999).

\bibitem{Volinsky2004}
Volinsky, A.~A., Moody, N.~R. \& Gerberich, W.~W.
\newblock {Nanoindentation of Au and Pt/Cu thin films at elevated
  temperatures}.
\newblock \emph{J. Mater. Res.} \textbf{19}, 2650--2657 (2004).

\bibitem{Beegan2003}
Beegan, D., Chowdhury, S. \& Laugier, M.~T.
\newblock {A nanoindentation study of copper films on oxidised silicon
  substrates}.
\newblock \emph{Surf. Coatings Technol.} \textbf{176}, 124--130 (2003).

\bibitem{Lu2005}
Lu, L. \emph{et~al.}
\newblock {Nano-sized twins induce high rate sensitivity of flow stress in pure
  copper}.
\newblock \emph{Acta Mater.} \textbf{53}, 2169--2179 (2005).

\bibitem{Beegan2007}
Beegan, D., Chowdhury, S. \& Laugier, M.~T.
\newblock {Comparison between nanoindentation and scratch test hardness
  (scratch hardness) values of copper thin films on oxidised silicon
  substrates}.
\newblock \emph{Surf. Coatings Technol.} \textbf{201}, 5804--5808 (2007).

\bibitem{Chang2007}
Chang, S.-Y. \& Chang, T.-K.
\newblock {Grain size effect on nanomechanical properties and deformation
  behavior of copper under nanoindentation test}.
\newblock \emph{J. Appl. Phys.} \textbf{101}, 033507 (2007).

\bibitem{Nili2014}
Nili, H., Walia, S., Bhaskaran, M. \& Sriram, S.
\newblock {Nanoscale electro-mechanical dynamics of nano-crystalline platinum
  thin films: An in situ electrical nanoindentation study}.
\newblock \emph{J. Appl. Phys.} \textbf{116}, 163504 (2014).

\bibitem{Richner2016a}
Richner, P. \emph{et~al.}
\newblock {Printable Nanoscopic Metamaterial Absorbers and Images with
  Diffraction-Limited Resolution}.
\newblock \emph{ACS Appl. Mater. Interfaces} \textbf{8}, 11690--11697 (2016).

\bibitem{Podsiadlo2010}
Podsiadlo, P. \emph{et~al.}
\newblock {The role of order, nanocrystal size, and capping ligands in the
  collective mechanical response of three-dimensional nanocrystal solids}.
\newblock \emph{J. Am. Chem. Soc.} \textbf{132}, 8953--8960 (2010).

\bibitem{Yan2013}
Yan, C., Arfaoui, I., Goubet, N. \& Pileni, M.~P.
\newblock {Soft supracrystals of Au nanocrystals with tunable mechanical
  properties}.
\newblock \emph{Adv. Funct. Mater.} \textbf{23}, 2315--2321 (2013).

\bibitem{Greer2007}
Greer, J.~R. \& Street, R.~A.
\newblock {Mechanical characterization of solution-derived nanoparticle silver
  ink thin films}.
\newblock \emph{J. Appl. Phys.} \textbf{101} (2007).

\bibitem{Lee2010}
Lee, D.~J. \& Oh, J.~H.
\newblock {Inkjet printing of conductive Ag lines and their electrical and
  mechanical characterization}.
\newblock \emph{Thin Solid Films} \textbf{518}, 6352--6356 (2010).

\bibitem{Dou2010}
Dou, R., Xu, B. \& Derby, B.
\newblock {High-strength nanoporous silver produced by inkjet printing}.
\newblock \emph{Scr. Mater.} \textbf{63}, 308--311 (2010).

\bibitem{Hashin1983}
Hashin, Z.
\newblock {Analysis of Composite Materials—A Survey}.
\newblock \emph{J. Appl. Mech.} \textbf{50}, 481 (1983).

\bibitem{Roberts2000}
Roberts, A.~P. \& Garboczi, E.~J.
\newblock {Elastic properties of model porous ceramics}.
\newblock \emph{J. Am. Ceram. Soc.} \textbf{83}, 3041--3048 (2000).

\bibitem{Hodge2007}
Hodge, A.~M. \emph{et~al.}
\newblock {Scaling equation for yield strength of nanoporous open-cell foams}.
\newblock \emph{Acta Mater.} \textbf{55}, 1343--1349 (2007).

\bibitem{Jakus2015}
Jakus, A.~E., Taylor, S.~L., Geisendorfer, N.~R., Dunand, D.~C. \& Shah, R.~N.
\newblock {Metallic Architectures from 3D-Printed Powder-Based Liquid Inks}.
\newblock \emph{Adv. Funct. Mater.} \textbf{25}, 6985--6995 (2015).

\bibitem{Taylor2016}
Taylor, S.~L., Jakus, A.~E., Shah, R.~N. \& Dunand, D.~C.
\newblock {Iron and Nickel Cellular Structures by Sintering of 3D-Printed Oxide
  or Metallic Particle Inks}.
\newblock \emph{Adv. Eng. Mater.}  (2016).

\bibitem{Rothlisberger2016}
R{\"{o}}thlisberger, A., H{\"{a}}berli, S., Spolenak, R. \& Dunand, D.~C.
\newblock {Synthesis, structure and mechanical properties of ice-templated
  tungsten foams}.
\newblock \emph{J. Mater. Res.} \textbf{31}, 753--764 (2016).

\bibitem{Calvo2018}
Calvo, M., Jakus, A.~E., Shah, R.~N., Spolenak, R. \& Dunand, D.~C.
\newblock {Microstructure and Processing of 3D Printed Tungsten Microlattices
  and Infiltrated W-Cu Composites}.
\newblock \emph{Adv. Eng. Mater.} \textbf{20}, 1800354 (2018).

\bibitem{Kenel2019}
Kenel, C., Casati, N. P.~M. \& Dunand, D.~C.
\newblock {3D ink-extrusion additive manufacturing of CoCrFeNi high-entropy
  alloy micro-lattices}.
\newblock \emph{Nat. Commun.} \textbf{10}, 904 (2019).

\bibitem{Zenou2015b}
Zenou, M., Sa'ar, A. \& Kotler, Z.
\newblock {Laser Transfer of Metals and Metal Alloys for Digital
  Microfabrication of 3D Objects}.
\newblock \emph{Small} \textbf{11}, 4082--4089 (2015).

\bibitem{Feinaeugle2018}
Feinaeugle, M., Pohl, R., Bor, T., Vaneker, T. \& R{\"{o}}mer, G.-w.
\newblock {Printing of complex free-standing microstructures via laser-induced
  forward transfer (LIFT) of pure metal thin films}.
\newblock \emph{Addit. Manuf.} \textbf{24}, 391--399 (2018).

\bibitem{Fogel2019}
Fogel, O. \emph{et~al.}
\newblock {An investigation of the influence of thermal process on the
  electrical conductivity of LIFT printed Cu structures}.
\newblock \emph{J. Phys. D. Appl. Phys.} \textbf{52}, 285303 (2019).

\bibitem{Suryavanshi2006}
Suryavanshi, A.~P. \& Yu, M.-F.
\newblock {Probe-based electrochemical fabrication of freestanding Cu nanowire
  array}.
\newblock \emph{Appl. Phys. Lett.} \textbf{88}, 083103 (2006).

\bibitem{Suryavanshi2007}
Suryavanshi, A.~P. \& Yu, M.-F.
\newblock {Electrochemical fountain pen nanofabrication of vertically grown
  platinum nanowires}.
\newblock \emph{Nanotechnology} \textbf{18}, 105305 (2007).

\bibitem{Huth2018}
Huth, M., Porrati, F. \& Dobrovolskiy, O.
\newblock {Focused electron beam induced deposition meets materials science}.
\newblock \emph{Microelectron. Eng.} \textbf{185-186}, 9--28 (2018).

\bibitem{Jiang1989a}
Jiang, X., Reichelt, K. \& Stritzker, B.
\newblock {The hardness and Young's modulus of amorphous hydrogenated carbon
  and silicon films measured with an ultralow load indenter}.
\newblock \emph{J. Appl. Phys.} \textbf{66}, 5805--5808 (1989).

\bibitem{Weiler1996}
Weiler, M. \emph{et~al.}
\newblock {Preparation and properties of highly tetrahedral hydrogenated
  amorphous carbon}.
\newblock \emph{Phys. Rev. B} \textbf{53}, 1594--1608 (1996).

\bibitem{Li2007c}
Li, J., Bresin, M., Dunn, K. \& Thiel, B.
\newblock {Determination of Impurity Carbon sp2/sp3 Bond Ratio in Electron Beam
  Deposited Tungsten Nanostructures using Electron Energy Loss Spectroscopy}.
\newblock \emph{Microsc. Microanal.} \textbf{13}, 1472--1473 (2007).

\bibitem{Ding2005}
Ding, W. \emph{et~al.}
\newblock {Mechanics of hydrogenated amorphous carbon deposits from
  electron-beam-induced deposition of a paraffin precursor}.
\newblock \emph{J. Appl. Phys.} \textbf{98}, 014905 (2005).

\bibitem{Cho2005}
Cho, S., Chasiotis, I., Friedmann, T.~A. \& Sullivan, J.~P.
\newblock {Young's modulus, Poisson's ratio and failure properties of
  tetrahedral amorphous diamond-like carbon for MEMS devices}.
\newblock \emph{J. Micromechanics Microengineering} \textbf{15}, 728--735
  (2005).

\bibitem{Bret2005a}
Bret, T., Mauron, S., Utke, I. \& Hoffmann, P.
\newblock {Characterization of focused electron beam induced carbon deposits
  from organic precursors}.
\newblock \emph{Microelectron. Eng.} \textbf{78-79}, 300--306 (2005).

\bibitem{Post2011}
Post, P.~C., Mohammadi-Gheidari, A., Hagen, C.~W. \& Kruit, P.
\newblock {Parallel electron-beam-induced deposition using a multi-beam
  scanning electron microscope}.
\newblock \emph{J. Vac. Sci. Technol. B, Nanotechnol. Microelectron. Mater.
  Process. Meas. Phenom.} \textbf{29}, 06F310 (2011).

\bibitem{Lee2008c}
Lee, J.-S. \emph{et~al.}
\newblock {Design and evaluation of a silicon based multi-nozzle for
  addressable jetting using a controlled flow rate in electrohydrodynamic jet
  printing}.
\newblock \emph{Appl. Phys. Lett.} \textbf{93}, 243114 (2008).

\bibitem{Pan2017}
Pan, Y., Chen, X., Zeng, L., Huang, Y. \& Yin, Z.
\newblock {Fabrication and evaluation of a protruding Si-based printhead for
  electrohydrodynamic jet printing}.
\newblock \emph{J. Micromechanics Microengineering} \textbf{27}, 125004 (2017).

\bibitem{Auyeung2015}
Auyeung, R. C.~Y., Kim, H., Mathews, S. \& Piqu{\'{e}}, A.
\newblock {Spatially modulated laser pulses for printing electronics}.
\newblock \emph{Appl. Opt.} \textbf{54}, F70 (2015).

\bibitem{Zheng2006}
Zheng, N., Fan, J. \& Stucky, G.~D.
\newblock {One-step one-phase synthesis of monodisperse noble-metallic
  nanoparticles and their colloidal crystals}.
\newblock \emph{J. Am. Chem. Soc.} \textbf{128}, 6550--6551 (2006).

\bibitem{Iwata2009}
Iwata, F., Kaji, M., Suzuki, A., Ito, S. \& Nakao, H.
\newblock {Local electrophoresis deposition of nanomaterials assisted by a
  laser trapping technique}.
\newblock \emph{Nanotechnology} \textbf{20}, 235303 (2009).

\bibitem{Charipar2018}
Charipar, K.~M., Diaz-Rivera, R.~E., Charipar, N.~A. \& Piqu{\'{e}}, A.
\newblock {Laser-induced forward transfer (LIFT) of 3D microstructures}.
\newblock In H.~Helvajian, A.~Piqu{\'{e}} \& B.~Gu, editors, \emph{Laser 3D
  Manuf. V}, February 2018, page~29,  (SPIE2018).

\bibitem{Zenou2016}
Zenou, M. \& Kotler, Z.
\newblock {Printing of metallic 3D micro-objects by laser induced forward
  transfer}.
\newblock \emph{Opt. Express} \textbf{24}, 1431 (2016).

\bibitem{Wheeler2013}
Wheeler, J.~M. \& Michler, J.
\newblock {Elevated temperature, nano-mechanical testing in situ in the
  scanning electron microscope}.
\newblock \emph{Rev. Sci. Instrum.} \textbf{84}, 045103 (2013).

\bibitem{KOSTER2014}
K{\"{o}}ster, W. \& Franz, H.
\newblock {Poisson's ratio for metals and alloys}.
\newblock \emph{Metall. Rev.} \textbf{6}, 1--56 (1961).

\bibitem{Maier2011}
Maier, V. \emph{et~al.}
\newblock {Nanoindentation strain-rate jump tests for determining the local
  strain-rate sensitivity in nanocrystalline Ni and ultrafine-grained Al}.
\newblock \emph{J. Mater. Res.} \textbf{26}, 1421--1430 (2011).

\bibitem{Leitner2016}
Leitner, A., Maier-Kiener, V. \& Kiener, D.
\newblock {Extraction of Flow Behavior and Hall-Petch Parameters Using a
  Nanoindentation Multiple Sharp Tip Approach}.
\newblock \emph{Adv. Eng. Mater.} \textbf{19}, 1600669 (2017).

\bibitem{Sneddon1965}
Sneddon, I.~N.
\newblock {The relation between load and penetration in the axisymmetric
  boussinesq problem for a punch of arbitrary profile}.
\newblock \emph{Int. J. Eng. Sci.} \textbf{3}, 47--57 (1965).

\bibitem{Pique2013}
Piqu{\'{e}}, A. \emph{et~al.}
\newblock {Laser transfer of reconfigurable patterns with a spatial light
  modulator}.
\newblock \emph{Proc. SPIE} \textbf{8608}, 86080K--86080K--9 (2013).

\bibitem{Breckenfeld2015}
Breckenfeld, E. \emph{et~al.}
\newblock {Laser-induced forward transfer of silver nanopaste for microwave
  interconnects}.
\newblock \emph{Appl. Surf. Sci.} \textbf{331}, 254--261 (2015).

\bibitem{Stanford2014}
Stanford, M.~G. \emph{et~al.}
\newblock {Purification of Nanoscale Electron-Beam-Induced Platinum Deposits
  via a Pulsed Laser-Induced Oxidation Reaction}.
\newblock \emph{ACS Appl. Mater. Interfaces} \textbf{6}, 21256--21263 (2014).

\bibitem{Geier2014}
Geier, B. \emph{et~al.}
\newblock {Rapid and Highly Compact Purification for Focused Electron Beam
  Induced Deposits: A Low Temperature Approach Using Electron Stimulated H 2 O
  Reactions}.
\newblock \emph{J. Phys. Chem. C} \textbf{118}, 14009--14016 (2014).

\bibitem{Shawrav2016}
Shawrav, M.~M. \emph{et~al.}
\newblock {Highly conductive and pure gold nanostructures grown by electron
  beam induced deposition}.
\newblock \emph{Sci. Rep.} \textbf{6}, 34003 (2016).

\bibitem{DosSantos2018}
{Puydinger dos Santos}, M.~V. \emph{et~al.}
\newblock {Comparative study of post-growth annealing of Cu(hfac) 2 , Co 2 (CO)
  8 and Me 2 Au(acac) metal precursors deposited by FEBID}.
\newblock \emph{Beilstein J. Nanotechnol.} \textbf{9}, 91--101 (2018).

\bibitem{Pablo-Navarro2018}
Pablo-Navarro, J., Mag{\'{e}}n, C. \& de~Teresa, J.~M.
\newblock {Purified and Crystalline Three-Dimensional Electron-Beam-Induced
  Deposits: The Successful Case of Cobalt for High-Performance Magnetic
  Nanowires}.
\newblock \emph{ACS Appl. Nano Mater.} \textbf{1}, 38--46 (2018).

\bibitem{Choi2004a}
Choi, T.~Y., Poulikakos, D. \& Grigoropoulos, C.~P.
\newblock {Fountain-pen-based laser microstructuring with gold nanoparticle
  inks}.
\newblock \emph{Appl. Phys. Lett.} \textbf{85}, 13--15 (2004).

\bibitem{Buffat1976}
Buffat, P. \& Borel, J.-P.
\newblock {Size effect on the melting temperature of gold particles}.
\newblock \emph{Phys. Rev. A} \textbf{13}, 2287--2298 (1976).

\bibitem{Tabor1970}
Tabor, D.
\newblock {The hardness of solids}.
\newblock \emph{Rev. Phys. Technol.} \textbf{1}, 145--179 (1970).

\bibitem{Jaeger1996}
Jaeger, H.~M., Nagel, S.~R. \& Behringer, R.~P.
\newblock {Granular solids, liquids, and gases}.
\newblock \emph{Rev. Mod. Phys.} \textbf{68}, 1259--1273 (1996).

\bibitem{Emery2003}
Emery, R.~D. \& Povirk, G.~L.
\newblock {Tensile behavior of free-standing gold films. Part I. Coarse-grained
  films}.
\newblock \emph{Acta Mater.} \textbf{51}, 2067--2078 (2003).

\bibitem{Emery2003a}
Emery, R.~D. \& Povirk, G.~L.
\newblock {Tensile behavior of free-standing gold films. Part II. Fine-grained
  films}.
\newblock \emph{Acta Mater.} \textbf{51}, 2079--2087 (2003).

\bibitem{Chen2016}
Chen, C.~Y. \emph{et~al.}
\newblock {Pulse electroplating of ultra-fine grained Au films with high
  compressive strength}.
\newblock \emph{Electrochem. commun.} \textbf{67}, 51--54 (2016).

\bibitem{Yanagida2017}
Yanagida, S. \emph{et~al.}
\newblock {Microelectronic Engineering Tensile tests of micro-specimens
  composed of electroplated gold}.
\newblock \emph{Microelectron. Eng.} \textbf{174}, 6--10 (2017).

\bibitem{Fang2003}
Fang, T.-H. \& Chang, W.-J.
\newblock {Nanomechanical properties of copper thin films on different
  substrates using the nanoindentation technique}.
\newblock \emph{Microelectron. Eng.} \textbf{65}, 231--238 (2003).

\bibitem{Hong2005}
Hong, S.~H. \emph{et~al.}
\newblock {Characterization of elastic moduli of Cu thin films using
  nanoindentation technique}.
\newblock \emph{Compos. Sci. Technol.} \textbf{65}, 1401--1408 (2005).

\bibitem{Dao2006}
Dao, M., Lu, L., Shen, Y.~F. \& Suresh, S.
\newblock {Strength, strain-rate sensitivity and ductility of copper with
  nanoscale twins}.
\newblock \emph{Acta Mater.} \textbf{54}, 5421--5432 (2006).

\bibitem{Jang2011}
Jang, D., Cai, C. \& Greer, J.~R.
\newblock {Influence of homogeneous interfaces on the strength of 500 nm
  diameter Cu nanopillars}.
\newblock \emph{Nano Lett.} \textbf{11}, 1743--1746 (2011).

\bibitem{Mieszala2016}
Mieszala, M. \emph{et~al.}
\newblock {Orientation-dependent mechanical behaviour of electrodeposited
  copper with nanoscale twins}.
\newblock \emph{Nanoscale} \textbf{8}, 15999--16004 (2016).

\bibitem{Darling1966}
Darling, A.~S.
\newblock {The Elastic and Plastic Properties of the Platinum Metals}.
\newblock \emph{Res. Lab. Johnson Matthey Co Ltd.} pages 14--19 (1966).

\bibitem{Merker2001}
Merker, J., Lupton, D., T{\"{o}}pfer, M. \& Knake, H.
\newblock {High temperature mechanical properties of the platinum group metals:
  Elastic properties of platinum, rhodium and iridium and their alloys at high
  temperatures}.
\newblock \emph{Platin. Met. Rev.} \textbf{45}, 74--82 (2001).

\end{thebibliography}

\begin{figure*}[b] 
   \centering
   \includegraphics[width=178mm]{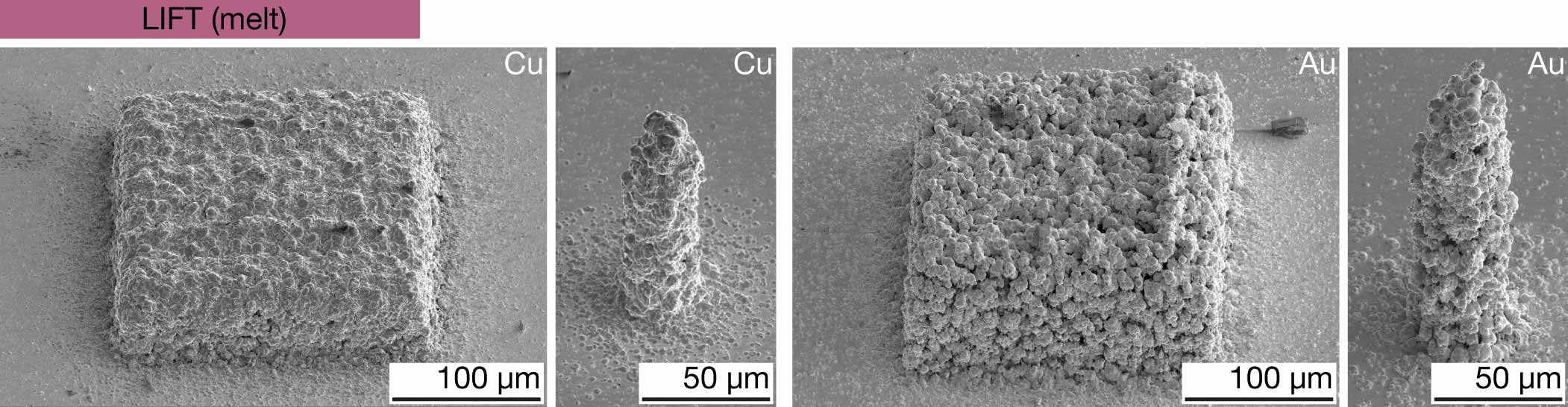} 
   \caption{\textbf{Morphology of LIFT-printed Cu and Au.} SE micrographs of representative pads and pillars printed from either a Cu (left) or a Au (right) donor thin film. Image tilts: \SI{45}{\degree} for pads and \SI{55}{\degree} for pillars.}
\label{fig:SI_Morphology_LIFTmelt}
\end{figure*}

\newpage

\section{Supplementary Information}
\subsection{Limitations of the study}\label{sec:SI_limitations}
Three factors constrain the interpretation of the reported results: first, the materials tested in this study are representative materials synthesized in the authors' laboratories, but were not optimized for high mechanical strength. Hence, within the limitations of each technique, higher modulus and strength can be expected once printing and annealing processes are optimized. Second, the range of metals presently most commonly studied is very limited and different for each technique: Ag or Au for ink-based techniques, Cu for electrochemical deposition, and Pt, Co, Fe or W in the case of FEBID / FIBID. As the study had to comply to these limits, the obtained data cannot entirely be decoupled from the fact that different elements were used. Consequently, the results cannot be linked exclusively to the process characteristics of the individual AM principles. However, general microstructural features directly related to the deposition principles typically remain unaffected by the chemical nature (porosity in Au and Ag inks, inter-droplet gaps in LIFT printed Au and Cu structures, Figure~\ref{fig:SI_LIFT}).} Third, an accurate measurement of mechanical properties requires well defined geometries of the test specimens. In this respect, many of the used small-scale AM methods struggle to deposit samples with ideal geometries. Nanoindentation analysis was sometimes limited by a high surface roughness $\gg$\SI{100}{\nano\meter} (LIFT (melt)), a sample thickness $<$1~\==~\SI{2}{\micro\meter} (FIBID, FEBID), or a distance from an indent to the edge of a pad $<10\times$ the indentation depth (EHDP, LAEPD, FluidFM, EHD-RP, FIBID and FEBID: pad width only 10~\==~\SI{20}{\micro\meter}). Microcompression analysis was complicated by inhomogeneous pillar diameters and pillar tilt (LAEPD, EHDP (annealed only), MCED, cryo-FEBID). For some AM methods, these factors result in a pronounced scatter between individual samples: standard deviations typically range between 10~\==~\SI{20}{\percent} but can be as high as 40~\==~\SI{80}{\percent} for a few techniques, especially for nanoindentation data (Section~\ref{sec:SI_mech}).

\begin{figure*}[htpb] 
   \centering
   \includegraphics[width=178mm]{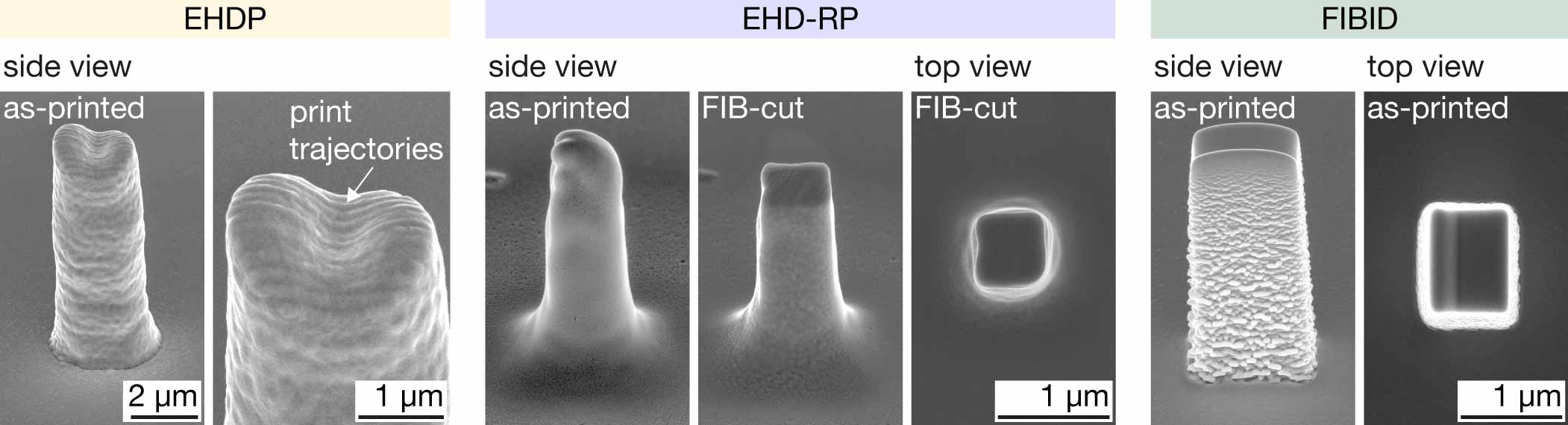} 
   \caption{\textbf{High-resolution in-plane hatching.} The techniques with a resolution $\ll$\SI{1}{\micro\meter} enable shape control of printed pillars' cross-sections. In the case of EHDP, pillars are round, but the visible print trajectories indicate the in-plane hatching. Pillars by EDH-RP and FIBID (as well as FEBID, not shown) are square due to the quadratic hatch patterns used. A FIB-cut pillar is shown for EHD-RP to highlight the quadratic cross-section. Tilt of side-view images: \SI{55}{\degree}.}
\label{fig:SI_Morphology_squarepillars}
\end{figure*}

\subsection{Morphology} \label{sec:SI_Morphology}
\subsubsection{LIFT (melt)}
Figure~\ref{fig:SI_Morphology_LIFTmelt} shows LIFT structures printed from a Cu as well as a Au donor film.

\subsubsection{Distortion upon annealing}
Au pillars and pads printed by EHDP and LAEPD both underwent pronounced inhomogeneous shrinkage that resulted in wrinkled pillars and often poor adhesion of the annealed pads. In contrast, DIW and LIFT (ink) deposited Ag structures, especially pillars, showed a more isotropic volume loss (although cracking is evident in annealed DIW (N.) pads). The precise reason for the observed difference is unknown. The most obvious distinctions between inhomogeneously and homogeneously shrunken pillars are their size ($\approx$\SI{2}{\micro\meter} dia. for the inhomogeneous pillars versus $\approx$10~\==~\SI{40}{\micro\meter} for the homogeneous pillars) and their material (Au and Ag, respectively). While the metal itself is probably secondary, the size might have a very simple effect: in small structures, small-scale inhomogeneities upon densification are proportionally larger compared to the size of the whole structure, and have thus a more profound effect on the resulting overall geometry. In larger structures, the same inhomogeneities are probably averaged or simply less obvious. However, as structures were printed by different methods and the as-deposited densities probably differ as well, no general conclusion should be drawn from this observation. In any case, the management of shrinkage and accompanied distortions will likely be a challenge for all ink-based methods in more complex geometries than those shown in this study.

\subsubsection{Effect of different printing strategies}
Printing strategies differ between the individual techniques. For example, pillars printed by DIW, LAEPD, MCED and the FluidFM were deposited with a width of a single voxel, {i.e.} without in-plane hatching but simple out-of-plane growth. In contrast, pillars printed by EHDP, LIFT (melt), EHD-RP, FIBID, FEBID and cryo-FEBID are built from hatched layers. With a resolution $\ll$\SI{1}{\micro\meter} this strategy enables control of the micropillar's cross-section (Figure \ref{fig:SI_Morphology_squarepillars}).
\par
Most pads were printed with multiple hatched layers, but pads by LIFT (ink) and MCED are single entities (LIFT (ink): a single sheet of ink, MCED: a deposit grown with a large capillary). The advantage of adapting to various shapes without the need for hatching within single layers is interesting: tuning the shape and size of single voxels to accommodate certain geometries with as few voxels as possible is an advantage which simplifies and accelerates the printing of the here demanded geometries and typically decreases the surface roughness (see pads by LIFT (ink) or MCED).  Nevertheless, the surface finish and overall fidelity of real-life structures built from many voxels cannot be interpolated from these results. Especially in case of MCED, this approach is clearly limited to geometries as simple as the ones presented here. In contrast, LIFT (ink) has demonstrated the transfer of complexly shaped, large sheets that enable smooth layers also for intricate layer designs\cite{Pique2013}. {Furthermore, smooth joining of multiple voxels has been demonstrated\cite{Breckenfeld2015}.}

\subsection{Microstructure and chemical composition}\label{sec:SI_microstructure}
\begin{figure*}[htbp] 
   \centering
   \includegraphics[width=178mm]{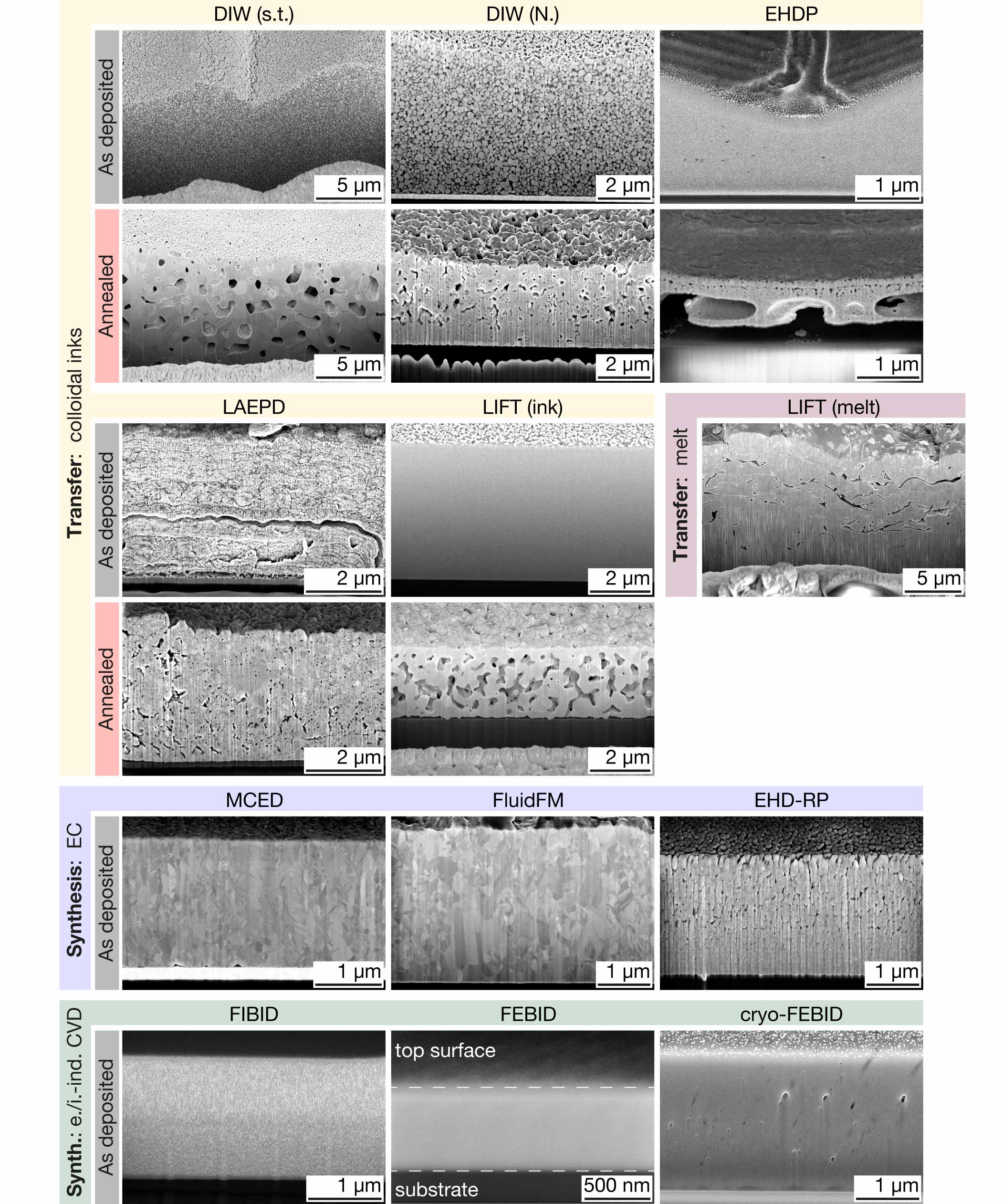} 
   \caption{\textbf{Microstructure of printed pads.} Representative cross-section micrographs of printed pads at lower magnification. The as-printed and annealed pads of EHDP and LIFT (ink) had different as-deposited thicknesses (hence no shrinkage can be concluded from these images). The cross-section of the as-deposited EHDP pad was made after indentation (hence the triangular indent at its surface). All images are tilt-corrected.}
\label{fig:SI_MicrostructurePads}
\end{figure*}

\begin{figure*}[htbp] 
   \centering
   \includegraphics[width=178mm]{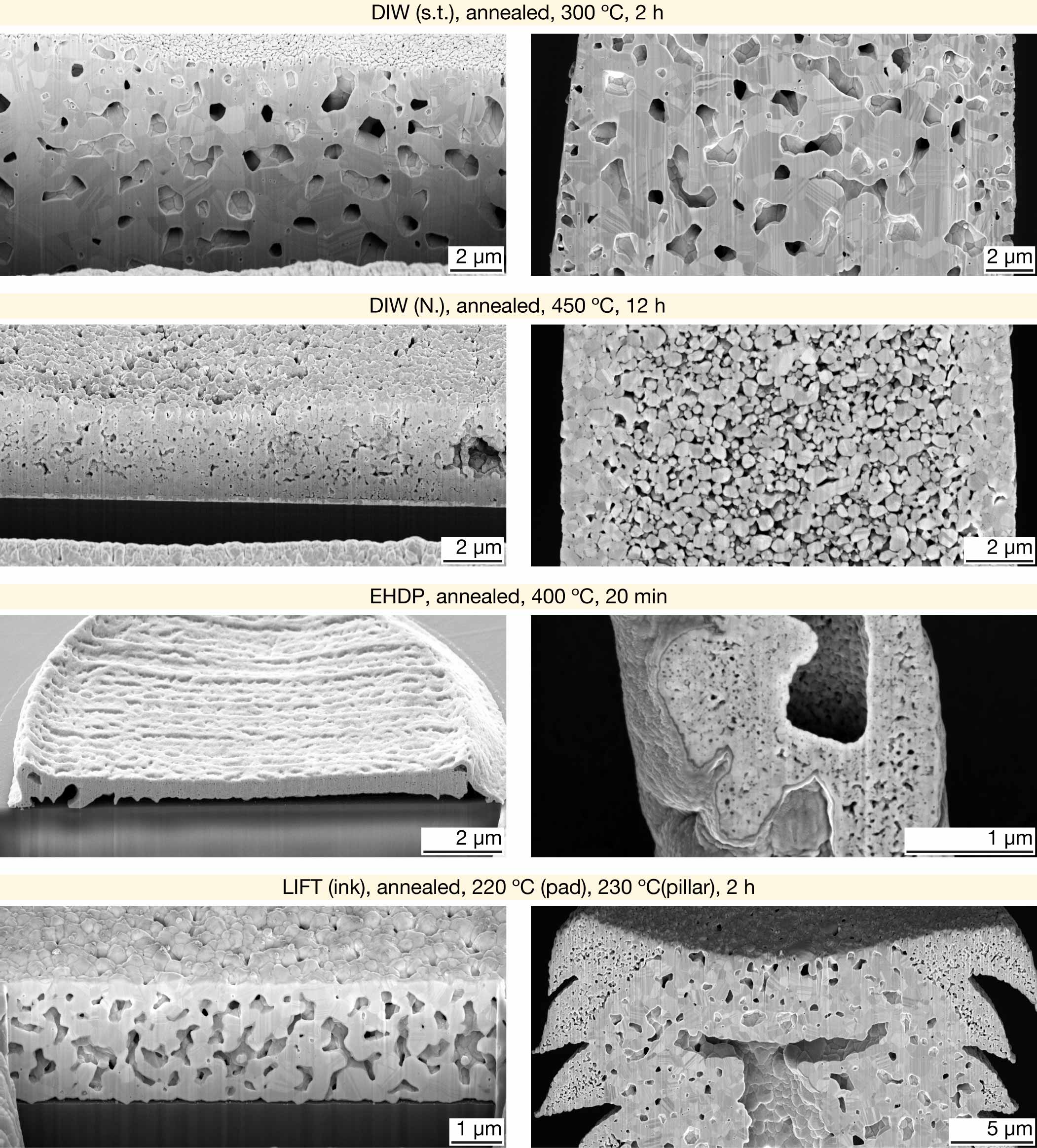} 
   \caption{\textbf{Microstructure of annealed pads and pillars printed from colloidal inks / suspensions.} FIB-cut cross-sections of representative pads and pillars printed with various colloid transfer techniques. Pads and pillars from the same techniques were annealed with the same annealing procedure.}
\label{fig:SI_Microstructure_padspillars}
\end{figure*}

Figure \ref{fig:SI_MicrostructurePads} presents cross-sections of pads imaged at low magnification. Figure \ref{fig:SI_Microstructure_padspillars} compares the microstructures of annealed inks in different geometries. For each method, the shown pads and pillars were annealed with the same annealing procedures. While some microstructures compare well (DIW (N.), EHDP), others differ between pads and pillars, presumably due to differences in local temperatures and local mechanical constraints. All shown pillars feature radial gradients in porosity, with some pillars even being hollow (EHDP, LIFT (ink)). In contrast, the microstructure of pads is typically more homogeneous (at least in the center of the pads, where the cross-sections were cut~\==~we have not studied the edges of the large DIW and LIFT (ink) pads). As an exception, the center of the EHDP pads delaminated upon annealing.
\par

\begin{figure*}[htbp] 
   \centering
   \includegraphics[width=178mm]{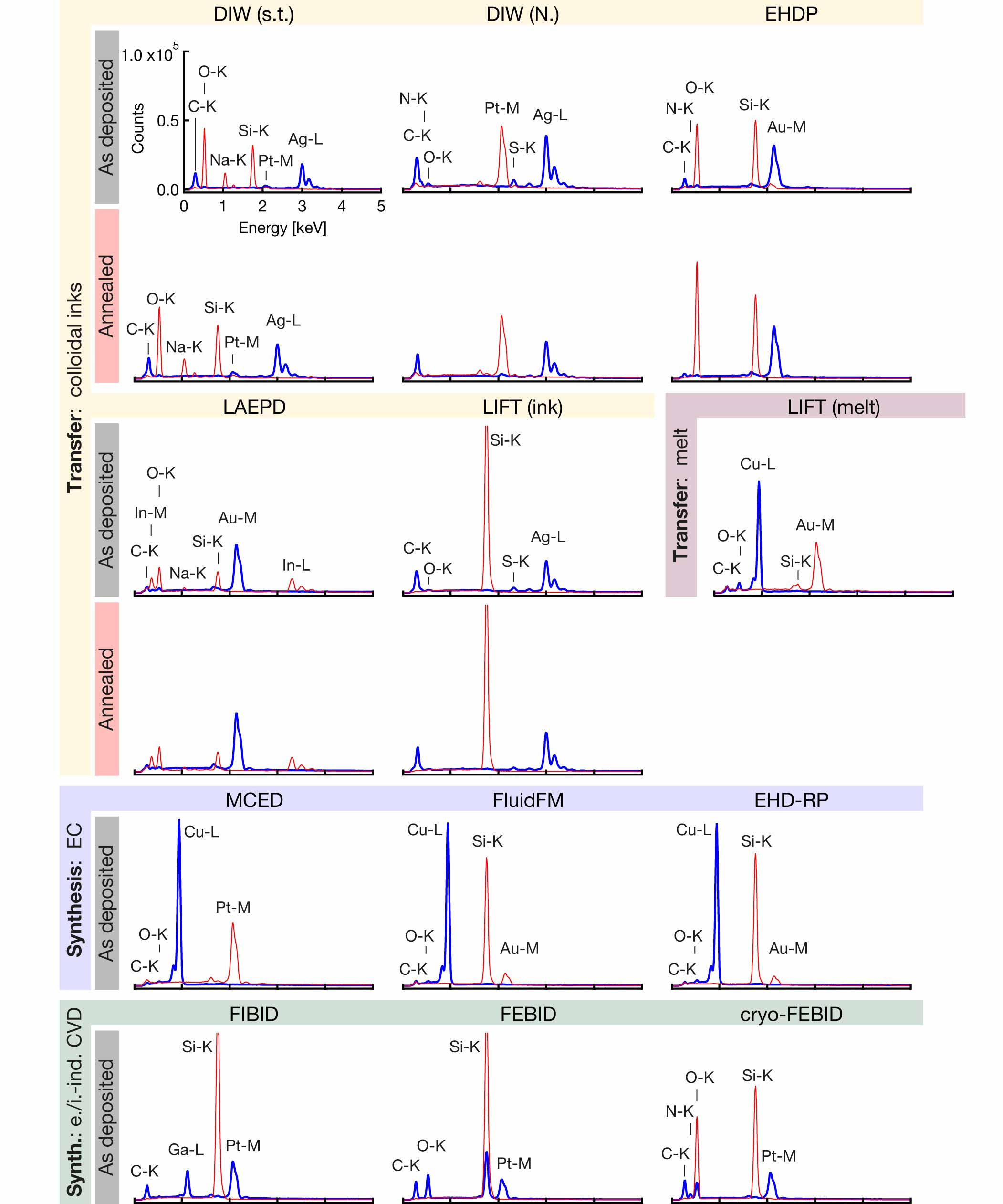} 
   \caption{\textbf{EDX analysis of printed pads.} EDX spectra of printed pads (blue) and respective substrates (red) acquired with identical parameters. The scale of all graphs is identical to the axes in the DIW (s.t.) graph.}
\label{fig:SI_EDX}
\end{figure*}

Figure \ref{fig:SI_EDX} summarizes the SEM EDX analysis performed on printed pads. EDX spectra were recorded with identical acquisition conditions from a printed pad (blue, same batch of samples presented in Figure~\ref{fig:SI_MicrostructurePads}) and the substrate (red). We refrain from a quantitative analysis of the presented data~\==~especially of the carbon and oxygen content~\==~for the following reasons: first, quantification of the light elements (C, N, O in our case) by EDX is unreliable. Second, C is always co-deposited by the electron beam during analysis. Third, the variation in geometry and density of the pads does not allow for a direct comparison between the techniques, and additionally, makes consistent decoupling signals from the printed geometries and the substrates impossible. Nevertheless, we cite some numbers in the following paragraphs~\==~please note that these are at best approximate values and that the signals from the substrate were not subtracted for this analysis.
\par
The following qualitative statements can be made: all as-deposited colloidal inks contain carbon. Upon annealing, this amount is often not significantly reduced (from $\approx$50~at.\% to $\approx$40~at.\% for both DIW inks annealed at 300 and \SI{450}{\celsius}, respectively; from $\approx$70~at.\% to $\approx$45 at.\% for EHDP pads annealed at \SI{400}{\celsius} in O$_\text{2}$ atmosphere; no reduction upon annealing for LAEPD pads ($\approx$50~at.\%)). Thus, post-print procedures that specifically target the removal of carbon should be studied if a pronounced reduction in carbon content is desired. Similarly, as-deposited FIBID and FEBID pads contain large amounts of carbon ($\approx$55~\==~65~at.\%). Additionally, FIBID structures contain Ga ($\approx$15~at.\%), and FEBID pads contain oxygen ($\approx$35~at.\%). Routines for purification of FEBID and FIBID structures exist\cite{Stanford2014,Geier2014,Shawrav2016,DosSantos2018,Pablo-Navarro2018} and need to be applied with most precursor compounds if high metal content is required in the final deposits. In contrast, electrochemically grown pads are mostly pure Cu~\==~a C-K peak is detected for pads printed by MCED, FluidFM and EHD-RP, but in all cases, the same peak is present in the spectra of the substrates, indicating a general C contamination of the samples rather than a specific contamination of the printed structures.

\subsubsection{Laser-induced fusing of Au particles in LAEPD}\label{sec:SI_microstructureLAEPD}
{We hypothesize that electrophoretically printed Au nanoparticles are fused upon laser-assisted deposition, resulting in the high strength observed for as-deposited LAEPD samples. The hypothesis is supported by the previous demonstration of laser-induced sintering of Au particles of the same size in air at lower power densities than those used by LAEPD\cite{Choi2004a}. Yet, in LAEPD one should expect a mediation of the local temperature by the surrounding liquid which would forbid temperatures that surpass the boiling point of water (as boiling was not observed). Because such low temperatures would inhibit the fusion of particles despite the melting point depression in small Au particles\cite{Buffat1976}, the details of the microstructure as well as the related deposition mechanism will need closer attention in future studies.}

\subsection{Mechanical data for all tested samples of all techniques}\label{sec:SI_mech}
\begin{figure*}[htbp] 
   \centering
   \includegraphics[width=178mm]{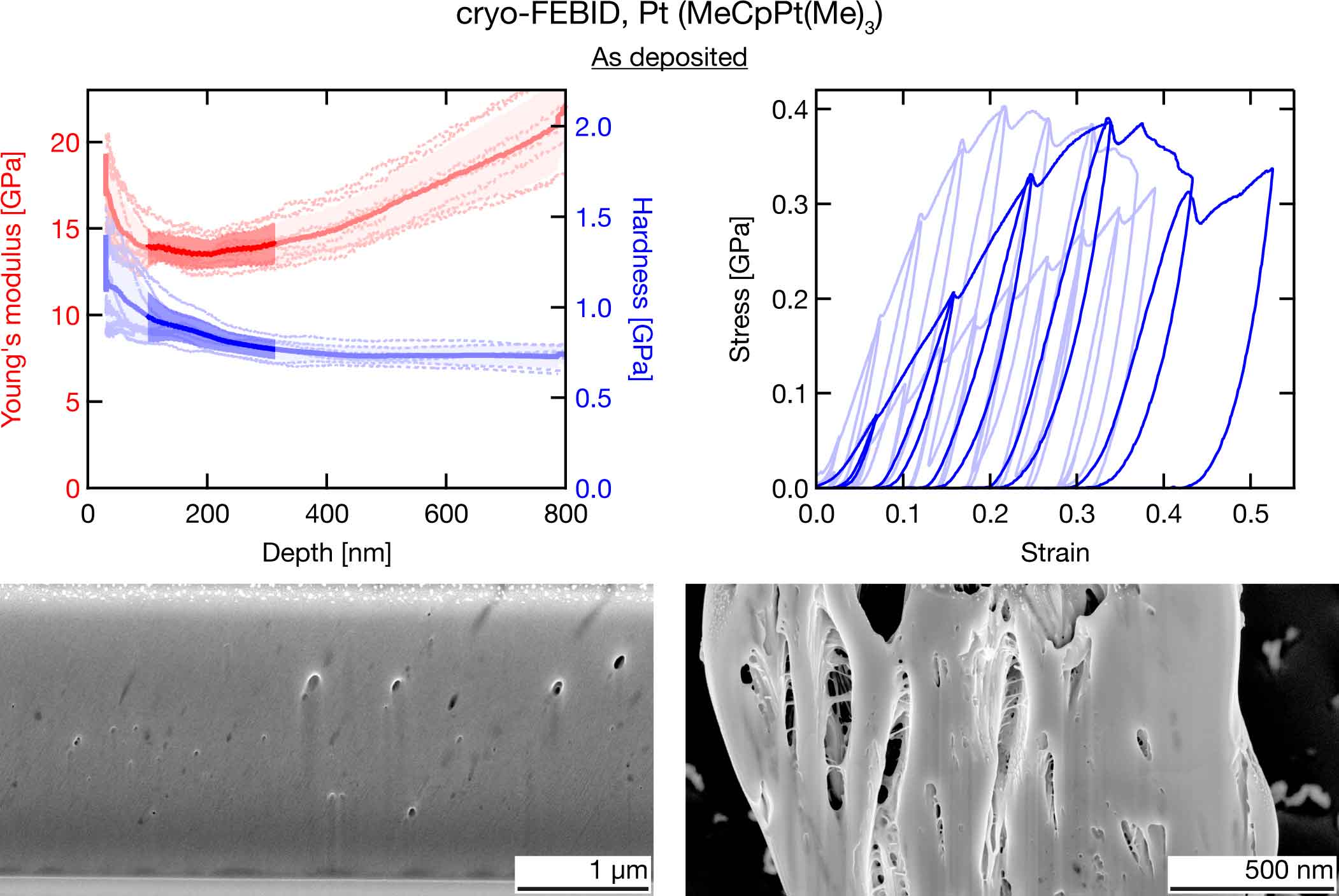} 
   \caption{Focused electron beam-induced deposition at cryogenic temperatures (cryo-FEBID) of Pt (MeCpPt(Me)$_3$).}
\label{fig:SI_KADunn}
\end{figure*}

Table \ref{tab:EH} lists the averaged values of Young's modulus $E$, hardness $H$ and yield stress $\sigma_\text{0.07}$ of the metals printed by all studied techniques. In addition, Figures \ref{fig:SI_Nanjia1}~\==~\ref{fig:SI_KADunn} summarize the mechanical data collected for all tested AM methods presented in this manuscript. The data is organized as follows (Figure~\ref{fig:SI_Nanjia1}): one or two figures per method present all analyzed mechanical data and representative micrographs of the materials' microstructures. In each figure, the data is grouped by annealing state. The left column presents nanoindentation data collected from printed pads, the right column microcompression data from printed pillars. Each graph is accompanied by a representative cross-section micrograph from one of the pads and pillars from the respective batch of samples. The nanoindentation graph shows Young's modulus (red) and hardness (blue) measured as a function of depth. The graph plots an averaged curve (bold line, with the shaded area representing the standard deviation) of all recorded curves (dashed lines). $E$ and $H$ values reported in the manuscript were extracted from the depth range highlighted in the average curves. The microcompression graph displays all measured stress-strain curves (using the average diameter of the deformed portion of the pillars). One representative curve is highlighted for better visibility.

\subsubsection{Hardness versus strength of as-deposited DIW inks}\label{sec:SI_mech_DIW}
\begin{figure*}[htbp] 
   \centering
   \includegraphics[width=178mm]{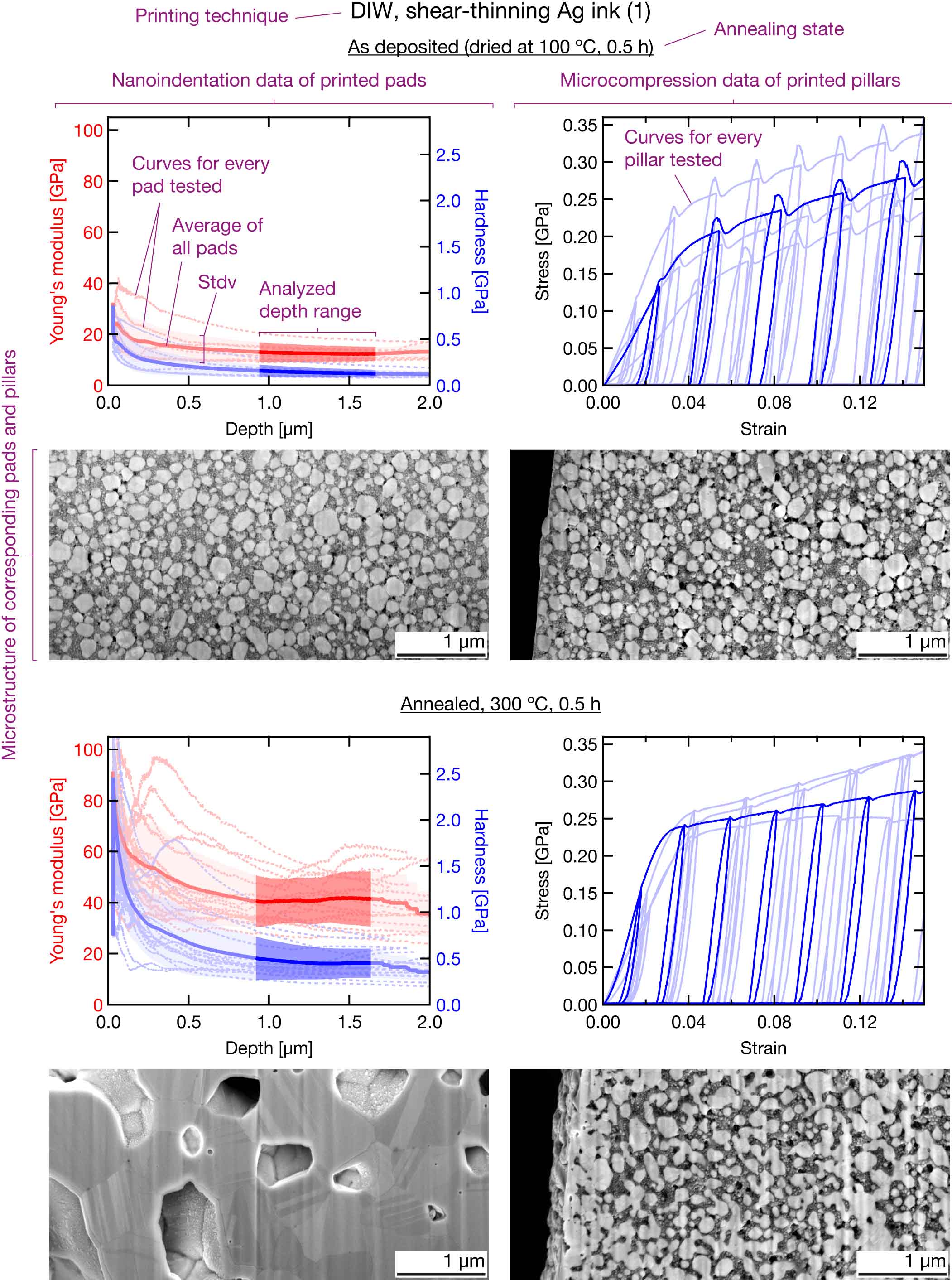} 
   \caption{Direct ink writing, shear-thinning Ag ink (DIW (s.t.)).}
\label{fig:SI_Nanjia1}
\end{figure*}

\begin{figure*}[htbp] 
   \centering
   \includegraphics[width=178mm]{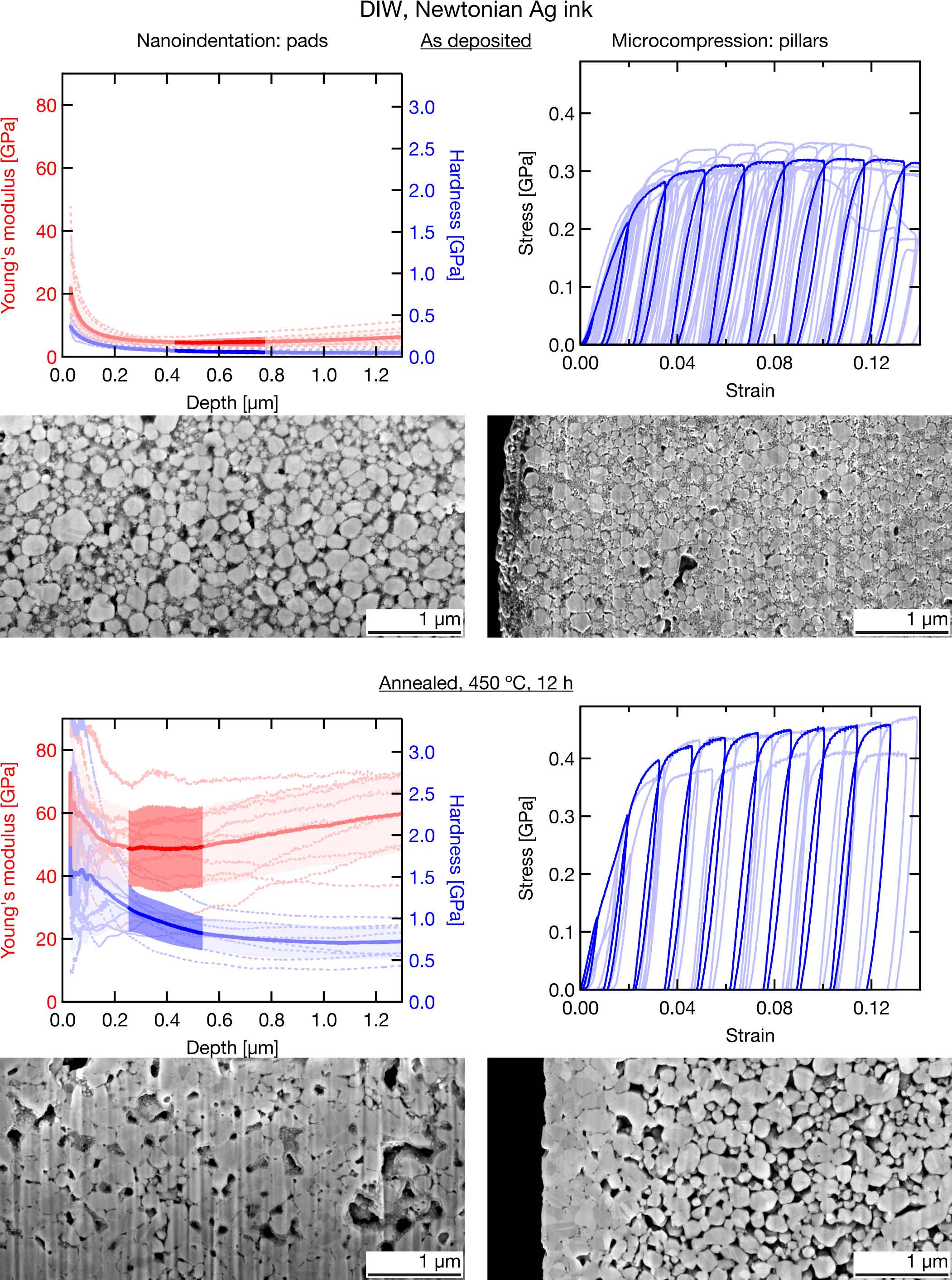} 
   \caption{Direct ink writing, Newtonian Ag ink (DIW (N.)).}
\label{fig:SI_DIWSH}
\end{figure*}

As a deviation from the expected relation of hardness to strength in solids, we measure a pronouncedly higher strength of DIW inks upon microcompression compared to the hardness derived from nanoindentation of the same materials ($\sigma_\text{0.07}$:~0.256~\==~\SI{0.315}{\giga\pascal} and $H$:~0.062~\==~\SI{0.141}{\giga\pascal}, Figures \ref{fig:SI_Nanjia1} and \ref{fig:SI_DIWSH}). For dense solids, indentation hardness $H$ is $\approx$3 times the yield strength $\sigma$\cite{Tabor1970}. For porous solids, the constraint factor can reduce to unity, {i.e.} $H=\sigma$, but $H$ never becomes smaller than $\sigma$. However, as-printed inks are clearly not solids, but rather granular media, as they are composed of discrete, noncohesive particles. Thus, strength is not necessarily related to plastic deformation of a solid, but rather the flow of individual particles under shear loading. While an analysis of the mechanics of these materials is outside the scope of this paper, we can speculate about the origins of the observed difference. Due to the interlocking of colloids, the necessary yield stresses for deformation is a function of the applied normal stresses\cite{Jaeger1996}. As the stress fields below a flat punch (microcompression) and a sharp pyramid (indentation) are very different, the resulting contact pressures detected with the two methods might differ. Additionally, the probe size in nanoindentation is comparable to the particle size and inter-particle spacing of the probed medium (indenter radius $\approx$\SI{50}{\nano\meter}, large particles $\approx$\SI{250}{\nano\meter}), whereas it is larger in microcompression (punch diameter $>$ pillar diameter, {i.e.} $>$\SI{10}{\micro\meter}). Thus, the two methods may potentially probe different deformation mechanisms. Interestingly, this difference is only observed for DIW inks, hinting towards an influence of the particle shape, the respective binder phase, or the particle size distribution.

\begin{table*}[htbp]
\caption[Study of mechanical properties: Measured $E$, $H$ and $\sigma_\text{0.07}$]{\textbf{$E$, $H$ and $\sigma_\text{0.07}$ data.} Averaged values for Young's modulus $E$, hardness $H$ and yield stress $\sigma_\text{0.07}$ {for as-deposited (ad.) and annealed samples}.}\label{tab:EH}
\centering
\begin{tabular}{l | c | c | c | c}
\toprule
Technique						& \multicolumn{2}{c}{Nanoindentation} 	\vline										& \multicolumn{2}{c}{Microcompression} 		\\
							& $E$ [GPa]							& $H$ [GPa] 						& $E$ [GPa]						& $\sigma_\text{0.07}$ [GPa] 		\\ \hline
\rule{0pt}{1.2em}\textbf{DIW (s.t.)}			&									&								&								&		\\	 	
\hspace{1em}
ad. (\SI{100}{\celsius}, \SI{0.5}{\hour})	& \num[separate-uncertainty]{12.6\pm3.2}	 		& \num[separate-uncertainty]{0.141\pm0.051}	 	& \num[separate-uncertainty]{33.3\pm4.1}	 	& \num[separate-uncertainty]{0.265\pm0.039}	 	\\ 
\hspace{1em}
\SI{300}{\celsius}, \SI{0.5}{\hour}		& \num[separate-uncertainty]{41.1\pm9.5}	 		& \num[separate-uncertainty]{0.461\pm0.181}	 	& \num[separate-uncertainty]{35.7\pm0.7}	 	& \num[separate-uncertainty]{0.289\pm0.027}	 	\\ 
\hspace{1em}
\SI{300}{\celsius}, \SI{2}{\hour}		& \num[separate-uncertainty]{31.1\pm10}	 		& \num[separate-uncertainty]{0.435\pm0.123}	 	& \num[separate-uncertainty]{42.1\pm3.8}	 		& \num[separate-uncertainty]{0.168\pm0.008}	 	\\ 
\hline
\rule{0pt}{1.2em}\textbf{DIW (N.)}	 	&									&								&								&		\\
\hspace{1em}
ad.							& \num[separate-uncertainty]{4.62\pm1.19}			& \num[separate-uncertainty]{0.0617\pm0.0144}	 	& \num[separate-uncertainty]{35.7\pm2.7}	 	& \num[separate-uncertainty]{0.315\pm0.015}	 	\\ 
\hspace{1em}
\SI{450}{\celsius}, \SI{12}{\hour}		& \num[separate-uncertainty]{48.8\pm13}	 		& \num[separate-uncertainty]{0.952\pm0.239}	 	& \num[separate-uncertainty]{65.7\pm2.7}	 	& \num[separate-uncertainty]{0.422\pm0.028}	 	\\ 
\hline
\rule{0pt}{1.2em}\textbf{EHDP}	 	&									&								&								&		\\
\hspace{1em}
ad.							& \num[separate-uncertainty]{0.521\pm0.16}			& \num[separate-uncertainty]{0.048\pm0.0154}	 	& \num[separate-uncertainty]{1.09\pm0.14}	 	& \num[separate-uncertainty]{0.0168\pm0.0017}	 	\\ 
\hspace{1em}
\SI{400}{\celsius}, \SI{20}{\minute}		& not analyzed	 		& not analyzed	 	& \num[separate-uncertainty]{31.3\pm0.4}	 	& \num[separate-uncertainty]{0.246\pm0.058}	 	\\ 
\hspace{1em}
\SI{400}{\celsius}, \SI{20}{\minute}, pillars \diameter$\approx$\SI{350}{\nano\meter}	& N/A	 		& N/A				& \num[separate-uncertainty]{39.8\pm8.8}	 	& \num[separate-uncertainty]{0.411\pm0.057}	 	\\ 

\hline
\rule{0pt}{1.2em}\textbf{LAEPD}	 	&									&								&								&		\\
\hspace{1em}
ad.							& \num[separate-uncertainty]{9.09\pm5.95}			& \num[separate-uncertainty]{0.369\pm0.321}	 	& \num[separate-uncertainty]{7.29\pm4.06}	 	& \num[separate-uncertainty]{0.146\pm0.104}	 	\\ 
\hspace{1em}
\SI{300}{\celsius}, \SI{1}{\hour}		& \num[separate-uncertainty]{54.2\pm12.5}	 		& \num[separate-uncertainty]{1.42\pm0.41}	 	& \num[separate-uncertainty]{47.1\pm8.5}	 	& \num[separate-uncertainty]{0.366\pm0.056}	 	\\ 

\hline
\rule{0pt}{1.2em}\textbf{LIFT (ink)}	 	&									&								&								&		\\
\hspace{1em}
ad.							& \num[separate-uncertainty]{1.94\pm0.07}			& \num[separate-uncertainty]{0.108\pm0.004}	 	& \num[separate-uncertainty]{1.16\pm0.01}	 	& \num[separate-uncertainty]{0.0125\pm0.0014}	 	\\ 
\hspace{1em}
\SI{200}{\celsius}, \SI{1}{\hour}		& \num[separate-uncertainty]{36.4\pm2.5}	 		& \num[separate-uncertainty]{0.661\pm0.074}	 	& \num[separate-uncertainty]{12.8\pm0.9} 	& \num[separate-uncertainty]{0.088\pm0.00008}	 	\\ 
\hspace{1em}
\SI{220}{} (pads) / \SI{230}{\celsius} (pillars), \SI{2}{\hour}		& \num[separate-uncertainty]{31.0\pm3.5}	 		& \num[separate-uncertainty]{0.400\pm0.072}	 	& \num[separate-uncertainty]{26.9\pm0.4}		& \num[separate-uncertainty]{0.118\pm0.004}	 	\\ 

\hline
\rule{0pt}{1.2em}\textbf{LIFT (melt)} 	&									&								&								&		\\
\hspace{1em}
Cu								& \num[separate-uncertainty]{73.2\pm10.4}			& \num[separate-uncertainty]{1.66\pm0.37}	 	& \num[separate-uncertainty]{49.8\pm7.4}	 	& \num[separate-uncertainty]{0.415\pm0.028}	 	\\
\hspace{1em}
Au								& \num[separate-uncertainty]{24.3\pm10.1}			& \num[separate-uncertainty]{0.293\pm0.235}	 	& \num[separate-uncertainty]{28.3\pm3.5}	 	& \num[separate-uncertainty]{0.186\pm0.020}	 	\\

\hline
\rule{0pt}{1.2em}\textbf{MCED}	 	&									&								&								&		\\
\hspace{1em}
ad.							& \num[separate-uncertainty]{121.8\pm7.4}			& \num[separate-uncertainty]{2.71\pm0.36}	 	& \num[separate-uncertainty]{114.2\pm3.6}	 	& \num[separate-uncertainty]{0.774\pm0.104}	 	\\ 

\hline
\rule{0pt}{1.2em}\textbf{FluidFM}	 	&									&								&								&		\\
\hspace{1em}
ad.							& \num[separate-uncertainty]{138.4\pm18.7}			& \num[separate-uncertainty]{2.28\pm0.45}	 	& \num[separate-uncertainty]{134\pm11}	 	& \num[separate-uncertainty]{0.962\pm0.026}	 	\\ 

\hline
\rule{0pt}{1.2em}\textbf{EHD-RP}	 	&									&								&								&		\\
\hspace{1em}
ad.							& \num[separate-uncertainty]{80.4\pm17.9}			& \num[separate-uncertainty]{2.22\pm0.82}	 	& \num[separate-uncertainty]{81.7\pm8.4}	 	& \num[separate-uncertainty]{1.10\pm0.12}	 	\\ 
\hspace{1em}
ad., pillars \diameter$\approx$\SI{170}{\nano\meter}& N/A			& N/A		& \num[separate-uncertainty]{88.4}		&	\num[separate-uncertainty]{1.38\pm0.06}	\\
\hline
\rule{0pt}{1.2em}\textbf{FIBID}	 	&									&								&								&		\\
\hspace{1em}
ad.							& \num[separate-uncertainty]{140\pm7.9}			& \num[separate-uncertainty]{9.42\pm0.11}	 	& \num[separate-uncertainty]{95.3\pm9.5}	 	& \num[separate-uncertainty]{2.64\pm0.25}	 	\\ 

\hline
\rule{0pt}{1.2em}\textbf{FEBID}	 	&									&								&								&		\\
\hspace{1em}
ad.							& \num[separate-uncertainty]{75.5\pm1.6}			& \num[separate-uncertainty]{6.01\pm0.32}	 	& \num[separate-uncertainty]{59.3\pm3.9}	 	& \num[separate-uncertainty]{2.65\pm0.04}	 	\\ 

\hline
\rule{0pt}{1.2em}\textbf{cryo-FEBID}	 	&									&								&								&		\\
\hspace{1em}
ad.							& \num[separate-uncertainty]{13.8\pm1.0}			& \num[separate-uncertainty]{0.843\pm0.074}	 	& \num[separate-uncertainty]{3.85\pm0.93}	 	& \num[separate-uncertainty]{0.100\pm0.059}	 	\\

\bottomrule
\end{tabular}
\end{table*}

\end{multicols}

\begin{figure*}[htbp] 
   \centering
   \includegraphics[width=178mm]{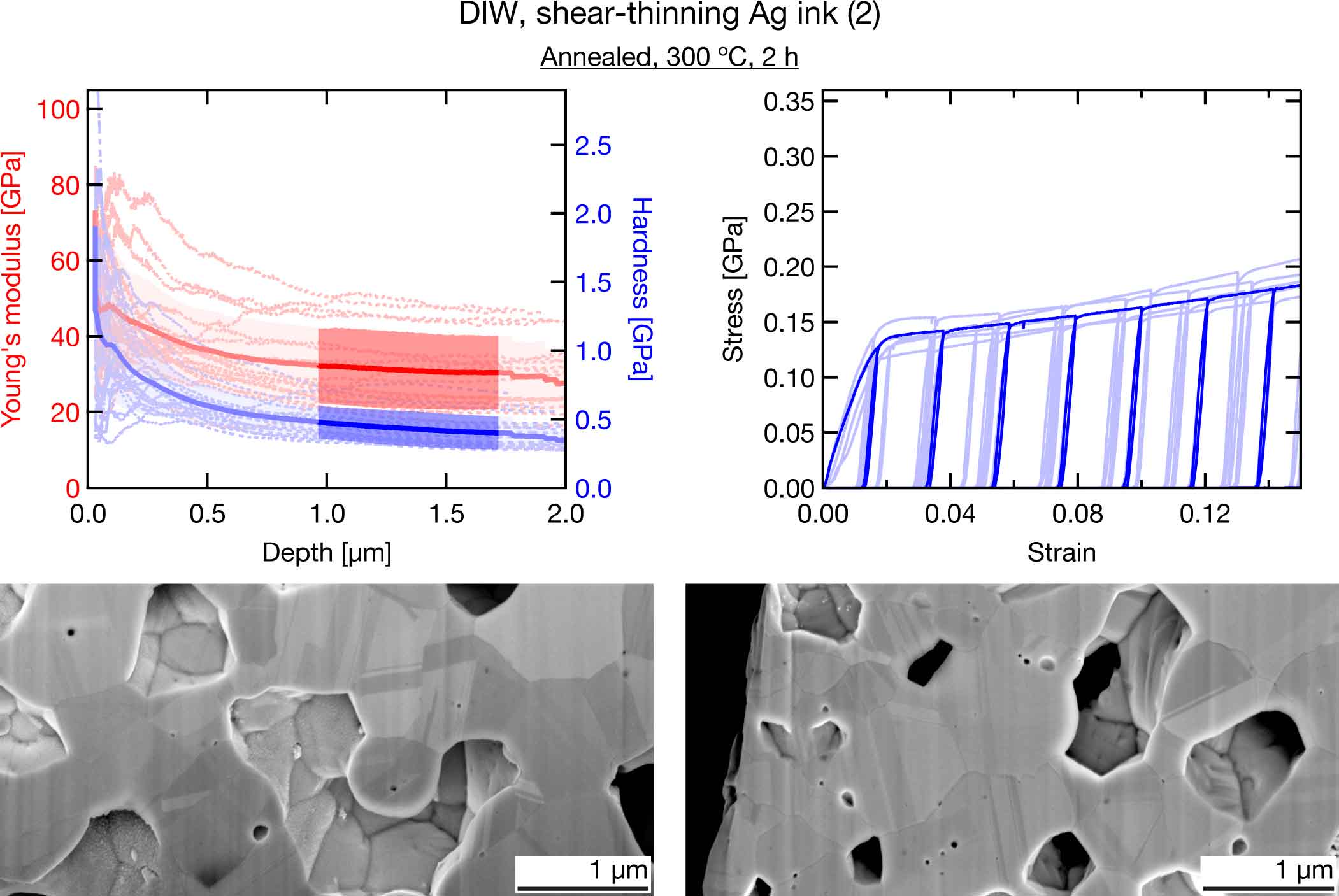} 
   \caption{Direct ink writing, shear-thinning Ag ink (DIW (s.t.)).}
\label{fig:SI_Nanjia2}
\end{figure*}

\begin{figure*}[htbp] 
   \centering
   \includegraphics[width=178mm]{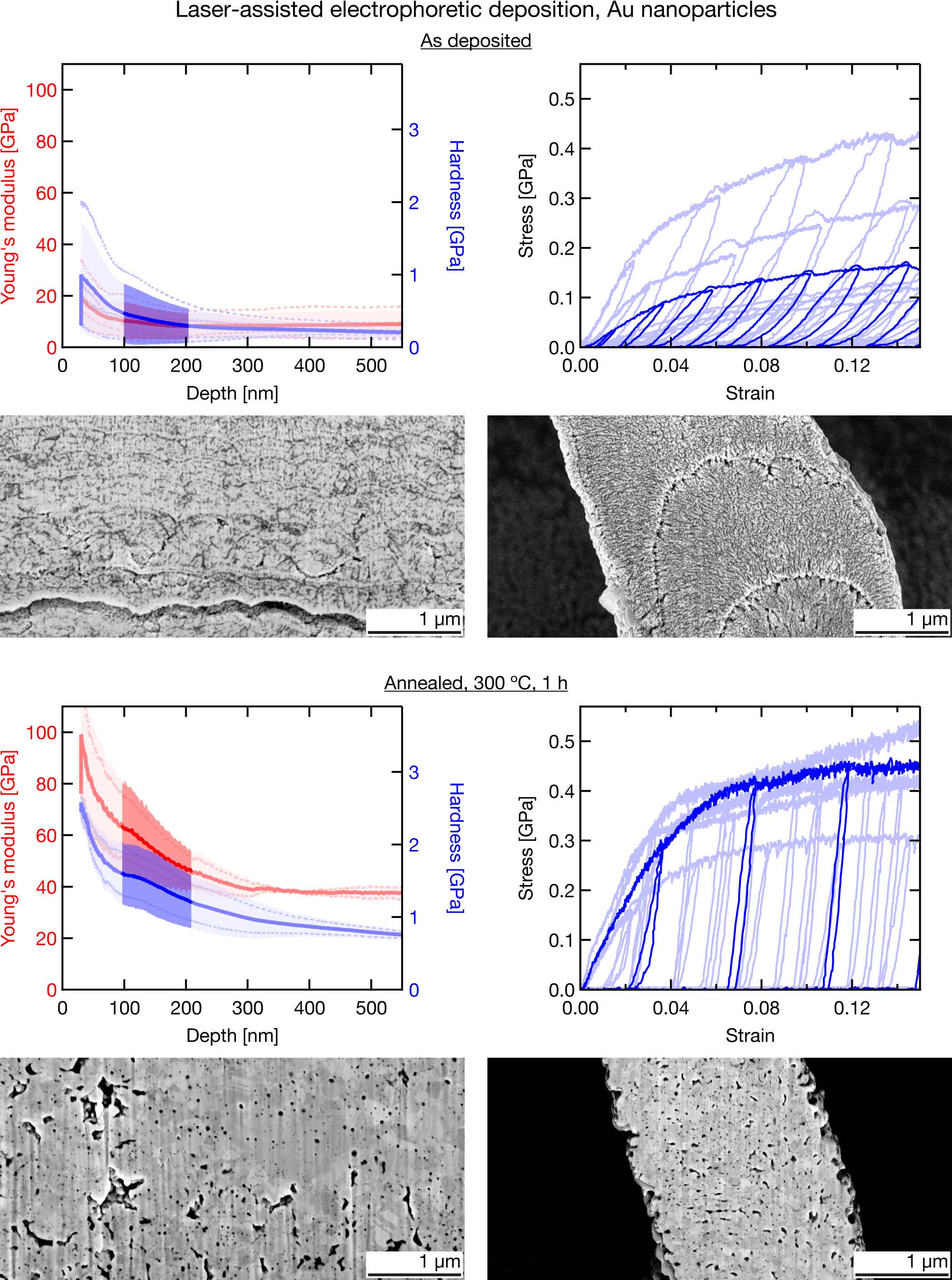} 
   \caption{Laser-assisted electrophoretic deposition (LAEPD), Au nanoparticles.}
\label{fig:SI_Iwata}
\end{figure*}

\begin{figure*}[htbp] 
   \centering
   \includegraphics[width=178mm]{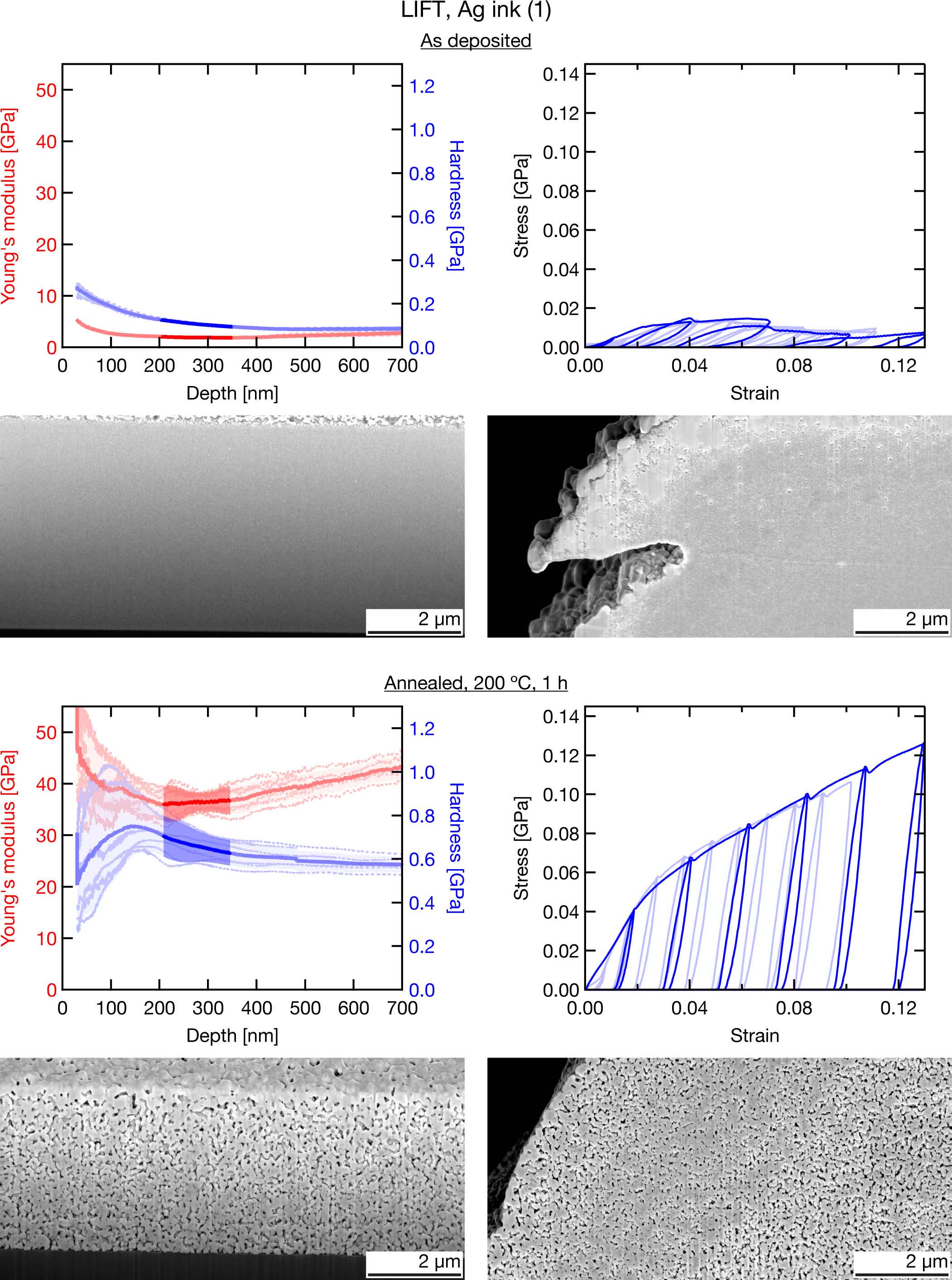} 
   \caption{Laser-induced forward transfer of Ag ink (LIFT (ink)).}
\label{fig:SI_Pique1}
\end{figure*}

\begin{figure*}[htbp] 
   \centering
   \includegraphics[width=178mm]{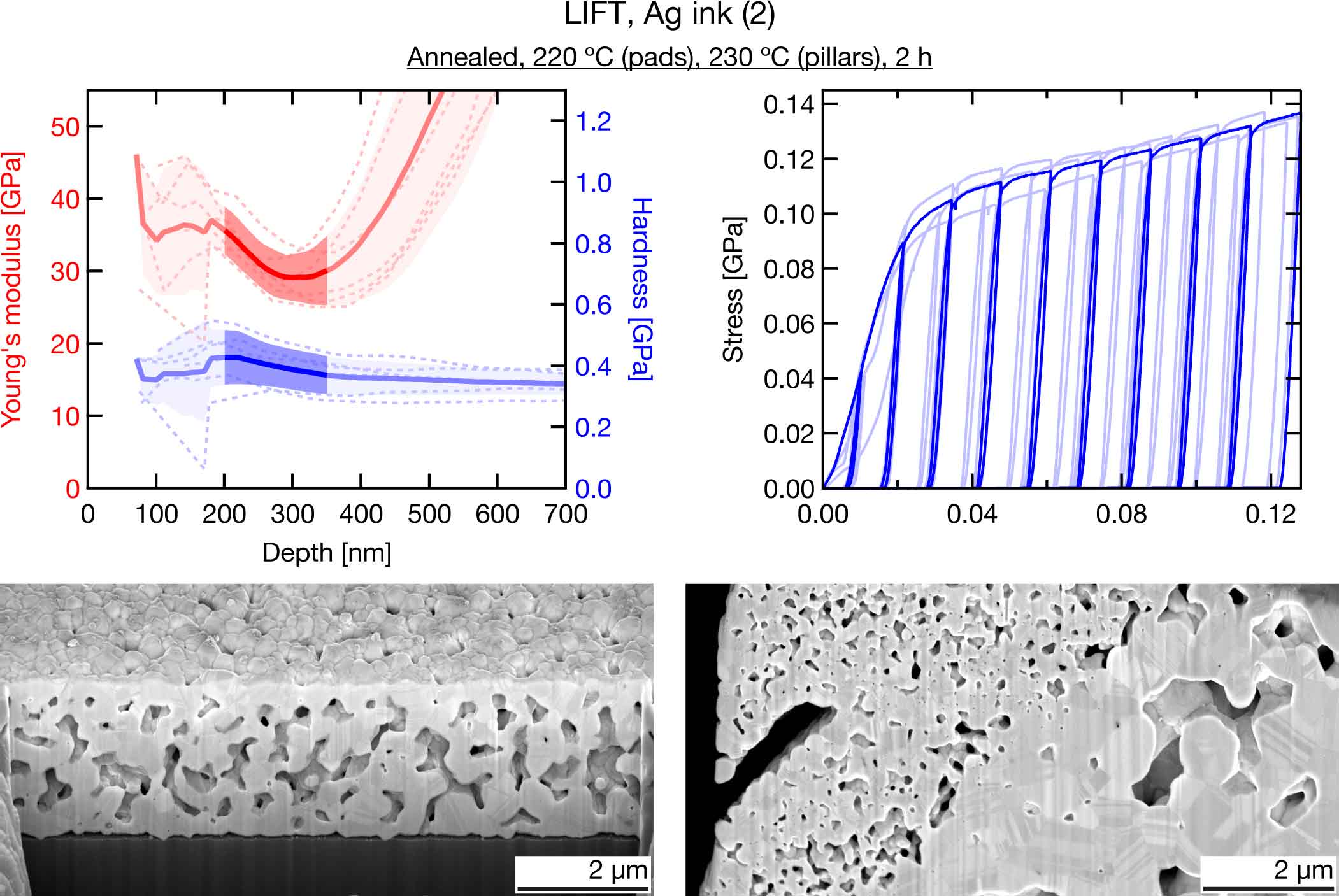} 
   \caption{Laser-induced forward transfer of Ag ink (LIFT (ink)).}
\label{fig:SI_Pique2}
\end{figure*}

\begin{figure*}[htbp] 
   \centering
   \includegraphics[width=178mm]{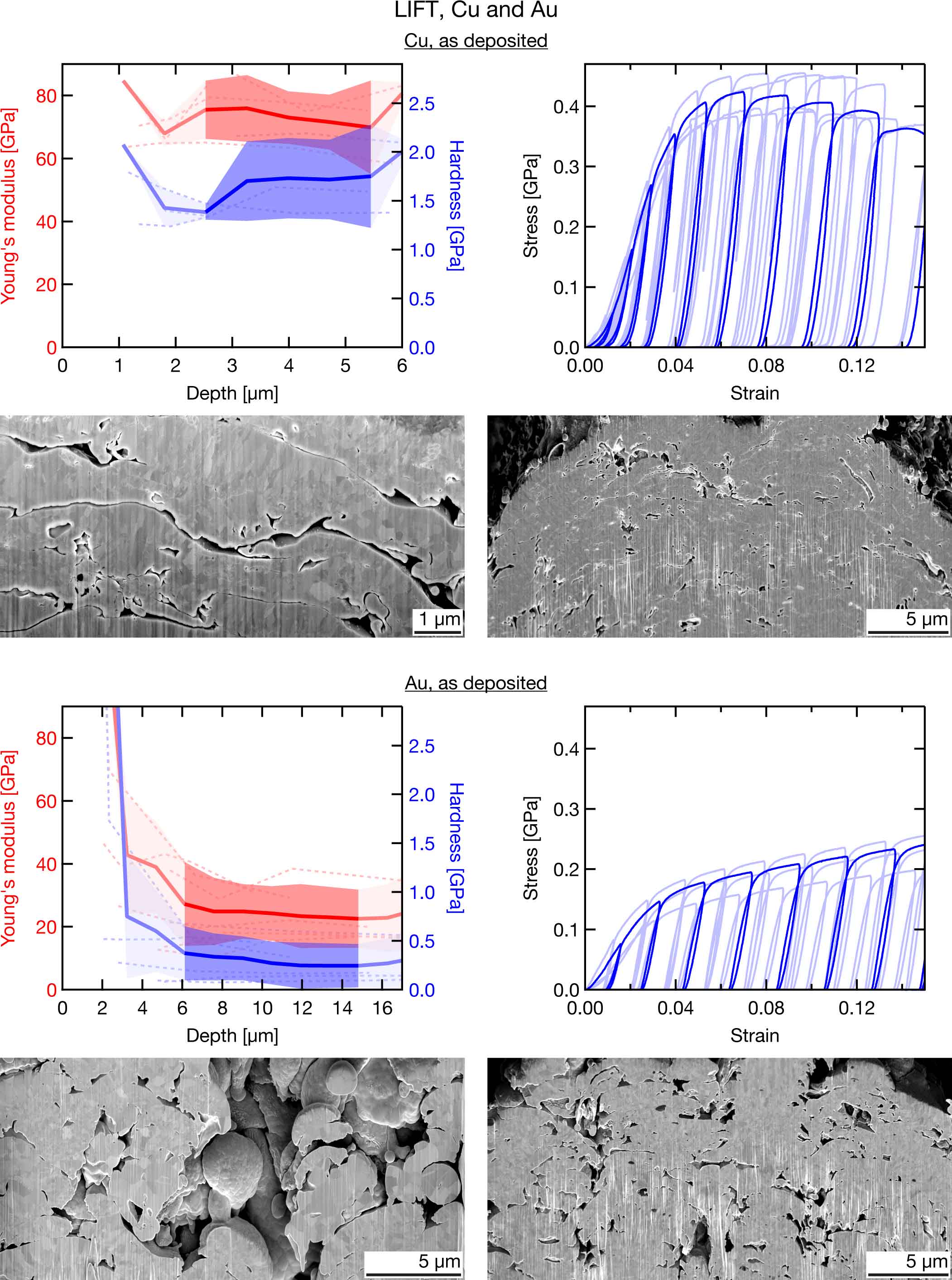} 
   \caption{Laser-induced forward transfer of Cu and Au (LIFT).}
\label{fig:SI_LIFT}
\end{figure*}

\begin{figure*}[htbp] 
   \centering
   \includegraphics[width=178mm]{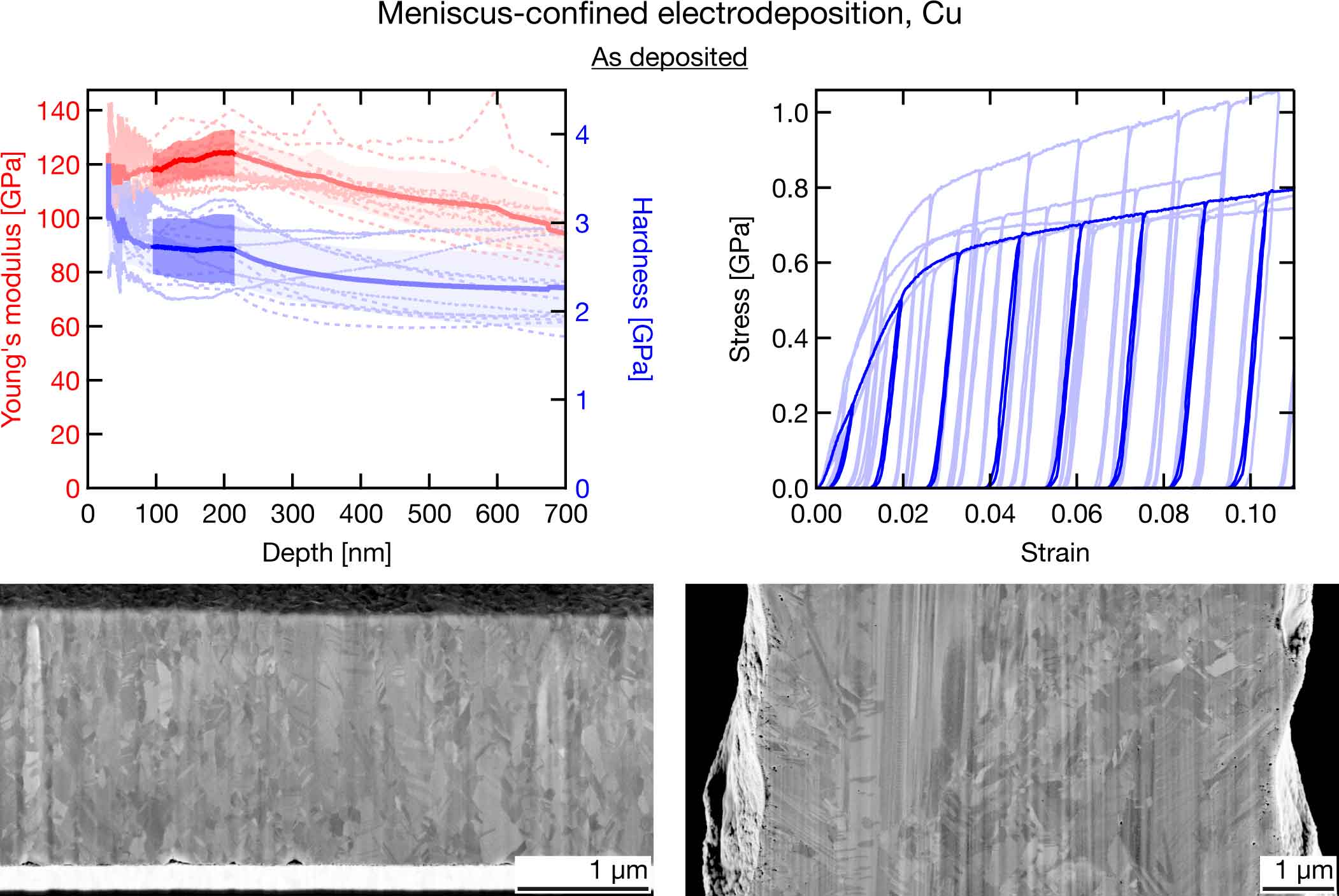} 
   \caption{Meniscus-confined electrodeposition (MCED) of Cu.}
\label{fig:SI_MC}
\end{figure*}

\begin{figure*}[htbp] 
   \centering
   \includegraphics[width=178mm]{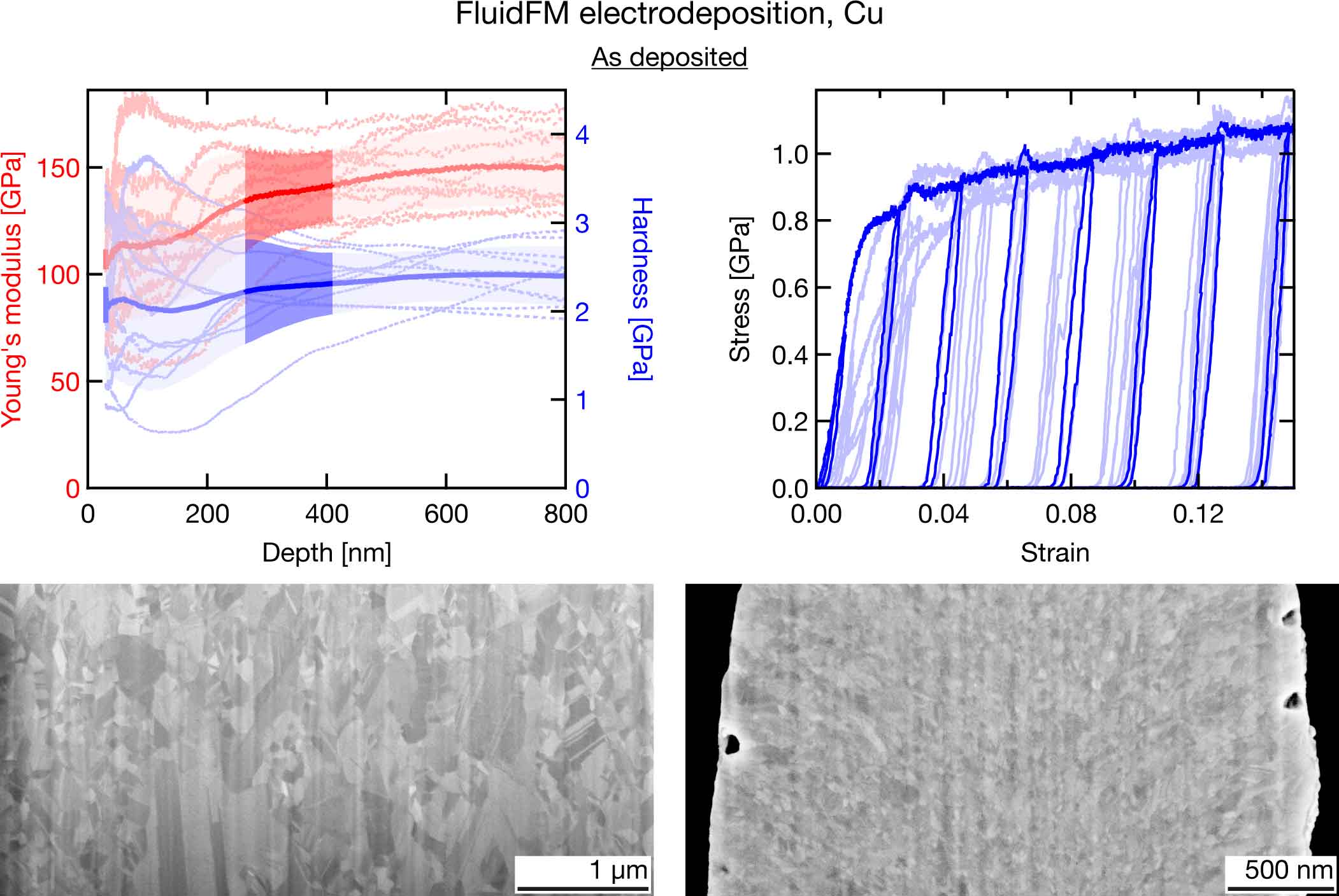} 
   \caption{FluidFM electrodeposition of Cu.}
\label{fig:SI_FluidFM}
\end{figure*}

\begin{figure*}[htbp] 
   \centering
   \includegraphics[width=178mm]{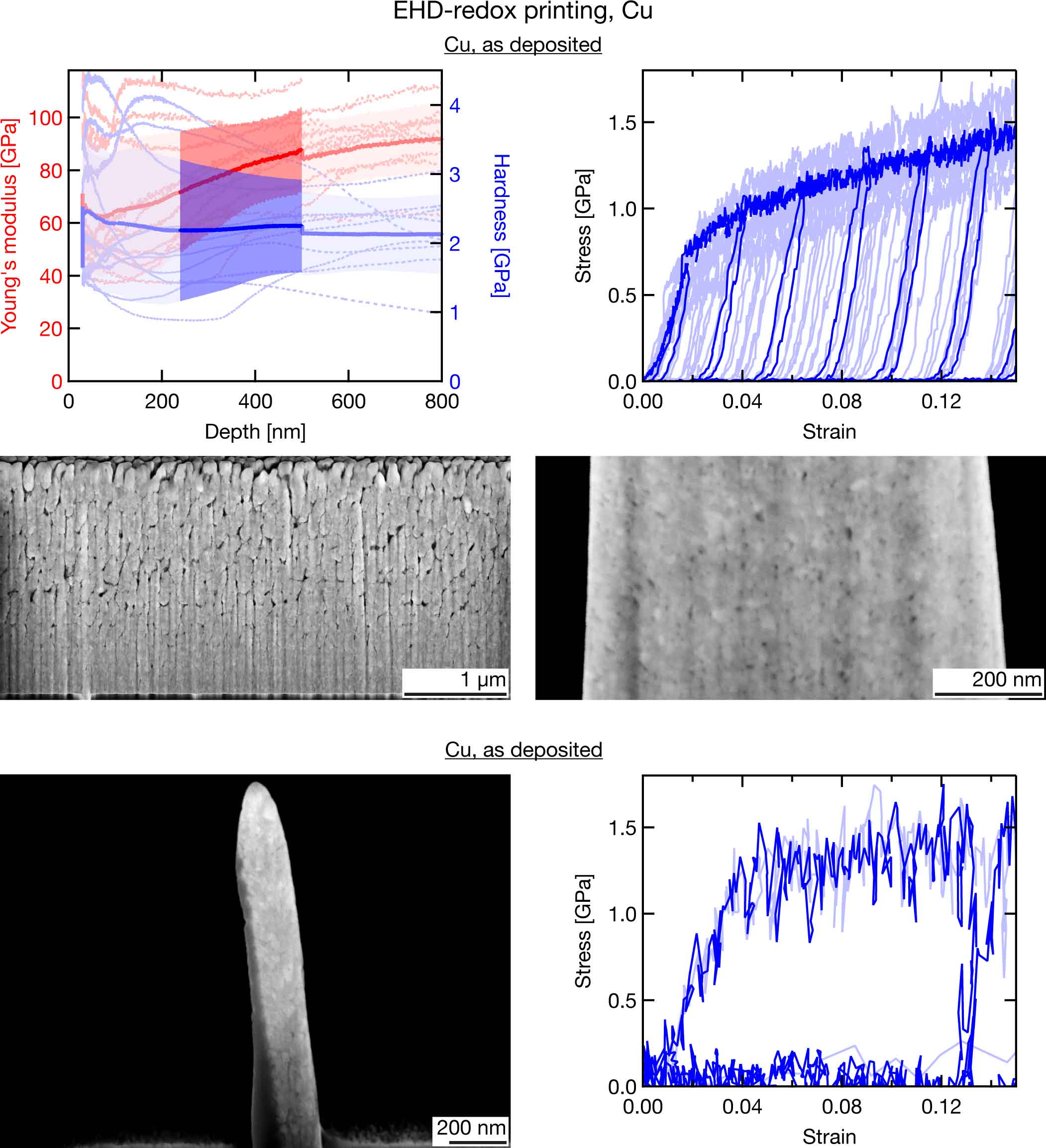} 
   \caption[Study of mechanical properties: EHD-RP]{Electrohydrodynamic redox printing (EHD-RP) of Cu. Bottom: representative cross-section and microcompression data for two Cu pillars of 160~\==~\SI{177}{\nano\meter} in diameter.}
\label{fig:SI_EHDRP}
\end{figure*}

\begin{figure*}[htbp] 
   \centering
   \includegraphics[width=178mm]{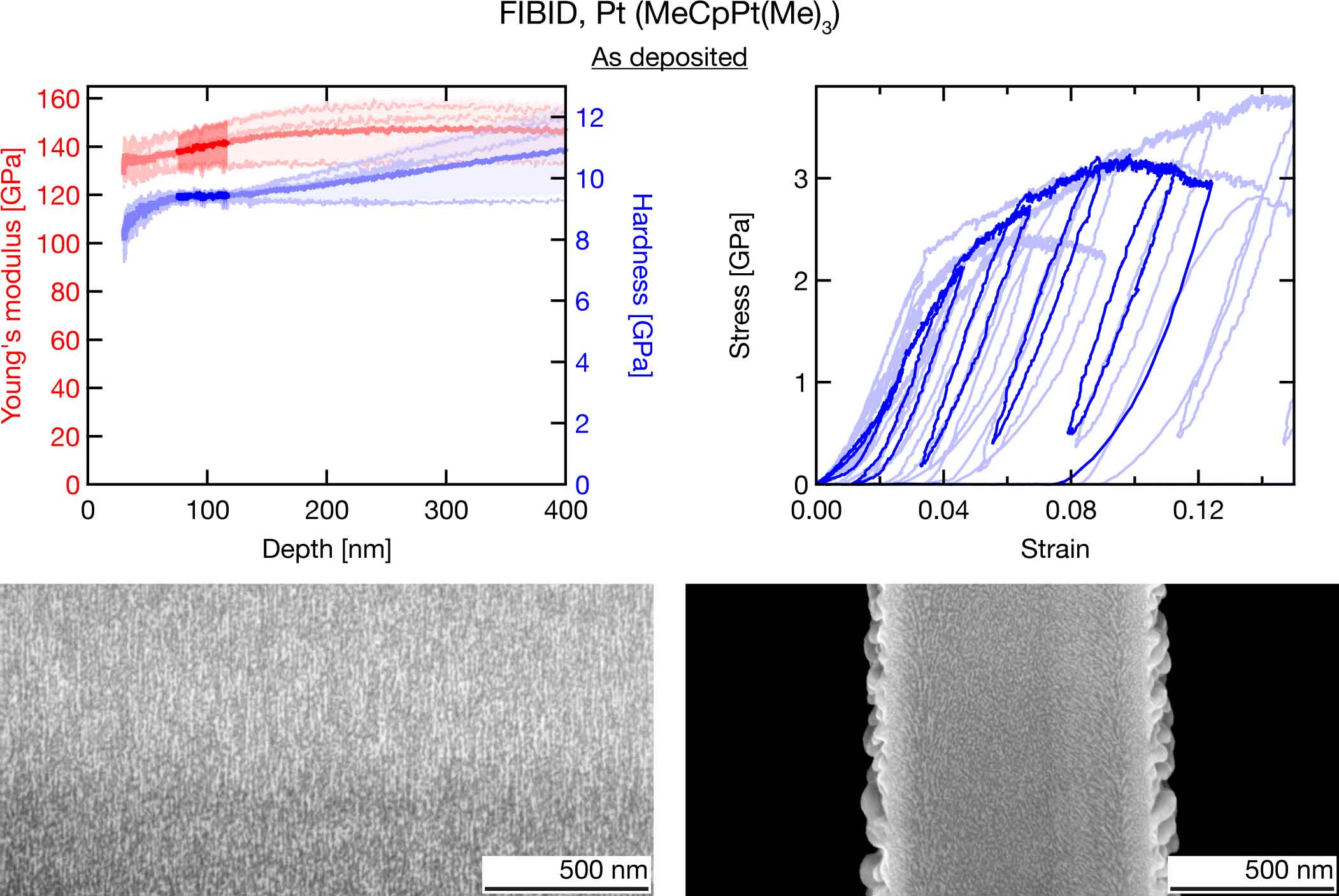} 
   \caption{Focused ion beam-induced deposition (FIBID) of Pt (MeCpPt(Me)$_3$).}
\label{fig:SI_FIBID}
\end{figure*}

\begin{figure*}[htbp] 
   \centering
   \includegraphics[width=178mm]{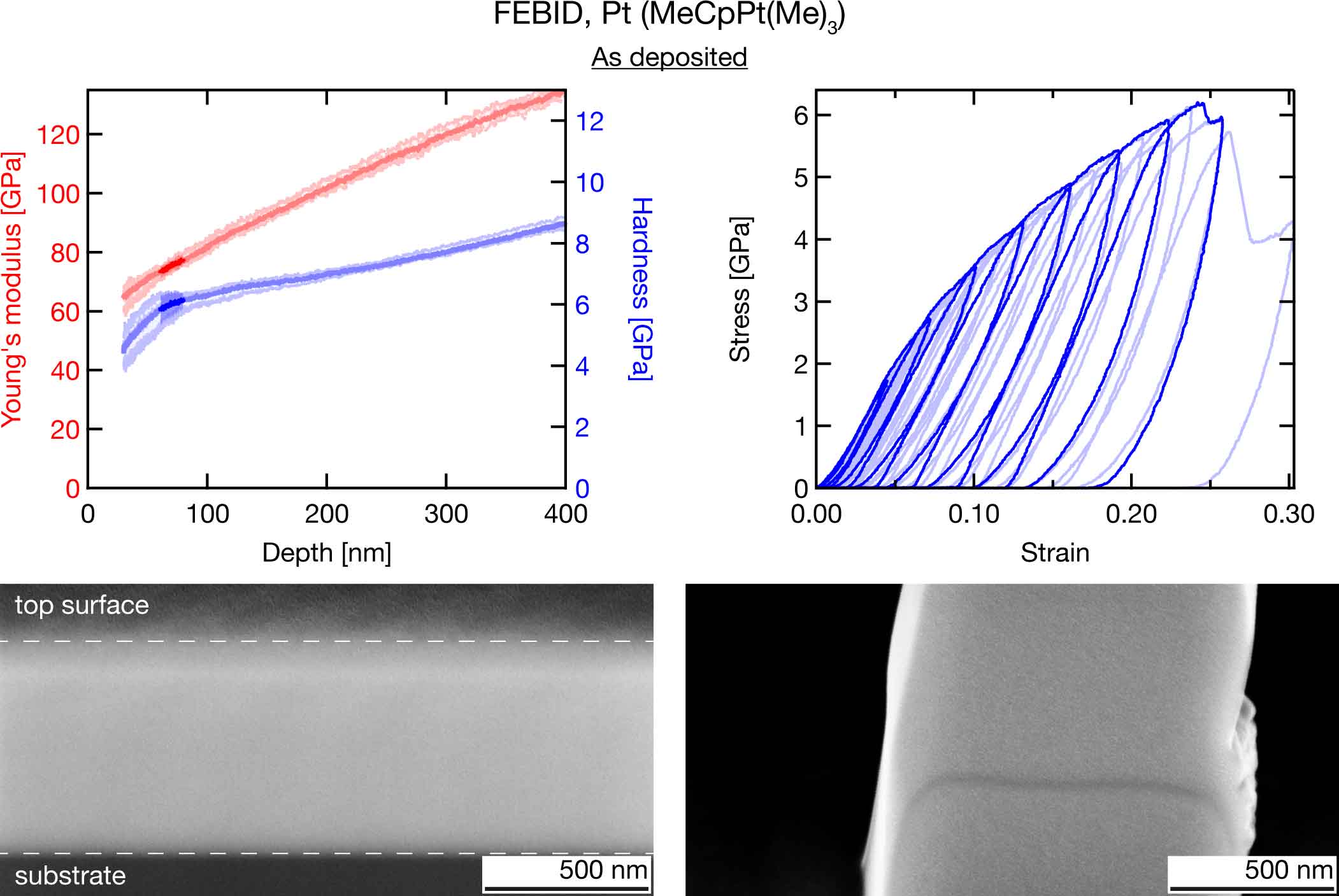} 
   \caption{Focused electron beam-induced deposition (FEBID) of Pt (MeCpPt(Me)$_3$). {The line feature seen in the pillar cross section is due to an interruption of the automated exposure pattern to mark the smooth growth front at this moment.}}
\label{fig:SI_FEBID}
\end{figure*}

\clearpage
\subsection{Mechanical literature data of thin films}
\begin{table}[htbp]
\caption[Study of mechanical properties: mechanical literature data of thin films]{\textbf{Mechanical literature data of thin films.} Literature values for $E$, $H$ (measured by nanoindentation, Berkovich indenter) and $\sigma_\text{y}$ (measured by microcompression and microtension) of polycrystalline thin films fabricated by traditional deposition techniques (PVD and electrodeposition). The data measured for the materials deposited by the studied small-scale AM methods was normalized to the values highlighted in \textbf{bold}.}\label{tab:SI_litval}
\centering
\begin{tabular}{l | c | c | c }
\toprule
Material							& $E$ [GPa]									& $H$ [GPa] 								& $\sigma_\text{y}$ [GPa] 	\\ \hline
\rule{0pt}{1.2em}\textbf{Ag}			& \textbf{80.5}\cite{KOSTER2014} (bulk), $\sim$80\cite{Shugurov2004a,Panin2005}		& 0.7~\==~1.5\cite{Shugurov2004a,Panin2005}	, \textbf{1.2}	& N/A					\\	 	
\hline
\rule{0pt}{1.2em}\textbf{Au}			& \textbf{80.2}\cite{KOSTER2014} (bulk), 55~\==~88.6\cite{Okuda1999,Volinsky2004}	& 1~\==~2\cite{Okuda1999,Volinsky2004}, \textbf{1.2}	& 0.04~\==~0.8\cite{Emery2003,Emery2003a,Chen2016,Yanagida2017}	\\	
\hline
\rule{0pt}{1.2em}\textbf{Cu}			& \textbf{125}\cite{KOSTER2014} (bulk), 88~\==~135\cite{Fang2003,Lu2005,Hong2005,Chang2007}		& 1.6~\==~3.5\cite{Beegan2003,Lu2005,Beegan2007,Chang2007}, \textbf{2}		& 0.2~\==~1.2\cite{Dao2006,Jang2011,Mieszala2016}	\\
\hline
\rule{0pt}{1.2em}\textbf{Pt}	& \textbf{174}\cite{KOSTER2014} (bulk), 150~\==~178\cite{Darling1966,Merker2001,Nili2014}	& 1.5~\==~8.6\cite{Nili2014,Mall2009,Mencik2006}, \textbf{4} 	& N/A\\
\hline
\rule{0pt}{1.2em}\textbf{amorphous C}	& 132 (bulk), 40~\==~760\cite{Jiang1989a,Weiler1996,Cho2005}	& 5~\==~30\cite{Jiang1989a,Weiler1996}		& 7\cite{Cho2005}	\\
		
\bottomrule
\end{tabular}
\end{table}

\end{document}